\documentclass[english]{article}
\usepackage{lmodern}
\usepackage[T1]{fontenc}
\usepackage[latin9]{inputenc}
\usepackage[letterpaper]{geometry}
\usepackage{color}

\usepackage{textcomp}

\usepackage[authoryear]{natbib}
\usepackage{subfigure}

\usepackage{amsmath}
\usepackage{amssymb}
\usepackage{amsfonts}
\usepackage{multirow}
\usepackage{array,epsfig,fancyheadings,rotating}
\usepackage[]{hyperref}  %<----modified by Ivan
\usepackage{sectsty, secdot}

\usepackage{graphicx}
\usepackage{amsthm}
\usepackage{mathtools}
\usepackage{booktabs}
\usepackage{lscape}
\usepackage{xcolor}
\usepackage{tabularx}
\usepackage{url}
\usepackage{threeparttable}
\usepackage{pdflscape}
\usepackage{bbm}
\usepackage{marginnote}
\usepackage{lipsum}
\usepackage{lscape}
\usepackage{bbm}
\usepackage{latexsym}
\usepackage{algorithm}
\usepackage{algorithmic}
\usepackage{mathrsfs}
\usepackage{changes}

\usepackage{indentfirst}
\makeatletter
\@ifundefined{definecolor}
 {\usepackage{color}}{}

\newtheorem{theorem}{Theorem}
\newtheorem{lemma}{Lemma}

\newtheorem{proposition}{Proposition}

\newtheorem{remark}{Remark}
\newtheorem{assum}{Assumption}

\newcommand{\lyxmathsym}[1]{\ifmmode\begingroup\def\b@ld{bold}
  \text{\ifx\math@version\b@ld\bfseries\fi#1}\endgroup\else#1\fi}
\makeatother

\usepackage{babel}

\newcommand{\be}{\mbox{\boldmath $e$}}

\begin{document}

\renewcommand{\baselinestretch}{1.5}
\fontsize{12}{14pt plus.8pt minus .6pt}\selectfont \vspace{0.8pc}

\begin{center}
 \large\bf
Quasi-maximum likelihood estimation of break point in high-dimensional factor models
\end{center}

\vspace{.4cm} \centerline{ Jiangtao Duan\textsuperscript{1}, Jushan Bai\textsuperscript{2}, Xu Han\textsuperscript{3}  } \vspace{.4cm} \centerline{\it
\textsuperscript{1}Northeast Normal University, \textsuperscript{2}Columbia University and \textsuperscript{3}City University of Hong Kong}
%\begin{center}
% \larg
%\end{center}
%\begin{center}
%\textsuperscript{1}Northeast Normal University\\
%\textsuperscript{2}Columbia University \\
%\textsuperscript{3}City University of Hong Kong
%\end{center}
 \vspace{.55cm} \fontsize{9}{11.5pt plus.8pt minus
.6pt}\selectfont

\begin{quotation}
\noindent {\it Abstract:}

This paper estimates the break point for large-dimensional factor models with a single structural break in
factor loadings at a common unknown date. We propose a quasi-maximum likelihood (QML) estimator of
the change point based on the second moments of factors, which are estimated by a single principal component analysis.
We show that the QML estimator is consistent for the true break point  when the covariance matrix of the
pre- or post-break factor loading (or both) is singular. Consistency here means that the deviation of the estimated break date
from the actual break date $k_0$ converges to zero as the sample size grows. This is a much stronger result than the break fraction
$\hat k/T$ being $T$-consistent (super-consistent) for $k_0/T$. Also, singularity occurs for most types of
structural changes, except for a rotational change. Even for a notational change, the QML estimator is still $T$-consistent
in terms of the break fraction. Simulation results confirm the theoretical properties of this estimator, and in fact QML significantly outperforms existing estimators for change points in factor models. Finally we apply the method to estimate the break points in a U.S. macroeconomic dataset and a stock return dataset.

\vspace{9pt}
\noindent {\it Key words and phrases:}
Structural break,
High-dimensional factor models,
Factor loadings\par
\end{quotation}\par
\vspace{-1em}
\section{Introduction}

Large factor models assume that a few factors can capture the common driving forces of a large number of economic variables.
Although factor models are useful, practitioners have to be cautious about the potential structural changes. For example, either the number of factors or the factor loadings may change over time. This
concern is empirically relevant because parameter instability is pervasive in large-scale panel data.

So far, many methods have been developed to test structural breaks in factor models (e.g., \cite{Stock2008}, \cite{Breitung2011}, and \cite{Chen2014}). The rejection of the null hypothesis of no
structural change leads to the subsequent issues of how to estimate the change point, determine the numbers of pre- and post-break factors, and estimate the factor space. \cite{Chen2015} considers a
least-squares estimator of the break point and proves the consistency of the estimated break fraction (i.e., the break date $k$ divided by the full time series $T$, $\frac{k}{T}$).
\cite{Cheng2016} propose a shrinkage method to obtain a consistent estimator of the break fraction.
\cite{Baltagi2017} develop a least-squares estimator of the change point based on the second moments of the estimated pseudo-factors and show that the estimation error of the proposed estimator is
$O_p(1)$, which indicates the consistency of the estimated break fraction. A few recent studies also explore a consistent estimation of break points, which is technically more challenging.
\cite{Ma_Su2018} develop an adaptive fused group Lasso method to consistently estimate all break points under a multibreak setup.
\cite{Barigozzi2018} propose a method based on wavelet transformations to consistently estimate the number and locations of break points in the common and idiosyncratic components.
\cite{Bai2017} establish the consistency of the least-squares estimator of the break point in large factor models when factor loadings are subjected to a structural break and the size of the break is
shrinking as the sample size increases. Although the estimators proposed in these studies are consistent under certain assumptions, the simulation results show that they perform poorly when (1) the
number of factors changes after the break or (2) the loading matrix undergoes a rotational type of change.

According to the factor model literature, a factor model with a break in factor loadings is observationally equivalent to that with constant loadings and possibly more pseudo-factors (e.g.,
\cite{Han2015} and \cite{Bai2016}). Thus, the estimation of the change point of factor loadings can be converted into that of the change point of the second moment of the pseudo-factors. We propose a
quasi-maximum likelihood (QML) method to estimate the break point based on the second moment of the estimated pseudo-factors; therefore, the number of original factors is not required to be known for
computing our estimator. First, we estimate the number of pseudo-factors (defined as the factors in the equivalent representation that ignores the break), and then estimate the pre- and
post-break second moment matrices of the estimated pseudo-factors for all possible sample splits. The structural break date is estimated by minimizing the QML function among all possible split points.

This paper makes the following contributions to the literature. First, we establish the consistency of the QML break point estimator if the break leads to more pseudo-factors than the original pre- or
post-break factors. This occurs when the break augments the factor space or in the presence of disappearing or emerging factors. Under these circumstances, the covariance matrix of loadings on the pre-
or post-break pseudo-factors is singular, which is the key condition to establish the consistency of our QML estimator. To the best of our knowledge, this is the first study that links the consistency
of the break point estimator to the singularity of covariance matrices of loadings on pre- and post-break pseudo-factors. In addition, we prove that the difference between the estimated and true change
points is stochastically bounded when both pre- and post-break loadings on the pseudo-factors have nonsingular covariance matrices. In this case, the loading matrix only undergoes a rotational change,
and both the numbers of pre- and post-break original factors are equal to the number of pseudo-factors.

The aforementioned singularity leads to a technical challenge of analyzing the asymptotic property. The singular population covariance matrix of the pre(post)-break loadings has a zero determinant,
whose logarithm is undefined. To resolve this issue, we show that the estimated covariance matrices have nonzero determinants and a well-defined inverse for any given sample size, by obtaining the
convergence rate of the lower bound of their smallest eigenvalues. This ensures that the objective function based on the estimated covariance is appropriately defined in any finite sample.

Our second major contribution is that the QML method allows a change in the number of factors. Namely, it allows for disappearing or emerging factors after the break. This is an advantage over the
methods developed by \cite{Ma_Su2018} and \cite{Bai2017}, who assume that the number of factors remains constant after the break. Our simulation result indicates that the estimator proposed by
\cite{Bai2017} is inconsistent when some factors disappear and the remaining factors have time-invariant loadings. \cite{Baltagi2017} allow a change in the number of factors; however, their estimation
error was only stochastically bounded. In contrast, our QML estimator remains consistent under a varying number of factors.

Finally, the QML method has a substantial computational advantage over the estimators that iteratively implement high-dimensional principal component analysis (PCA). For example, the estimator proposed
by \cite{Bai2017} runs PCA for pre- and post-split sample covariance matrices for all possible split points. In comparison, our QML runs PCA for the entire sample only once, and thus, is computationally
more efficient, especially in large samples.

The rest of this paper is organized as follows. Section 2 introduces the factor model with a single break on the factor loading matrix and describes the QML estimator for the break date. Section 3
presents the assumptions made for this model. Section 4 presents the consistency and asymptotic distribution of the QLM estimator for the break date.
Section 5 investigates the finite-sample properties of the QML estimator through simulations. Section 6 implements the proposed method to estimate the break points in a monthly macroeconomic dataset of
the United States and a dataset of weekly stock returns of Nasdaq 100 components. Section 7 concludes the study.

The following notations will be used throughout the paper. Let $\rho_i(\mathbb{B})$ denote the $i$-th eigenvalue of an $n\times n$ symmetric matrix $\mathbb{B}$, and $\rho_1(\mathbb{B})\geq \rho_2(\mathbb{B})\geq \cdots \geq \rho_n(\mathbb{B})$.
For an $m\times n$ real matrix $\mathbb{A}$, we denote its Frobenius norm as $\|\mathbb{A}\|= [tr(\mathbb{A}\mathbb{A}^{'})]^{1/2}$, its MP inverse as $\mathbb{A}^{-}$, its $i$-th singular value as $\sigma_{i}(\mathbb{A})$, and its adjoint matrix as $\mathbb{A}^{\#}$ when $m=n$. Let $\mathrm{Proj}(\mathbb{A}|\mathbb{Z})$ denote the projection of matrix $\mathbb{A}$ onto the columns of matrix $\mathbb{Z}$. For a real number $x$, $[x]$ represents the integer part of $x$.

\vspace{-1em}
\section{Model and estimator}

Let us consider the following factor model with a common break at $k_0$ in the factor loadings for $i=1,\cdots,N$:
\begin{eqnarray}
x_{it}=
\begin{cases}
\lambda_{i1}f_{t}+e_{it} & for~~  t=1,2,\cdots,k_0(T) \cr
\lambda_{i2}f_{t}+e_{it} & for~~  t=k_0(T)+1,\cdots,T,
\end{cases}
\label{model_1}
\end{eqnarray}
where $f_t$ is an $r-$dimensional vector of unobserved common factors; $r$ is the number of pseudo-factors; $k_0(T)$ is the unknown break date; $\lambda_{i1}$ and $\lambda_{i2}$ are the pre- and
post-break factor loadings, respectively; and $e_{it}$ is the error term allowed to have serial and cross-sectional dependence as well as heteroskedasticity.
$\tau_0\in (0,1)$ is a fixed constant and $[x]$ represents the integer part of $x$. For notational simplicity, hereinafter, we suppress the dependence of $k_0$ on $T$. Note that the
dimension of $f_t$ is the same as that of the pseudo-factors (to be defined soon) instead of the original underlying factors. This formulation simplifies the representation of various types of breaks in a unified framework, which
will be clarified in the examples below.
%we formulate the model using pseudo-factors instead of the original underlying factors.

In vector form, model (\ref{model_1}) can be expressed as
\begin{eqnarray}
x_{t}=
\begin{cases}
\Lambda_{1}f_{t}+e_{t} & for~~ t=1,2,\cdots,k_0 \cr
\Lambda_{2}f_{t}+e_{t} & for~~ t=k_0+1,\cdots,T,
\end{cases}
\label{model_vector}
\end{eqnarray}
where $x_t=[ x_{1t},\cdots,x_{Nt} ]^{'}$, $e_t=[ e_{1t},\cdots,e_{Nt} ]^{'}$, $\Lambda_1=[ \lambda_{11},\cdots,\lambda_{N1} ]^{'}$, and $\Lambda_2=[ \lambda_{12},\cdots,\lambda_{N2} ]^{'}$.

For any $k=1,\cdots,T-1$, we define
$$X_{k}^{(1)}=[x_1,\cdots,x_k]^{'},X_{k}^{(2)}=[x_{k+1},\cdots,x_T]^{'},$$
$$F_{k}^{(1)}=[f_1,\cdots,f_k]^{'},F_k^{(2)}=[f_{k+1},\cdots,f_T],$$
$$\be_{k}^{(1)}=[e_1,\cdots,e_k]^{'},\be_k^{(2)}=[e_{k+1},\cdots,e_T],$$
where the subscript $k$ denotes the date at which the sample is to be split, and the superscripts $(1)$ and $(2)$ denote the pre- and post-$k$ data, respectively. We rewrite (\ref{model_vector}) using
the following matrix representation:
\begin{eqnarray}
\left[
\begin{array}{cccccccccc}
X_{k_0}^{(1)}\\
X_{k_0}^{(2)}\\
\end{array}
\right]
&=&\left[
\begin{matrix}
F_{k_0}^{(1)}\Lambda_1^{'}\\
F_{k_0}^{(2)}\Lambda_2^{'}
\end{matrix}
\right]+
\left[
\begin{array}{cccccccccc}
e_{k_0}^{(1)}\\
e_{k_0}^{(2)}\\
\end{array}
\right]
=\left[
\begin{matrix}
F_{k_0}^{(1)}(\Lambda B)^{'}\\
F_{k_0}^{(2)}(\Lambda C)^{'}
\end{matrix}
\right]+
\left[
\begin{array}{cccccccccc}
e_{k_0}^{(1)}\\
e_{k_0}^{(2)}\\
\end{array}
\right],\nonumber \\
&=&\left[
\begin{matrix}
F_{k_0}^{(1)}B^{'}\\
F_{k_0}^{(2)}C^{'}\\
\end{matrix}
\right]
\Lambda^{'}+
\left[
\begin{array}{cccccccccc}
e_{k_0}^{(1)}\\
e_{k_0}^{(2)}\\
\end{array}
\right], \nonumber\\
&=&G\Lambda^{'}+E,
\label{Baltagi}
\end{eqnarray}
where $F_{k_0}^{(1)}$ and $F_{k_0}^{(2)}$ have dimensions $k_0\times r$ and $(T-k_0)\times r$, respectively, and $\Lambda$ is an $N\times r$ matrix with full column rank. The pre- and post-break
loadings are modeled as $\Lambda_1=\Lambda B$ and $\Lambda_2=\Lambda C$, respectively, where $B$ and $C$ are some $r\times r$ matrices. Both $\Lambda_1$ and $\Lambda_2$ have dimension $N\times r$.
%Model (\ref{Baltagi}) is an equivalent model with constant factor loadings and a change in factor dynamics.
In this model, $r_1=rank(B)\leq r$ and $r_2=rank(C)\leq r$ denote the numbers of \emph{original factors} before and after the break, respectively. We refer to $G$ in (\ref{Baltagi}) as the \emph{pseudo-factors} because the last line of (\ref{Baltagi}) provides an observationally equivalent
representation without a change in the loadings matrix $\Lambda$. In other words, if the break is ignored in the estimation process, then the factors being estimated by a full-sample PCA are actually
the pseudo-factors $G$ in  (\ref{Baltagi}).
It is well known that the break can augment the factor space; thus, $r_1 \leq r$ and $r_2 \leq r$, with $rank(G)=r$.
Our representation in (\ref{Baltagi}) allows for changes in the factor loadings
and the number of factors. Below, several examples are provided
to illustrate that the pseudo-factor representation in (\ref{Baltagi}) is general
enough to cover three types of breaks.

\textbf{Type 1}. Both $B$ and $C$ are singular. In this case, the
number of original factors is strictly less than that of the pseudo-factors
both before and after the break (i.e., $r_{1}<r$ and $r_{2}<r$).
This means that the structural break in the factor loadings augments
the dimension of the factor space. Let us consider the following example.

Example (1): Let $\mathbb{F}_{k_{0}}^{(1)}$$(k_{0}\times r_{1})$
and $\mathbb{F}_{k_{0}}^{(2)}$$((T-k_{0})\times r_{2})$ denote the
original factors before and after the break, respectively, and $\Theta_{1}$
and $\Theta_{2}$ denote the pre- and post-break loadings on these
factors. Thus, this model can be represented and transformed as \begin{eqnarray}
\left[\begin{array}{c}
X_{k_{0}}^{(1)}\\
X_{k_{0}}^{(2)}\end{array}\right] & = & \left[\begin{array}{c}
\mathbb{F}_{k_{0}}^{(1)}\Theta_{1}^{\prime}\\
\mathbb{F}_{k_{0}}^{(2)}\Theta_{2}^{\prime}\end{array}\right]+e=\left[\begin{array}{cc}
\mathbb{F}_{k_{0}}^{(1)} & 0\\
0 & \mathbb{F}_{k_{0}}^{(2)}\end{array}\right]\left[\begin{array}{c}
\Theta_{1}^{\prime}\\
\Theta_{2}^{\prime}\end{array}\right]+e\nonumber \\
 & = & \left[\begin{array}{c}
[\mathbb{F}_{k_{0}}^{(1)}\;\vdots\;*]B^{\prime}\\
{}[*\;\vdots\;\mathbb{F}_{k_{0}}^{(2)}]C^{\prime}\end{array}\right]\Lambda^{\prime}+e=\underbrace{\left[\begin{array}{c}
F_{k_{0}}^{(1)}B^{\prime}\\
F_{k_{0}}^{(2)}C^{\prime}\end{array}\right]}_{G}\Lambda^{\prime}+e,\label{eq:ex1}\end{eqnarray}
where $\Lambda=[\Theta_{1},\Theta_{2}]$, $B=diag(I_{r_{1}},0_{r_{2}\times r_{2}})$,
$C=diag(0_{r_{1}\times r_{1}},I_{r_{2}})$, $F_{k_{0}}^{(1)}=[\mathbb{F}_{k_{0}}^{(1)}\;\vdots\;*]$,
$F_{k_{0}}^{(2)}=[*\;\vdots\;\mathbb{F}_{k_{0}}^{(2)}]$, and the asterisk
denotes some unidentified numbers such that all rows in $F_{k_{0}}^{(1)}$
and $F_{k_{0}}^{(2)}$ have the same variance (to satisfy Assumption
\ref{factors} in Section 3). (Note that the asterisk entries are cancelled due to multiplication by zero in $B$ and $C$.) In the special case of $r_{1}=r_{2}$, $\Lambda$
is of full rank $2r_{1}$ (i.e., the dimension of the pseudo-factor
space is twice that of the original factor space) if the shift in
the loading matrix $\Theta_{2}-\Theta_{1}$ is linearly independent
of $\Theta_{1}$. We refer to this special case as the shift type
of change, because the augmentation of the factor space is induced
by a linearly independent shift in the loading matrix. Hence, Type
1 covers the shift type of change.

\textbf{Type 2}. Only $B$ or $C$ is singular. In this case, emerging or disappearing factors are present in the model. Let us consider the following
example of disappearing factors.

Example (2): Without loss of generality, let us assume that $r_{2}<r_{1}$ and
$\Theta_{2}$ is equal to the first $r_{2}$ columns of $\Theta_{1}$;
thus, the last $r_{1}-r_{2}$ factors disappear after the break. Therefore,
we can obtain the pseudo-factors by using the following transformation
from the original factors $\mathbb{F}$: \begin{eqnarray}
\left[\begin{array}{c}
X_{k_{0}}^{(1)}\\
X_{k_{0}}^{(2)}\end{array}\right] & = & \left[\begin{array}{c}
\mathbb{F}_{k_{0}}^{(1)}\Theta_{1}^{\prime}\\
\mathbb{F}_{k_{0}}^{(2)}\Theta_{2}^{\prime}\end{array}\right]+e=\left[\begin{array}{c}
\mathbb{F}_{k_{0}}^{(1)}\Theta_{1}^{\prime}\\
{}[\mathbb{F}_{k_{0}}^{(2)}\;\vdots\;*]C^{\prime}\Theta_{1}^{\prime}\end{array}\right]+e\nonumber \\
 & = & \left[\begin{array}{c}
F_{k_{0}}^{(1)}\\
F_{k_{0}}^{(2)}C^{\prime}\end{array}\right]\Theta_{1}^{\prime}+e=\underbrace{\left[\begin{array}{c}
F_{k_{0}}^{(1)}\\
F_{k_{0}}^{(2)}C^{\prime}\end{array}\right]}_{G}\Lambda^{\prime}+e,\label{eq:ex2}\end{eqnarray}
where $F_{k_{0}}^{(1)}=\mathbb{F}_{k_{0}}^{(1)}$, $F_{k_{0}}^{(2)}=[\mathbb{F}_{k_{0}}^{(2)}\;\vdots\;*]$,
$C=diag(I_{r_{2}},0_{(r_{1}-r_{2})\times(r_{1}-r_{2})})$, $\Lambda=\Theta_{1}$,
and the asterisk is defined in a similar manner to that in \eqref{eq:ex1}.
In this example, $B=I_{r_{1}}$, $r=r_{1}$, and $r_{2}=\mathrm{rank}(C)<r$.
Symmetrically, if $B$ is singular and $C=I_{r_{2}}$, then $r_{2}=r$
and $r_{1}=\mathrm{rank}(B)<r$, which means that certain factors emerge after the break point. Type 2 changes are important in empirical
analysis. Please refer to \cite{Mcalinn2018} for empirical evidence
regarding the varying number of factors in the U.S. macroeconomic dataset. For
Types 1 and 2, we obtain a significant result that $P(\hat{k}\lyxmathsym{\textminus}k_{0}=0)\to1$
as $N,T\to\infty$.%
\footnote{Technically, Types 1 and 2 can be combined into one type that involves
singularity, which renders our QML estimator consistent. We consider
Type 2 separately to emphasize the case of emerging and disappearing
factors.%
}

\textbf{Type 3}. Both $B$ and $C$ are nonsingular. In this case,
the loadings on the original factors undergo a rotational change,
and the dimension of the original factors is the same as that of the pseudo-factors.

Example (3): Let us assume that $r_{2}=r_{1}$ and $\Theta_{2}=\Theta_{1}C$
for a nonsingular matrix $C$. The model with the original factors
$\mathbb{F}$ can be transformed into the following pseudo-factor representation: \begin{eqnarray}
\left[\begin{array}{c}
X_{k_{0}}^{(1)}\\
X_{k_{0}}^{(2)}\end{array}\right] & = & \left[\begin{array}{c}
\mathbb{F}_{k_{0}}^{(1)}\Theta_{1}^{\prime}\\
\mathbb{F}_{k_{0}}^{(2)}\Theta_{2}^{\prime}\end{array}\right]+e=\left[\begin{array}{c}
\mathbb{F}_{k_{0}}^{(1)}\Theta_{1}^{\prime}\\
\mathbb{F}_{k_{0}}^{(2)}C^{\prime}\Theta_{1}^{\prime}\end{array}\right]+e\nonumber \\
 & = & \left[\begin{array}{c}
F_{k_{0}}^{(1)}\\
F_{k_{0}}^{(2)}C^{\prime}\end{array}\right]\Theta_{1}^{\prime}+e=G\Lambda^{\prime}+e,\label{eq:ex3}\end{eqnarray}
where $F_{k_{0}}^{(1)}=\mathbb{F}_{k_{0}}^{(1)}$, $F_{k_{0}}^{(2)}=\mathbb{F}_{k_{0}}^{(2)}$,
and $\Lambda=\Theta_{1}$. In this example, $B=I_{r_{1}}$ and $r=r_{1}=r_{2}$,
and the factor dimension remains constant. In the observationally
equivalent pseudo-factor representation, the loading is time-invariant
and the original post-break factors $\mathbb{F}_{k_{0}}^{(2)}$ are
rotated by $C$. We refer to this as the rotation type of change.

The above examples show that a factor model with any of these three
types of change can be unified and reformulated by the representation
in (\ref{Baltagi}) with pseudo-factors. This representation controls the break
type by varying the settings for $B$ and $C$, and thus, is convenient
for our theoretical analysis.

\cite{Bai2017} rule out the rotation type of change because the break date is not identifiable by minimizing the sum of squared residuals.
\cite{Baltagi2017} allow changes in the number of factors and rotation type of change; however, the difference between their estimator and the true break point is only stochastically bounded (i.e.,
their estimator is not consistent).
Ma and Su's (2018) setup requires $r_1 = r_2$; thus, Type 2 is ruled out under their assumptions. Our simulation result shows that Ma and Su's estimator does not perform well under rotational changes
(Type 3), whereas our QML method can handle changes in all three types discussed above. We obtain a significant result that $\hat{k}-k_0=O_p(1)$ if both $B$ and $C$ are of full rank (i.e.,
Type 3) and $\hat{k}-k_0=o_p(1)$ if $B$ or $C$, or both, is singular (i.e., Type 1 and Type 2).

In this paper, we consider the QML estimator of the break date for model (\ref{Baltagi}):
\begin{eqnarray}
&&\hat{k}=\arg\min_{[\tau_1 T] \leq k \leq [\tau_2 T]} U_{NT}(k),
\label{obj_fun}
\end{eqnarray}
where $[\tau_1 T]$ and $[\tau_2 T]$ denote the prior lower and upper bounds for the real break point $k_0$ with $\tau_1,\tau_2\in (0,1)$ and $\tau_1 \leq \tau_0 \leq \tau_2$. The QML objective function
$U_{NT}(k)$ is equal to
\begin{eqnarray}
&&U_{NT}(k)=k\log( \det( \hat{\Sigma}_1  ) )+(T-k)\log( \det( \hat{\Sigma}_{2} ) ),
\label{obj}
\end{eqnarray}
where $\hat{\Sigma}_1$ and $\hat{\Sigma}_{2}$ are defined as
\begin{eqnarray}
\hat{\Sigma}_1&=&\frac{1}{k}\sum\limits_{t=1}^k \hat{g}_t \hat{g}_t^{'},\nonumber\\
\hat{\Sigma}_{2}&=&\frac{1}{T-k}\sum\limits_{t=k+1}^T \hat{g}_t \hat{g}_t^{'},
\label{Sig}
\end{eqnarray}
and $\hat{g}_t$ is the PCA estimator of $g_t$ (i.e., the transpose of the $t$-th row of $G$). We define $\Sigma_{G,1}=E(g_t g_t^{'})$ for $t\leq k_0$, $\Sigma_{G,2}=E(g_t g_t^{'})$ for $t>k_0$, and $\Sigma_G=\tau_0 \Sigma_{G,1}+(1-\tau_0)\Sigma_{G,2}$. We define $\Sigma_\Lambda$ as the covariance matrix of $\Lambda$. The PCA estimator $\hat{g}_t$ is asymptotically close to $H^{'}g_t$ for a rotation matrix $H$, and $H \xrightarrow{p} H_0=\Sigma_{\Lambda}^{1/2}\Phi V^{-1/2}$ as $(N,T)\rightarrow \infty$, where $V$ and
$\Phi$ are the eigenvalue and eigenvector matrices of $\Sigma_{\Lambda}^{1/2}\Sigma_G \Sigma_{\Lambda}^{1/2}$, respectively. Evidently, the second moment of $H_0 g_t$ shares the same change point as
that of $g_t$. Therefore, we proceed to estimate the pre- and post-break second moments of $g_t$ by using the estimated factors $\hat{g}_t$, and then use (\ref{obj_fun}) to obtain the QML break point
estimator $\hat{k}_{QML}$. Similar QML objective functions have been used for multivariate time series with observed data (e.g., \cite{Bai2000}).

\vspace{-1em}
\section{Assumptions}
\vspace{-1em}

 In this section, we state the assumptions made for establishing the consistency and asymptotic distribution of the QML estimator.

\noindent \begin{assum}\label{factors}

(i) $E\left\|f_t \right\|^4<M<\infty$, $E(f_tf_t^{'})=\Sigma_F$, where $\Sigma_F$ is positive definite, and
$\frac{1}{k_0}\sum_{t=1}^{k_0}f_tf_t^{'}\xrightarrow{p}\Sigma_F,\frac{1}{T-k_0}\sum_{t=k_0+1}^{T}f_tf_t^{'}\xrightarrow{p}\Sigma_F$;

(ii) There exists $d>0$ such that $\left\|\Delta\right\|\geq d>0$, where $\Delta =B\Sigma_FB^{'}- C\Sigma_FC^{'}$ and $B,C$ are $r\times r$ matrices.

%(iii) {\color{red}$\Sigma_G$ is positive definite matrix.}
\end{assum}

\noindent \begin{assum}\label{Factor_Loadings}
$\left\| \lambda_{\ell i} \right\|\leq \bar{\lambda}<\infty$ for $\ell=1,2$, $i=1,\cdots,N$, $\left\| \frac{1}{N}\Lambda^{'}\Lambda-\Sigma_{\Lambda} \right\|\rightarrow 0$ for some $r\times r$ positive
definite matrix $\Sigma_{\Lambda}$.
\end{assum}

\noindent \begin{assum}\label{Depen_and_Hetero}
There exists a positive constant $M<\infty$ such that

\begin{itemize}
\item[(i)] $E(e_{it})=0$ and $E|e_{it}|^8\leq M$ for all $i=1,\cdots,N$ and $t=1,\cdots,T$;
\item[(ii)] $E(\frac{e_s^{'}e_t}{N})=E(N^{-1}\sum_{i=1}^Ne_{is}e_{it})=\gamma_N(s,t)$ and $\sum_{s=1}^{T}|\gamma_N(s,t)|\leq M$ for every $t\leq T$;
\item[(iii)] $E(e_{it}e_{jt})=\tau_{ij,t}$ with $|\tau_{ij,t} |<\tau_{ij}$ for some $\tau_{ij}$ and for all $t=1,\cdots,T$ and $\sum_{j=1}^{N}|\tau_{ij}|\leq M$ for every $i\leq N$;
\item[(iv)] $E(e_{it}e_{js})=\tau_{ij,ts}$,
\begin{equation*} \frac{1}{NT}\sum\limits_{i,j,t,s=1}|\tau_{ij,ts}|\leq M;
\end{equation*}
\item[(v)] For every $(s,t)$, $E\left| N^{-1/2}\sum_{i=1}^{N}(e_{is}e_{it}-E[e_{is}e_{it}]) \right|^4\leq M$.
\end{itemize}
\end{assum}

\noindent \begin{assum}\label{Weak_Dependence}

There exists a positive constant $M<\infty$ such that
\begin{eqnarray*}
E(\frac{1}{N}\sum\limits_{i=1}^{N}\left\| \frac{1}{\sqrt{k_0}}\sum\limits_{t=1}^{k_0}f_te_{it} \right\|^2)&\leq& M,\\
E(\frac{1}{N}\sum\limits_{i=1}^{N}\left\| \frac{1}{\sqrt{T-k_0}}\sum\limits_{t=k_0+1}^{T}f_te_{it}\right\|^2)&\leq& M.
\end{eqnarray*}

\end{assum}

\noindent \begin{assum}\label{eigenvalues}
The eigenvalues of $\Sigma_G\Sigma_\Lambda$ are distinct.
\end{assum}

\noindent \begin{assum}\label{Hajek-Renyi}
Let us define $\epsilon_t=f_tf_t^{'}-\Sigma_F$. According to the data-generating process (DGP) of factors, the H\'{a}jek-R\'{e}nyi inequality applies to the processes $\{\epsilon_t,t=1,\cdots,k_0\}$,
$\{\epsilon_t,t=k_0,\cdots,1\}$, $\{\epsilon_t,t=k_0+1,\cdots,T\}$, and $\{\epsilon_t,t=T,\cdots,k_0+1\}$.
\end{assum}
\begin{remark}
Using the H\'{a}jek-R\'{e}nyi equality on $\epsilon_t$, we can ensure that $\max\limits_{k_0<k\leq [\tau_2 T]}\| \frac{1}{T-k} \sum\limits_{t=k+1}^{T} f_tf_t^{'}-\Sigma_F\|=O_p(\frac{1}{\sqrt{T}})$ in
Lemma \ref{differ} and $\max\limits_{[\tau_1 T]\leq k<k_0}\| \frac{1}{k_0-k} \sum\limits_{t=k+1}^{k_0} g_tg_t^{'} \|=O_p(1), \max\limits_{k_0<k\leq [\tau_2 T]}\| \frac{1}{k-k_0}
\sum\limits_{t=k_0+1}^{k} g_tg_t^{'} \|=O_p(1)$ in Lemmas \ref{differ} and \ref{differ2}.
\label{Hajek_Renyi_Generalized inequality}
\end{remark}

\noindent \begin{assum}\label{error_sup}
There exists an $M<\infty$ such that

(i) For each $s=1,\cdots,T$,
\begin{eqnarray*}
E(\max_{k<k_0}\frac{1}{k_0-k}\sum_{t=k+1}^{k_0}|\frac{1}{\sqrt{N}}\sum_{i=1}^{N}[e_{is}e_{it}-E(e_{is}e_{it})]|^2)&\leq& M,\\
E(\max_{k> k_0}\frac{1}{k-k_0}\sum_{t=k_0+1}^{k}|\frac{1}{\sqrt{N}}\sum_{i=1}^{N}[e_{is}e_{it}-E(e_{is}e_{it})]|^2)&\leq& M;
\end{eqnarray*}
(ii)
\begin{eqnarray*}
E(\max_{k<k_0}\frac{1}{k_0-k}\sum_{t=k+1}^{k_0}\left\|\frac{1}{\sqrt{N}}\sum_{i=1}^{N} \lambda_i e_{it}   \right\|^2)&\leq &M,\\
E(\max_{k>k_0}\frac{1}{k_0-k}\sum_{t=k_0+1}^{k}\left\|\frac{1}{\sqrt{N}}\sum_{i=1}^{N} \lambda_i e_{it}    \right\|^2)&\leq &M.
\end{eqnarray*}
\end{assum}

\noindent \begin{assum}\label{Central_Limit}
There exists an $M<\infty$ such that for all values of $N$ and $T$,

(i) for each $t$,
\begin{eqnarray*}
E\left(\max_{1 \leq k<k_0}\frac{1}{k_0-k}\sum_{t=k+1}^{k_0} \left\| \frac{1}{\sqrt{NT}}\sum_{s=1}^{T}\sum_{i=1}^Nf_s[e_{is}e_{it}-E(e_{is}e_{it})] \right\|^2\right)&\leq& M,\\
E\left(\max_{k_0<k\leq  T}\frac{1}{k-k_0}\sum_{t=k_0+1}^{k} \left\| \frac{1}{\sqrt{NT}}\sum_{s=1}^{T}\sum_{i=1}^Nf_s[e_{is}e_{it}-E(e_{is}e_{it})] \right\|^2\right)&\leq& M;
\end{eqnarray*}
(ii) the $r\times r$ matrix satisfies
\begin{eqnarray*}
E\left\| \frac{1}{\sqrt{NT}}\sum_{t=1}^{T}\sum_{i=1}^Nf_t\lambda_i^{'}e_{it} \right\|^2\leq M.
\end{eqnarray*}

\end{assum}

\vspace{-1em}
\section{Asymptotic properties of the QML estimator}
\vspace{-1em}
In this section, we derive the asymptotic properties of the QML estimator for various breaks. In the literature of structural breaks for a fixed-dimensional time series, conventional break point estimators, such as the least-squares (LS) estimator of \cite{Bai1997} or the QML estimator of
\cite{Qu_Perron2007}, are usually inconsistent. The estimation error of these conventional estimators is $O_p(1)$ when the break size is fixed. To reach consistency, the cross-sectional dimension of the
time series must be large (e.g., \cite{Bai2010} and \cite{Kim2011}).

Recall that the observationally equivalent representation in (\ref{Baltagi}) has time-invariant loadings and varying pseudo-factors. Hence, our problem converges to estimating the break point in the
$r$-dimensional time series $g_t$, where $r$ is fixed. Theorems \ref{bound_theorem} and \ref{distribution_theorem} below show that, for rotational breaks (Type 3), the convergence rate and limiting
distribution are similar to those available in the literature. However, for Type 1 and 2 breaks, Theorem \ref{consistency} derives a much more significant result than that available in the literature,
according to which our QML estimator is consistent even if our $g_t$ has only a fixed cross-sectional dimension $r$.

\noindent
\begin{theorem}
Under Assumptions \ref{factors}--\ref{Central_Limit}, when both $B$ and $C$ are of full rank, $\hat{k}-k_0=O_p(1)$.
\label{bound_theorem}
\end{theorem}

This theorem implies that the difference between the QML estimator and the true change point is stochastically bounded in model (\ref{eq:ex3}). Although the estimation errors of both \cite{Baltagi2017} and our QML
methods are bounded, the QML estimator has much better finite sample properties. To confirm this theoretical result, we conduct a simulation where the factor loadings have a rotational change (see
DGP 1.B in Section 5). Table \ref{rotation_full_rank} presents the MAEs and RMSEs of different estimators. The simulation result shows that the QML estimators have much smaller MAEs and RMSEs than other
methods. In addition, $\hat{k}$ does not collapse to $k_0$, leading to a nondegenerate distribution. We will state the limiting distribution in Theorem \ref{distribution_theorem}. Nevertheless, this
theorem shows that the break point can be appropriately estimated because $\hat{\tau}=\hat{k}/T$ is still consistent for $\tau_0$.

To make an inference regarding the change point when both $B$ and $C$ are of full rank, we derive the limiting distribution of $\hat{k}$. Let us define
\begin{eqnarray*}
\xi_t & = & H_{0}^{'}g_{t}g_{t}^{'}H_{0}-\Sigma_{1}\text{ for }t\leq k_{0},\\
\xi_t & = & H_{0}^{'}g_{t}g_{t}^{'}H_{0}-\Sigma_{2}\text{ for }t>k_{0},
\end{eqnarray*}
where $\Sigma_{1}=H_{0}^{'}\Sigma_{G,1}H_{0}$ and $\Sigma_{2}=H_{0}^{'}\Sigma_{G,2}H_{0}$
are the pre- and post-breaks of $H_{0}^{'}E(g_{t}g_{t}^{'})H_{0}$. The limiting distribution of $\hat{k}$ is given by the following theorem:
\noindent
\begin{theorem}
Under Assumptions \ref{factors}--\ref{Central_Limit}, when both $B$ and $C$ are of full rank,
\begin{eqnarray*}
\hat{k}-k_0\xrightarrow{d} \arg\min\limits_\ell W(\ell),
\end{eqnarray*}
where
\begin{eqnarray*}
&&W(\ell)=\sum\limits_{t=k_0+\ell}^{k_0-1}tr((\Sigma_2^{-1}-\Sigma_1^{-1})\xi_t)-\left( tr(\Sigma_1\Sigma_2^{-1})-r-\log|\Sigma_1\Sigma_2^{-1}| \right)\ell\\
&&\text{for }\ell=-1,-2,\cdots,\\
&&W(\ell)=0\text{ for }\ell=0,\\
&&W(\ell)=\sum\limits_{t=k_0+1}^{k_0+\ell}tr((\Sigma_1^{-1}-\Sigma_2^{-1})\xi_t)+\left( tr(\Sigma_1^{-1}\Sigma_2)-r-\log|\Sigma_1^{-1}\Sigma_2| \right)\ell\\
&&\text{for }\ell=1,2,\cdots.
\end{eqnarray*}
\label{distribution_theorem}
\end{theorem}

This result shows that the limiting distribution depends on $\xi_t$.
If $\xi_t$ is independent over time, then $W(\ell)$ is a two-sided random walk. If $f_t$ is stationary, then $\xi_t$ is stationary in each regime.
Here, the limiting distribution of the estimated break date is dependent on the generation processes of the unobserved factors, and thus, cannot be directly used to construct a confidence interval for a
true break point.
\cite{Bai2017} propose a bootstrap method to construct a confidence interval for $k_0$ when the change in the factor loading matrix shrinks as $N\to \infty$.
However, their bootstrap procedure lacks robustness in the cross-sectional correlation in the error terms.
In the current setup, the break magnitude $\left\| \Sigma_2-\Sigma_1 \right\|$ is fixed and we leave the case of shrinking break magnitude as a future topic.

Next, we establish a much stronger result than that available in the literature, which states that the QML estimator remains consistent when $B$ or $C$, or both, is singular. We make the following
additional assumptions.

\noindent \begin{assum}\label{as-.LeeL}
With probability approaching one (w.p.a.1), the following inequalities hold:
\begin{align*}
&0<\underline{c}\le \min_{[\tau_{1}T]\le k\le k_{0}}\rho_{j}\left(\frac{1}{Nk}\sum_{t=1}^{k}\Lambda^{\prime}e_{t}e_{t}^{\prime}\Lambda\right),\\
&0<\underline{c}\le \min_{k_{0}\le k\le[\tau_{2}T]}\rho_{j}\left(\frac{1}{N(T-k)}\sum_{t=k+1}^{T}\Lambda^{\prime}e_{t}e_{t}^{\prime}\Lambda\right),\,\text{for }j=1,\cdots,r;\\
&\rho_{1}\left(\frac{1}{NT}\sum_{t=1}^{T}\Lambda^{\prime}e_{t}e_{t}^{\prime}\Lambda\right)\le \overline{c}<+\infty,
\end{align*}
as $N,T\to\infty$, where $\underline{c}$ and $\overline{c}$ are some constants.
\noindent \end{assum}

\noindent \begin{assum}\label{invariance}

\begin{align*}
\max_{[\tau_{1}T]\le k\le k_{0}}\left\Vert \frac{1}{\sqrt{Nk}}\sum_{t=1}^{k}\sum_{i=1}^{N}f_{t}e_{it}\lambda_{i}^{\prime}\right\Vert  & =O_{p}(1),\\
\max_{k_{0}\le k\le[\tau_{2}T]}\left\Vert \frac{1}{\sqrt{N(T-k)}}\sum_{t=k+1}^{T}\sum_{i=1}^{N}f_{t}e_{it}\lambda_{i}^{\prime}\right\Vert  & =O_{p}(1).\end{align*}
\end{assum}\vspace{2em}

Assumption \ref{as-.LeeL} is useful to derive the lower bound of the smallest eigenvalue of $\hat{\Sigma}_1$ (or $\hat{\Sigma}_2$) if $B$ (or $C$) is a singular matrix. Assumption \ref{invariance} strengthens Assumption \ref{Central_Limit}(ii), which is similar to Assumption F2 of \cite{Bai2003}. Note that the summation $\sum\limits_{t=1}^k$ in Assumptions \ref{as-.LeeL}-\ref{invariance} involves a positive fraction of observations over time since the lower bound of $k$ is $\tau_1 T$, with $\tau_1\in (0,1)$.

Also, as the log of matrix determinant is involved in the QML function, a natural problem is that the
log determinant of a singular population covariance matrix is undefined when $B$ or $C$, or both, is singular. Fortunately, the determinants of $\hat{\Sigma}_1=\frac{1}{k}\sum\limits_{t=1}^k \hat{g}_t \hat{g}_t^{'}$ and
$\hat{\Sigma}_2=\frac{1}{T-k}\sum\limits_{t=k+1}^T \hat{g}_t \hat{g}_t^{'}$ are small but not equal to zero in finite samples, when $\Sigma_1$ and $\Sigma_2$ are
singular matrices. The following proposition develops a lower bound for the smallest
eigenvalues of $\hat{\Sigma}_1$ and $\hat{\Sigma}_2$.

\begin{proposition}\label{low_bound} Under Assumptions \ref{factors}--\ref{invariance}, for $k\ge k_{0}$
and $k\le[\tau_{2}T]$, if $C$ is singular and $\sqrt{N}/T\to0$
as $N,T\to\infty$, then there exist constants
$c_{U}\ge c_{L}>0$ such that \begin{align*}
P\left(\min_{k\in[k_{0},[\tau_{2}T]]}\rho_{j}(\hat{\Sigma}_{2})\ge\frac{c_{L}}{N}\right) & \to1,\\
P\left(\max_{k\in[k_{0},[\tau_{2}T]]}\rho_{j}(\hat{\Sigma}_{2})\le\frac{c_{U}}{N}\right) & \to1,\end{align*}
 for $j=r_{2}+1,...,r$.

\end{proposition}

In proposition \ref{low_bound}, the lower bound of the smallest eigenvalue of the estimated sample covariance matrix $\hat{\Sigma}_2$ is $c_L/N$ for a constant $c_L>0$ w.p.a.1. A similar lower bound for the smallest eigenvalue of $\hat{\Sigma}_1$  can be obtained when $B$ is singular under the same assumptions.
This ensures a lower bound for the determinants of the estimated sample covariance matrices. Proposition \ref{low_bound} provides a useful tool to establish the consistency of our QML estimator.
Although this technical result is a byproduct in our analysis, we believe that it is of independent interest and useful in other contexts.

\noindent \begin{assum}\label{B_C_full_rank_project}

(i) $[B,C]$ is of full row rank.

(ii) $C^{\#}Bf_{k_0}\neq 0$ when $r-1=r_2>0$; and $B^{\#}Cf_{k_0+1}\neq 0$ when $r-1=r_1>0$, where $\mathbb{A}^{\#}$ denotes the adjoint matrix for a singular matrix $\mathbb{A}$.

(iii) $\|Bf_{k_{0}}-\mathrm{Proj}(Bf_{k_{0}}|C)\|\ge d>0$ when $r-r_2\geq 2$ or $r_{2}=0$; and $\|Cf_{k_{0}+1}-\mathrm{Proj}(Cf_{k_{0}+1}|B)\|\ge d>0$ when $r-r_1\geq 2$ or $r_{1}=0$, where
$\mathrm{Proj}(\mathbb{A}|\mathbb{Z})$ denotes the projection
of $\mathbb{A}$ onto the columns of $\mathbb{Z}$, and $d$ is a constant.
\end{assum}

Assumption \ref{B_C_full_rank_project}(i) implies that $\Sigma_G$ is positive definite.$\footnote{Since \begin{eqnarray*}
rank(\Sigma_G)&=&rank\left(\left[
\begin{array}{cccccccccc}
\sqrt{\tau_0}B,\sqrt{1-\tau_0}C
\end{array}
\right]
diag\left(\Sigma_F,\Sigma_F\right)
\left[
\begin{array}{cccccccccc}
\sqrt{\tau_0}B,\sqrt{1-\tau_0}C
\end{array}
\right]^{'}\right)
=rank\left(\left[
\begin{array}{cccccccccc}
\sqrt{\tau_0}B,\sqrt{1-\tau_0}C
\end{array}
\right]\right)\\
&=&rank\left(
[B,C]diag\left(  \sqrt{\tau_0}I_r, \sqrt{1-\tau_0}I_r  \right)  \right)=rank\left([B,C]\right)
\end{eqnarray*}
and $1<\tau_0<1$, Assumption \ref{B_C_full_rank_project} implies that $\Sigma_G$ is a positive definite matrix.
}
$ Assumptions \ref{B_C_full_rank_project}(ii) implies that $B^{\#}C\neq 0$ when $r-1=r_1>0$, and $C^{\#}B\neq 0$ when $r-1=r_2>0$. It also excludes the possibility that $f_{k_0}$ and $f_{k_0+1}$ are in the null space of $C^{\#}B$ and $B^{\#}C$, respectively. Similarly, Assumption \ref{B_C_full_rank_project}(iii) rules out the cases that $Bf_{k_0}$ lies in the column space of $C$ when $r-r_2\geq 2$ or $r_2=0$ and that $Cf_{k_0+1}$ lies in the column space of $B$ when $r-r_1\geq 2$ or $r_1=0$.\footnote{Note that $r_2=0$ means $C=0$, so $B$ has to be nonsingular by  Assumption \ref{B_C_full_rank_project}(i). Thus, Assumption \ref{B_C_full_rank_project}(iii) implies that $f_{k_0}\neq 0$ when $C=0$.}
Assumption \ref{B_C_full_rank_project} is used to establish Lemma \ref{differ2}, which is useful for validating the consistency result that $Prob(\hat{k}-k = 0) \to 1$ in the proof of Theorem
\ref{consistency}. It ensures that the value of the objective function becomes larger even if $\hat{k}$ slightly deviates from the true break point in large samples.  Assumption \ref{B_C_full_rank_project} is flexible enough to allow various data generating processes for $f_t$. For example, if $f_{k_{0}}$ and $f_{k_{0}+1}$ have continuous probability distribution functions, then Assumptions \ref{B_C_full_rank_project}(ii)-(iii) just exclude a zero probability event since $C^{\#}B$ and $B^{\#}C$ are not equal to zero.

It is remarkable that existing estimators such as \cite{Baltagi2017} and \cite{Bai2017} are not consistent even if Assumptions \ref{as-.LeeL} -- \ref{B_C_full_rank_project} hold. In contrast, our QML estimator is shown to be consistent under these additional assumptions. The following theorem summarizes the result.

\begin{theorem}
Under Assumptions \ref{factors}--\ref{B_C_full_rank_project} and $\frac{N}{T}\to\kappa$, as
$N,T\to\infty$ for $0<\kappa<\infty$, when $B$ or $C$, or both, is singular, $Prob(\hat{k}-k = 0) \to 1.$
\label{consistency}
\end{theorem}

Theorem \ref{consistency} shows that the estimated change point converges to the true change point w.p.a.1 when $B$ or $C$, or both, is singular (Types 1 and 2 in Section 2). This result is much more
significant than that obtained by \cite{Baltagi2017}, who show that the distance between the estimated and true break dates is bounded for Types 1--3. Note that the case in which only $B$ (or $C$) is
singular corresponds to Type 2 with emerging (or disappearing) factors. Our QML estimator is consistent under this type of change, whereas \cite{Bai2017} and \cite{Ma_Su2018} rule out this type by
assumption. In empirical applications, the conditions of theorem \ref{consistency} are rather flexible and likely to hold and the consistency of the break date estimator is expected
in most economic data for the factor analysis.

\begin{remark}
An important contribution of Theorem \ref{consistency} is to link the consistency of the QML estimator with the singularity of the covariance matrices of the pre- or post-break factor loadings.
The singularity is generated by the special structure of the pseudo-factors $g_t$ shown in (\ref{eq:ex1}) and (\ref{eq:ex2}) in the presence of a structural change.
The PCA estimator $\hat{g}_t$ is consistent for $g_t$ (up to some rotation) for large $N$ and $T$, so the singularity structure is maintained in $\hat{\Sigma}_1$ and $\hat{\Sigma}_2$ and hence contributes to the consistency of our QML estimator. The result in Theorem \ref{consistency} is in contrast to
conventional break point estimators, which only have $O_p(1)$ estimation errors in multivariate time series with a small cross-sectional
dimension (e.g., \cite{Bai1997}; \cite{Qu_Perron2007}). Although our $\hat{g}_t$ has a fixed dimension, the divergence rate of the objective function depends on $N$. $\footnote{This is because the convergence rate of the smallest eigenvalue of $\hat{\Sigma}_2$ is $N^{-1}$ for $k = k_0$ when $C$ is singular. See Proposition 1.}$
In other words, our QML estimator still implicitly utilizes the information in the large cross-sectional dimension, which is the source of our consistency.

\label{consistency_remark}
\end{remark}
\begin{remark}
The conditions that $B$ or $C$, or both, is singular and $\frac{N}{T}\rightarrow \kappa\in (0,\infty)$ are likely to hold in many economic datasets for factor
analysis.
If both $B$ and $C$ are singular, the break occurs such that the number of pseudo-factors in the entire factor model is larger than that of the factors in the pre- and post-break subsamples. This can happen when the factor loadings undergo a shift type of change, as discussed in Example (1) for Type 1 changes.
If $B$ is of full rank and $C$ is singular, some factors become irrelevant, and thus, the loading coefficients attached to these disappearing factors become zero. For example, in the momentum portfolio,
some risks are not part of the firm's long-run structure as only sorting based on recent returns works; the reward is high but disappears within less than a year.
If $B$ is singular and $C$ is of full rank, some factors emerge after the break date, increasing the dimension of the post-break factor space. For example, changes in the technology or policy may
produce certain new factors.
\label{consistency_remark2}
\end{remark}
\begin{remark}
Theorem \ref{consistency} indicates that $U_{NT}(k)$ can be minimized to consistently estimate $k_0$. The intuition for this is that $U_{NT}(k)-U_{NT}(k_0)$ is always larger than zero, even if $k$
deviates only slightly from the true break point $k_0$, so that $\hat{k}$ must be equal to $k_0$ to minimize $U_{NT}(k)-U_{NT}(k_0)$.
For example, in Type 1, when both $B$ and $C$ are singular for $k<k_0$, we can decompose $\hat{\Sigma}_2$ as $\hat{\Sigma}_2=\frac{1}{T-k}\sum\limits_{t=k+1}^{k_0} \hat{g}_t
\hat{g}_t^{'}+\frac{1}{T-k}\sum\limits_{t=k_0+1}^T \hat{g}_t \hat{g}_t^{'}$, and the term $\frac{1}{T-k}\sum\limits_{t=k+1}^{k_0} \hat{g}_t \hat{g}_t^{'}$ results in a larger determinant of $\hat{\Sigma}_2$ than that of $\hat{\Sigma}_2^{0}=\frac{1}{T-k_0}\sum\limits_{t=k_0+1}^T \hat{g}_t \hat{g}_t^{'}$.
By symmetry, we obtain a similar result for $k>k_0$. (See Lemmas \ref{differ} and \ref{differ2} for more technical details.) Thus, $U_{NT}(k)-U_{NT}(k_0)>0$ w.p.a.1 as $N,T\to\infty$ if $k\neq k_0$.
\label{singular_remark}
\end{remark}
\begin{remark}
With the QML estimator, we do not need to know the numbers of original factors $r_1$ and $r_2$ before and after the break point, but only  the number of pseudo-factors in the entire sample.
\cite{Bai2017} and \cite{Ma_Su2018} require knowledge of the number of original factors, which is much more difficult to estimate due to the augmented factor space resulting from the break. In practice,
the number of pseudo-factors is much easier to estimate by using one of a number of estimators, such as the information criteria developed by \cite{Bai2002}.
\label{pseudo_true_fators}
\end{remark}

\vspace{-1em}
\section{Simulation}
\vspace{-1em}

In this section, we consider DGPs corresponding to Types 1--3 to evaluate the finite sample performance of the QML estimator. We compare the QML estimator with three other estimators. As shown below,
$\hat{k}_{BKW}$ is the estimator proposed by Baltagi, Kao, and Wang (2017, BKW hereafter); $\hat{k}_{BHS}$ is the estimator proposed by Bai, Han, and Shi (2020, BHS hereafter); $\hat{k}_{MS}$ is the
estimator proposed by Ma and Su (2018, MS hereafter); and $\hat{k}_{QML}$ is the QML estimator. \cite{Barigozzi2018} develops a change point estimator using wavelet transformation, which exhibits
similar performance to that of the estimator proposed by \cite{Ma_Su2018}. Hence, the comparison with the estimator proposed by \cite{Barigozzi2018} is not reported here, but the result is available
upon request.
The DGP roughly follows BKW, which can be used to examine various elements that may affect the finite sample performance of the estimators, and we use this DGP for model (\ref{Baltagi}).
We calculate the root mean square error (RMSE) and mean absolute error (MAE) of these change point estimators $\hat{k}_{BKW}$, $\hat{k}_{BHS}$, and $\hat{k}_{QML}$, and each experiment is repeated 1000
times, where RMSE$=\sqrt{\frac{1}{1000}\sum\limits_{s=1}^{1000}(\hat{k}_s-k_0)^2}$ and MAE$=\frac{1}{1000}\sum\limits_{s=1}^{1000}|\hat{k}_s-k_0|$.
When $T$ is small, there is a possibility that Ma and Su's (2018) method detects no break or multiple breaks; thus, the definition of the estimation error for a single break point in such cases is not
straightforward. For a comparison, we compute the RMSE and MAE of the MS estimator by only using the results obtained by the MS estimator when it successfully detects a single break.
As the computation of $\hat{k}_{BHS}$ and $\hat{k}_{MS}$ requires the number of original factors and that of $\hat{k}_{BKW}$ and $\hat{k}_{QML}$ requires the number of pseudo-factors, we set
$\hat{r}=r_0$ for $\hat{k}_{BHS}$ and $\hat{k}_{MS}$ and $\hat{r}=r$ for $\hat{k}_{QML}$ and $\hat{k}_{BKW}$, where $r_0$ is the number of original factors and $r$ is the number of pseudo-factors.

We generate factors and idiosyncratic errors using a DGP similar to that of BKW. Each factor is generated by the following AR(1) process:
\begin{eqnarray*}
f_{tp}=\rho f_{t-1,p}+u_{t,p},\quad for\quad t=2,\cdots,T;\quad p=1,\cdots,r_0,
\end{eqnarray*}
where $u_t=(u_{t,1},\cdots,u_{t,r_0})^{'}$ is i.i.d. $N(0,I_{r_0})$ for $t=2,\cdots,T$ and $f_1=(f_{1,1},\cdots,f_{1,r_0})^{'}$ is i.i.d. $N(0,\frac{1}{1-\rho^2}I_{r_0})$. The scalar $\rho$ captures the
serial correlation of factors, and the idiosyncratic errors are generated by
\begin{eqnarray*}
e_{i,t}=\alpha e_{i,t-1}+v_{i,t},\quad for\quad i=1,\cdots,N\quad t=2,\cdots,T,
\end{eqnarray*}
where $v_t=(v_{1,t},\cdots,v_{N,t})^{'}$ is i.i.d. $N(0,\Omega)$ for $t=2,\cdots,T$ and $e_1=(e_{1,1},\cdots,e_{N,1})^{'}$ is $N(0,\frac{1}{(1-\alpha^2) \Omega})$. The scalar $\alpha$ captures the
serial correlation of the idiosyncratic errors, and $\Omega$ is generated as $\Omega_{ij}=\beta^{|i-j|}$ so that $\beta$ captures the degree of cross-sectional dependence of the idiosyncratic errors. In
addition, $u_t$ and $v_t$ are mutually independent for all values of $t$. We set $r_0=3$ and $k_0=T/2$.
We consider the following DGPs for factor loadings and investigate the performance of the QML estimator for the three types of breaks discussed in Section 2.

\textbf{DGP 1.A} We first consider the case in which $C$ is singular, and set $C=[1,0,0;0,1,0;0,0,0]$. This setup aims to model (\ref{eq:ex2}). In the pre-break regime, all elements of $\lambda_{i,1}$
are i.i.d. $N(0,\frac{1}{r_0^2}I_{r_0})$ across $i$. In the post-break regime, $\Lambda_2=(\lambda_{1,2},\cdots,\lambda_{N,2})^{'}=\Lambda_1C$. This case corresponds to a Type 2 change with a
disappearing factor. The number of pseudo-factors is the same as $r_0$, so $r=3$, and the numbers of pre- and post-break factors are 3 and rank$(C)=2$, respectively.
Table \ref{rotation_singular} lists the RMSEs and MAEs of three estimators for different values of $(\rho, \alpha, \beta)$.
In all cases, $\hat{k}_{QML}$ has much smaller MAEs and RMSEs than $\hat{k}_{BKW}$ and $\hat{k}_{BHS}$.
Moreover, the MAEs and RMSEs of $\hat{k}_{QML}$ tend to decrease as $N$ and $T$ increase. This confirms the consistency of $\hat{k}_{QML}$ established in Theorem \ref{consistency}. In addition, the
RMSEs and MAEs of $\hat{k}_{BKW}$ do not converge to zero as $N$ and $T$ increase, which confirms that $\hat{k}_{BKW}$ has a stochastically bounded estimation error. $\hat{k}_{BHS}$ does not appear to
be consistent when a factor disappears after the break. Moreover, a larger AR(1) coefficient $\rho$ tends to deteriorate the performance of $\hat{k}_{BKW}$, but does not have much impact on our QML
estimator.

\textbf{DGP 1.B} We next consider the case in which $C$ is of full rank. We set $C$ as a lower triangular matrix. The diagonal elements are equal to $0.5$, $1.5$, and $2.5$, and the elements below these
diagonal elements are i.i.d. and drawn from a standard normal distribution. Under this DGP, we have $r = r_0$.
Table \ref{rotation_full_rank} reports the performance of three estimators for different values of $(\rho, \alpha, \beta)$.
In all cases, $\hat{k}_{BKW}$ and $\hat{k}_{QML}$ appear to have stochastically bounded estimation errors, which confirms Theorem 1 of BKW and Theorem \ref{bound_theorem} of this paper.
Both $\hat{k}_{QML}$ and $\hat{k}_{BKW}$ are inconsistent under this DGP; however, under all settings, our QML estimator tends to have much smaller RMSEs and MAEs than the estimator of BKW.
The MAEs and RMSEs of $\hat{k}_{BHS}$ appear to increase with the sample size; thus, the BHS method cannot handle this case.

\textbf{DGP 1.C} In this case, we set $C=[1,0,0;2,1,0;3,2,m]$ and $m\in \{1,0.8,0.5,0.1,0\}$. As $m$ decreases to zero, the matrix $C$ changes from full rank to singular. We still consider serial
correlation in factors and serial correlation and cross-sectional dependence in idiosyncratic errors simultaneously with $N=100, T=100$. Table \ref{full_rank_to_singular} shows that the MAEs and RMSEs
of $\hat{k}_{QML}$ monotonically decrease with $m$, which confirms our findings in Theorems \ref{bound_theorem} and \ref{consistency}.
In addition, the RMSEs and MAEs of $\hat{k}_{BKW}$ and $\hat{k}_{BHS}$ are much larger than those of $\hat{k}_{QML}$, and do not tend toward zero as $m$ decreases. For each value of $m$, the experiment
is repeated 10000 times to more accurately estimate and compare the RMSEs (MAEs) of our QML estimator across different values of $m$.

\textbf{DGP 1.D} This DGP considers a Type 1 break. In the first regime, the last elements of $\lambda_{i,1}$ are zeros for all $i$, and the first two elements of $\lambda_{i,1}$ are both i.i.d.
$N(0,\frac{1}{2}I_{r_0})$. In the second regime, $\lambda_{i,2}$ is i.i.d. $N(0,\frac{1}{3}I_{r_0})$ across $i$. As $\lambda_{i,1}$ and $\lambda_{i,2}$ are independent, the numbers of factors in the two
regimes are $r_1=2$ and $r_2=3$, respectively, and the number of pseudo-factors is $r=5$. Because the numbers of pre- or post-break factors are smaller than that of the pseudo-factors, both $\Sigma_1$ and
$\Sigma_2$ are singular matrices.
Table \ref{singular_pre_post_break} reports the MAEs and RMSEs of $\hat{k}_{QML}$, $\hat{k}_{BHS}$, and $\hat{k}_{BKW}$ under this DGP.
Table \ref{singular_pre_post_break} shows the performances of both $\hat{k}_{BHS}$ and our $\hat{k}_{QML}$. Their MAEs (RMSEs) are less than 0.05 (0.25) for all combinations of $N$, $T$,
$\rho$, $\alpha$, and $\beta$. Although $\hat{k}_{BHS}$ is consistent under this DGP, our QML estimator still has smaller RMSEs than $\hat{k}_{BHS}$ in most cases reported in Table
\ref{singular_pre_post_break}. In addition, $\hat{k}_{BKW}$ performs better under this DGP than DGPs 1.A--1.C. However, its estimation error is much larger than that of our QML estimator. This is not
surprising because $\hat{k}_{BKW}$ is not consistent. Finally, a larger AR(1) coefficient $\rho$ tends to yield a larger bias for $\hat{k}_{BKW}$, but does not have much effect on the performances of
$\hat{k}_{BHS}$ and $\hat{k}_{QML}$.

In summary, Tables \ref{rotation_singular} and \ref{rotation_full_rank} show that the QML estimator performs much better than $\hat{k}_{BHS}$ under Type 2 and 3 breaks, which are ruled out under the
assumptions of \cite{Bai2017}. Table \ref{singular_pre_post_break} shows that the QML estimator often slightly outperforms $\hat{k}_{BHS}$, even though the latter is known to be consistent and has excellent finite-sample performance under Type
1 breaks.  Note that the strength of the BHS method is the consistent estimation of break point for Type 1 break, especially when the size of the break is shrinking as the sample size increases. Under settings with a shrinking break size, the QML method will lose its power because the dimension
of $G$ (determined by the IC criterion in \cite{Bai2002}) will not be augmented, which means that the singularity does not show up in the covariance if breaks are small enough.

\begin{table}[H]
\renewcommand{\arraystretch}{0.8}
\caption{Simulated mean absolute errors (MAEs) and root mean squared errors (RMSEs) of $\hat{k}_{BKW}$, $\hat{k}_{BHS}$, and $\hat{k}_{QML}$ under DGP 1.A.}
\centering
\label{rotation_singular}
\begin{center}
\begin{tabular}{l l r r r r r r} \hline
$N,T$         &  & \multicolumn{2}{c}{$\hat{k}_{BKW}$} &  \multicolumn{2}{c}{$\hat{k}_{BHS}$}  & \multicolumn{2}{c}{$\hat{k}_{QML}$} \\
&  & MAE & RMSE & MAE & RMSE & MAE & RMSE \\ \hline
     &     &  &               &$\rho=0$     &$\alpha=0$    &$\beta=0$  & \\\hline
100,100  &    &6.3130  &8.9546   &5.4600   &7.7325   &1.6070  &2.9293 \\
100,200  &    &7.0230  &11.9053  &7.9580   &12.4801  &1.2990  &2.3206 \\
200,200  &    &5.6730  &9.9774   &6.7150   &10.8610  &0.7960  &1.5218 \\
200,500  &    &4.6940  &8.5732   &10.0960  &17.9778  &0.7340  &1.3799 \\
500,500  &    &4.4580  &8.5789   &8.6770   &15.6509  &0.3890  &0.8597 \\\hline
 &     &  &                  &$\rho=0.7$    &$\alpha=0$    &$\beta=0$  & \\\hline
100,100  &    &9.7200  &12.0612   &4.5670   &6.9270   &1.3570  &2.7592 \\
100,200  &    &14.3410 &19.5941   &7.0110   &11.1559  &1.0470  &2.2070 \\
200,200  &    &13.6260 &19.1151   &6.7760   &10.9099  &0.5840  &1.2394 \\
200,500  &    &15.4880 &27.5716   &10.5450  &18.7350  &0.5190  &1.1406 \\
500,500  &    &16.9890 &29.5463   &8.2030   &15.1581  &0.3210  &0.7944 \\\hline
 &     &  &                 &$\rho=0$       &$\alpha=0.3$  &$\beta=0$  & \\\hline
100,100  &    &6.5060  &9.1533   &6.1520   &8.6248   &2.3740  &4.0635 \\
100,200  &    &7.5490  &12.4416  &8.7150   &13.4473  &1.6920  &3.1464 \\
200,200  &    &6.2890  &10.8337  &8.4910   &13.2894  &1.0230  &1.9409 \\
200,500  &    &5.1220  &10.1068  &11.3960  &19.4945  &0.8110  &1.5156 \\
500,500  &    &4.7580  &9.5055   &10.3660  &18.7453  &0.4570  &0.9407 \\\hline
 &     &  &                 &$\rho=0$     &$\alpha=0$      &$\beta=0.3$  & \\\hline
100,100  &    &6.6620  &9.2573   &4.7300   &6.9593   &1.7580  &3.1183 \\
100,200  &    &7.8200  &12.5561  &6.1740   &10.2069  &1.4930  &2.6943 \\
200,200  &    &6.4500  &10.9881  &5.8020   &9.7340   &0.7480  &1.4276 \\
200,500  &    &4.9340  &10.3110  &5.9390   &10.6041  &0.7020  &1.3900 \\
500,500  &    &4.0550  &7.7718   &5.8820   &11.0830  &0.3660  &0.8567 \\\hline
 &     &  &                 &$\rho=0.7$   &$\alpha=0.3$    &$\beta=0.3$  & \\\hline
100,100  &    &9.9510  &12.3063  &5.4430   &7.6969   &1.8080  &3.4531 \\
100,200  &    &14.2890 &19.5804  &7.1810   &11.7141  &1.3250  &2.5367 \\
200,200  &    &14.8820 &20.3572  &7.3080   &11.8072  &0.7450  &1.6592 \\
200,500  &    &17.0330 &29.3210  &9.2010   &17.0803  &0.6680  &1.3461 \\
500,500  &    &14.7130 &26.3587  &10.3800  &19.2727  &0.3540  &0.8331 \\\hline
\end{tabular}
\end{center}
\end{table}

\begin{table}[H]
\renewcommand{\arraystretch}{0.8}
\caption{ Simulated mean absolute errors (MAEs) and root mean squared errors (RMSEs) of $\hat{k}_{BKW}$, $\hat{k}_{BHS}$, and $\hat{k}_{QML}$ under DGP 1.B.}
\centering
\label{rotation_full_rank}
\begin{center}
\begin{tabular}{l l r r r r r r} \hline
$N,T$         &  & \multicolumn{2}{c}{$\hat{k}_{BKW}$} &  \multicolumn{2}{c}{$\hat{k}_{BHS}$}  & \multicolumn{2}{c}{$\hat{k}_{QML}$} \\
&  & MAE & RMSE & MAE & RMSE & MAE & RMSE \\ \hline
     &     &  &               &$\rho=0$,     &$\alpha=0$,    &$\beta=0$  & \\\hline
100,100  &    &4.1610  &6.6934   &8.7430    &11.0347    &1.2180  &2.3259 \\
100,200  &    &4.4450  &8.4477   &18.5660   &22.9913    &0.9960  &1.8799 \\
200,200  &    &4.9160  &8.9420   &19.4440   &23.6923    &0.9060  &1.7082 \\
200,500  &    &4.4530  &8.8368   &49.3330   &59.4865    &0.9130  &1.7085 \\
500,500  &    &3.9420  &7.2061   &51.9270   &61.5507    &0.8370  &1.5959 \\\hline
 &     &  &               &$\rho=0.7$     &$\alpha=0$    &$\beta=0$  & \\\hline
100,100  &    &6.4570  &9.4427   &10.1710   &12.3371    &1.9460  &3.7691 \\
100,200  &    &9.1750  &14.8115  &21.3380   &25.1834    &1.8480  &3.6362 \\
200,200  &    &9.6310  &15.0080   &21.5560  &25.2723    &1.7850  &3.5901 \\
200,500  &    &11.4150 &21.3302   &51.9850  &61.9028    &1.6750  &3.4218 \\
500,500  &    &9.5430  &18.4598   &53.6060  &62.7128    &1.6490  &3.5501 \\\hline
 &     &  &               &$\rho=0$     &$\alpha=0.3$    &$\beta=0$  & \\\hline
100,100  &    &3.9840  &6.4778   &7.9990    &10.5485    &1.0910  &2.1824 \\
100,200  &    &4.6820  &8.6151   &17.6010   &22.5002    &1.0360  &1.9432 \\
200,200  &    &4.6350  &8.4454   &21.9190   &26.0996    &0.8770  &1.7306 \\
200,500  &    &4.2690  &8.2870   &50.1790   &61.5307    &0.8600  &1.6474 \\
500,500  &    &4.2040  &8.3094   &54.8050   &64.8615    &0.8040  &1.5492 \\\hline
 &     &  &               &$\rho=0$     &$\alpha=0$    &$\beta=0.3$  & \\\hline
100,100  &    &4.3220  &6.9244   &7.7560    &10.0601    &1.0510  &1.9802 \\
100,200  &    &4.7150  &8.6248   &14.5640   &19.4154    &0.9830  &1.8571 \\
200,200  &    &4.5300  &8.2421   &18.7950   &23.1307    &0.9090  &1.7587 \\
200,500  &    &3.9080  &7.3553   &42.6850   &54.8098    &0.8900  &1.6199 \\
500,500  &    &4.3570  &8.5140   &49.6030   &59.6292    &0.8250  &1.6843 \\\hline
 &     &  &               &$\rho=0.7$     &$\alpha=0.3$    &$\beta=0.3$  & \\\hline
100,100  &    &6.6990  &9.6327    &9.1750    &11.4037   &2.0750  &3.9735 \\
100,200  &    &9.4990  &15.0852   &18.7450   &23.3085   &2.0590  &4.5305 \\
200,200  &    &9.2240  &14.6721   &20.4670   &24.5054   &1.8140  &3.8021 \\
200,500  &    &12.8110 &23.1517   &51.0760   &61.1890   &1.7200  &3.5844 \\
500,500  &    &10.0590 &19.2453   &52.5400   &62.2628   &1.7000  &3.6521 \\\hline
\end{tabular}
\end{center}
\end{table}

\begin{table}[H]
\renewcommand{\arraystretch}{0.7}
\caption{ Simulated mean absolute errors (MAEs) and root mean squared errors (RMSEs) of $\hat{k}_{BKW}$, $\hat{k}_{BHS}$, and $\hat{k}_{QML}$ under DGP 1.C with $N=100,T=100$ among 10000 replications.}
\centering
\label{full_rank_to_singular}
\begin{center}
\begin{tabular}{l l r r r r r r} \hline
$m$        &  & \multicolumn{2}{c}{$\hat{k}_{BKW}$} &  \multicolumn{2}{c}{$\hat{k}_{BHS}$}  & \multicolumn{2}{c}{$\hat{k}_{QML}$} \\
&  & MAE & RMSE & MAE & RMSE & MAE & RMSE \\ \hline
     &     &  &               &$\rho=0$     &$\alpha=0$    &$\beta=0$  & \\\hline
1    &  &3.9228  &6.4437  &7.2579  &9.7141  &0.6562  &1.2903 \\
0.8  &  &3.9425  &6.4624  &6.6330  &9.1145  &0.6348  &1.2559 \\
0.5  &  &3.7847  &6.2319  &5.4950  &7.9789  &0.5420  &1.0814 \\
0.1  &  &3.8469  &6.2895  &4.6050  &6.9212  &0.5093  &1.0568 \\
0    &  &3.8310  &6.2414  &4.4915  &6.8352  &0.4969  &1.0315 \\\hline
 &     &  &               &$\rho=0.7$     &$\alpha=0$    &$\beta=0$  & \\\hline
1    &  &6.0404  &9.0280  &9.3131  &11.5733  &0.9478  &2.0680 \\
0.8  &  &6.0063  &9.0017  &8.5168  &10.9192  &0.8547  &1.8960 \\
0.5  &  &5.9803  &8.9390  &6.5127  &9.0641  &0.6925  &1.5752 \\
0.1  &  &5.9300  &8.8833  &4.6894  &7.0610  &0.5178 &1.2335 \\
0    &  &6.0197  &8.9771  &4.5440  &6.9049  &0.5070 &1.2057 \\\hline
 &     &  &               &$\rho=0$     &$\alpha=0.3$    &$\beta=0$  & \\\hline
1    &  &3.8349  &6.2423  &7.1824  &9.7359  &0.6727  &1.3234 \\
0.8  &  &3.8234  &6.2331  &6.6551  &9.2338  &0.6535  &1.2963 \\
0.5  &  &3.8345  &6.3110  &5.8040  &8.3371  &0.6152  &1.2362 \\
0.1  &  &3.9127  &6.4083  &5.0645  &7.4846  &0.5895 &1.1644 \\
0    &  &3.9188  &6.4124  &4.9815  &7.3974  &0.5813 &1.1551  \\\hline
 &     &  &               &$\rho=0$     &$\alpha=0$    &$\beta=0.3$  & \\\hline
1    &  &3.8250  &6.3150  &6.2535  &8.7224 &0.6622  &1.3039 \\
0.8  &  &3.8135  &6.2932  &5.6808  &8.1438 &0.6259  &1.2379 \\
0.5  &  &3.8253  &6.3061  &4.6189  &6.9328  &0.5619 &1.1171 \\
0.1  &  &3.9120  &6.4147  &3.9299  &6.0820  &0.5424 &1.0949 \\
0    &  &3.8176  &6.2881  &3.8963  &6.0564  &0.5199 &1.0497 \\\hline
 &     &  &               &$\rho=0.7$     &$\alpha=0.3$    &$\beta=0.3$  & \\\hline
1    &  &6.0745  &9.0347  &8.0648  &10.5304  &1.0515  &2.2669 \\
0.8  &  &6.0041  &8.9542  &7.3126  &9.8433  &0.9338  &2.0173 \\
0.5  &  &6.0519  &9.0124  &5.8471  &8.4490  &0.7798  &1.7537 \\
0.1  &  &6.0120  &8.9694  &4.6376  &7.1447  &0.6100 &1.4401 \\
0    &  &6.0379  &8.9861  &4.5336  &7.0337  &0.5850 &1.3509 \\\hline
\end{tabular}
\end{center}
\end{table}

\begin{table}[H]
\renewcommand{\arraystretch}{0.8}
\caption{Simulated mean absolute errors (MAEs) and root mean squared errors (RMSEs) of $\hat{k}_{BKW}$, $\hat{k}_{BHS}$, and $\hat{k}_{QML}$ under DGP 1.D.}
\centering
\label{singular_pre_post_break}
\begin{center}
\begin{tabular}{l l r r r r r r} \hline
$N,T$         &  & \multicolumn{2}{c}{$\hat{k}_{BKW}$} &  \multicolumn{2}{c}{$\hat{k}_{BHS}$}  & \multicolumn{2}{c}{$\hat{k}_{QML}$} \\
&  & MAE & RMSE & MAE & RMSE & MAE & RMSE \\ \hline
     &     &  &               &$\rho=0$,     &$\alpha=0$,    &$\beta=0$  & \\\hline
100,100  &    &0.4330  &1.3494   &0.0370   &0.1975   &0.0260  &0.1673 \\
100,200  &    &0.3380  &1.0900   &0.0300   &0.1732   &0.0240  &0.1549 \\
200,200  &    &0.2780  &0.7668   &0.0180   &0.1342   &0.0130  &0.1140 \\
200,500  &    &0.2850  &0.8155   &0.0070   &0.0837   &0.0100  &0.1000 \\\hline
 &     &  &               &$\rho=0.7$     &$\alpha=0$    &$\beta=0$  & \\\hline
100,100  &    &1.8760  &4.8750   &0.0120   &0.1095   &0.0110  &0.1049 \\
100,200  &    &1.1140  &4.0007   &0.0150   &0.1225   &0.0110  &0.1140 \\
200,200  &    &0.8700  &3.5000   &0.0050   &0.0707   &0.0020  &0.0447 \\
200,500  &    &0.4070  &1.3435   &0.0030   &0.0548   &0.0010  &0.0316 \\\hline
 &     &  &               &$\rho=0$     &$\alpha=0.3$    &$\beta=0$  & \\\hline
100,100  &    &0.4400  &1.4519   &0.0450   &0.2302   &0.0410  &0.2258 \\
100,200  &    &0.3590  &1.3802   &0.0440   &0.2145   &0.0340  &0.1897 \\
200,200  &    &0.3080  &0.8438   &0.0150   &0.1225   &0.0140  &0.1265 \\
200,500  &    &0.2150  &0.6656   &0.0160   &0.1265   &0.0120  &0.1095 \\\hline
 &     &  &               &$\rho=0$     &$\alpha=0$    &$\beta=0.3$  & \\\hline
100,100  &    &0.3710  &1.0747   &0.0380   &0.1949   &0.0360  &0.1897 \\
100,200  &    &0.2850  &0.7918   &0.0340   &0.1897   &0.0220  &0.1483 \\
200,200  &    &0.3150  &0.8972   &0.0100   &0.1000   &0.0110  &0.1049 \\
200,500  &    &0.2380  &0.6885   &0.0120   &0.1183   &0.0050  &0.0707 \\\hline
 &     &  &               &$\rho=0.7$     &$\alpha=0.3$    &$\beta=0.3$  & \\\hline
100,100  &    &1.9420  &4.8557   &0.0260   &0.1612   &0.0180  &0.1414 \\
100,200  &    &1.0170  &3.6438   &0.0220   &0.1549   &0.0090  &0.0949 \\
200,200  &    &0.9750  &3.9242   &0.0060   &0.0775   &0.0080  &0.0894 \\
200,500  &    &0.6390  &2.5879   &0.0080   &0.0894   &0.0050  &0.0707 \\\hline
\end{tabular}
\end{center}
\end{table}
Tables \ref{rotation_singular_correct_Pro}--\ref{singular_pre_post_correct_Pro} present the probabilities of the correct estimation of the break date.
The results are consistent with those displayed in Tables \ref{rotation_singular}--\ref{singular_pre_post_break}: the QML estimator $\hat{k}_{QML}$ can detect the true break date with higher
probabilities than others regardless of the values of $(\rho,\alpha,\beta)$.
The MS method sometimes detects more than one or no break; hence, we only compute its probability of correctly estimating $k_0$ under the condition that it detects a single break. The probabilities of a
correct estimation of the QML method increase with the sample sizes $N$ and $T$ in Tables \ref{rotation_singular_correct_Pro}, \ref{rotation_full_rank_correct_Pro}, and
\ref{singular_pre_post_correct_Pro}.

Table \ref{full_rank_to_singular_correct_Pro} shows that the probabilities of correct estimation of the QML estimators increase as $m$ decreases. A smaller $m$ means that $C$ is closer to a singular
matrix. Table \ref{full_rank_to_singular_correct_Pro} is consistent with Table \ref{full_rank_to_singular}, and confirms Theorems \ref{bound_theorem} and \ref{consistency}. To explore in more detail the
effect of changes in $m$ on the QML estimator, we vary the value of $m$ using finer grids and find a similar pattern to that shown in Table \ref{full_rank_to_singular_correct_Pro}. The results are
reported in the supplementary appendix.

Figures \ref{1A_NT100} and \ref{1A_NT500} show the frequency of the estimated change points under DGP 1.A for $N=100,T=100$ and $N=500,T=500$ for 1000 replications. According to these figures, the QML
estimators exhibit the highest frequency around the true break under different settings. When we increase the $(N,T)$ value from $100$ to $500$, the frequency at the true break point increases and the
simulated distribution becomes tighter. This indicates that the QML estimators are highly likely to identify the true break point. This is consistent with our theory. However, the other three methods
are found to have much larger variation and substantially lower probabilities to correctly estimate the break point. Thus, the QML estimators are advantageous in this case.
Moreover, the simulation result indicates that for a sample size exceeding $N = 5000, T = 1000$, the probabilities of correctly estimating the QML estimator exceed $90\%$.

\begin{figure}[htbp]
\centering
\subfigure[$(\rho,\alpha,\beta)=(0.7,0,0)$]{
\includegraphics[width=7.0cm]{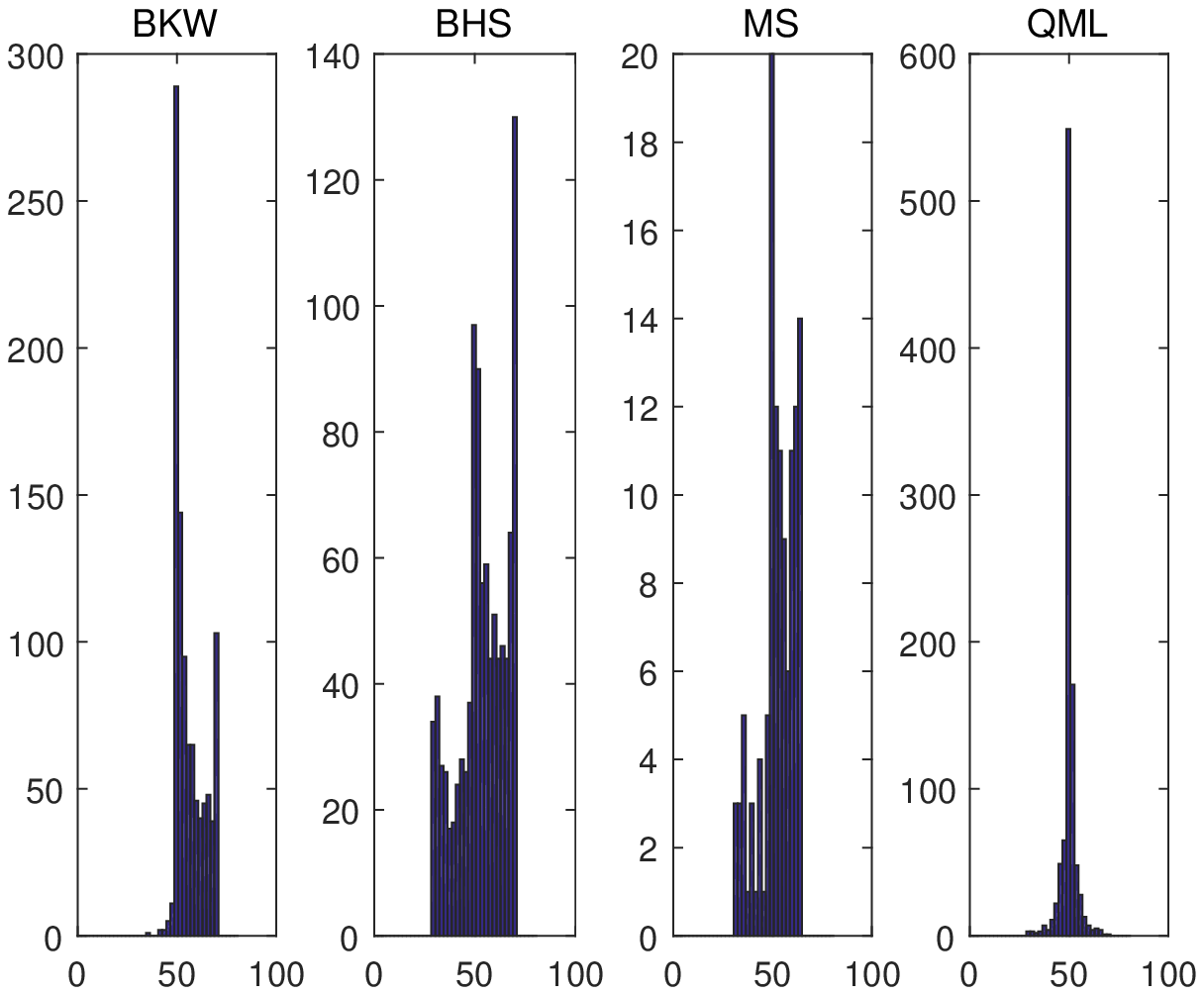}
%\caption{fig1}
}
\quad
\subfigure[$(\rho,\alpha,\beta)=(0,0.3,0)$]{
\includegraphics[width=7.0cm]{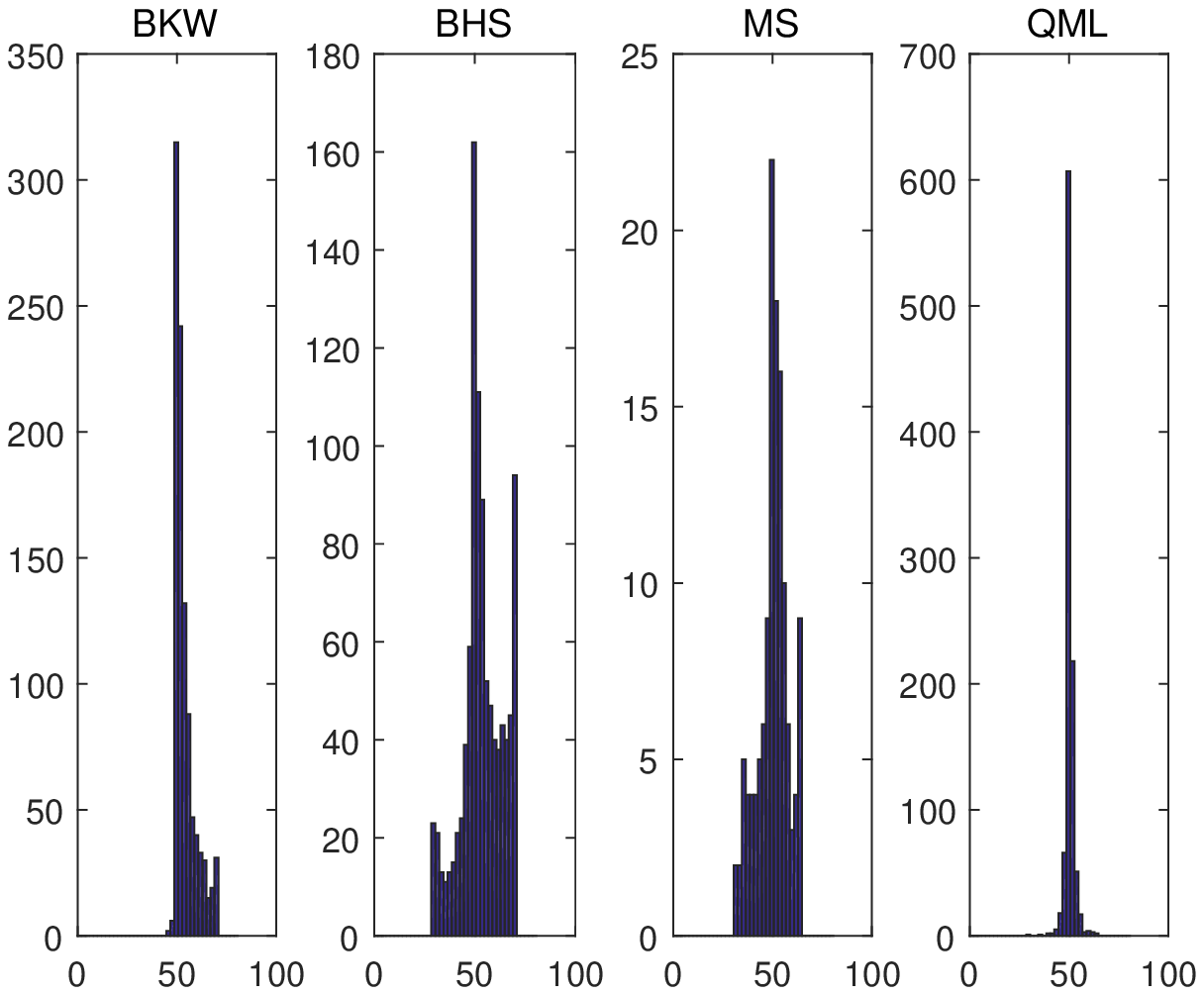}
}
\quad
\subfigure[$(\rho,\alpha,\beta)=(0,0,0.3)$]{
\includegraphics[width=7.0cm]{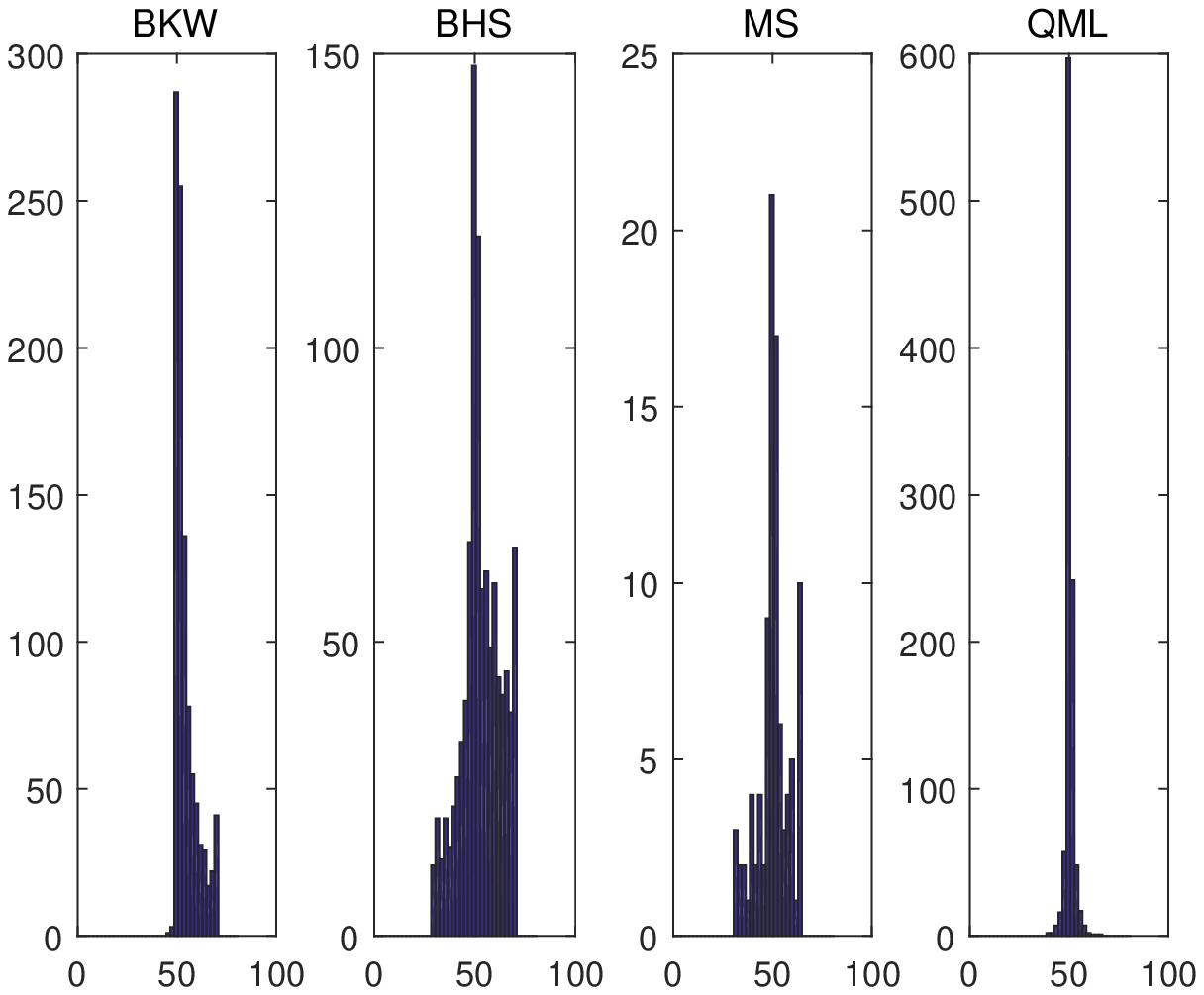}
}
\quad
\subfigure[$(\rho,\alpha,\beta)=(0.7,0.3,0.3)$]{
\includegraphics[width=7.0cm]{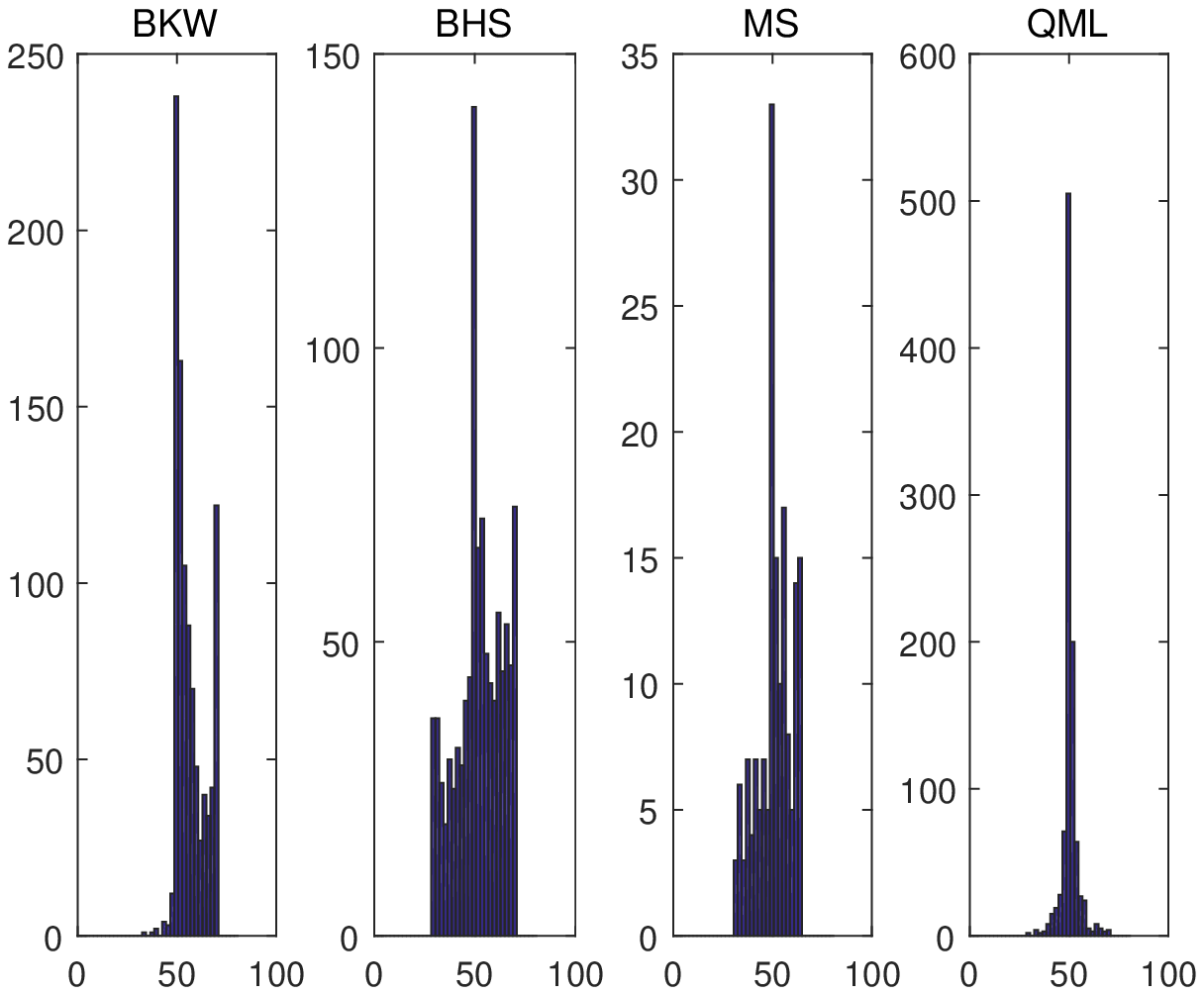}
}
\caption{Plots of the frequency of the estimated break points among 1000 replications for DGP 1.B and $N=100,T=100$.}
\label{1B_NT100}
\end{figure}

\begin{figure}[htbp]
\centering
\subfigure[$(\rho,\alpha,\beta)=(0.7,0,0)$]{
\includegraphics[width=7.0cm]{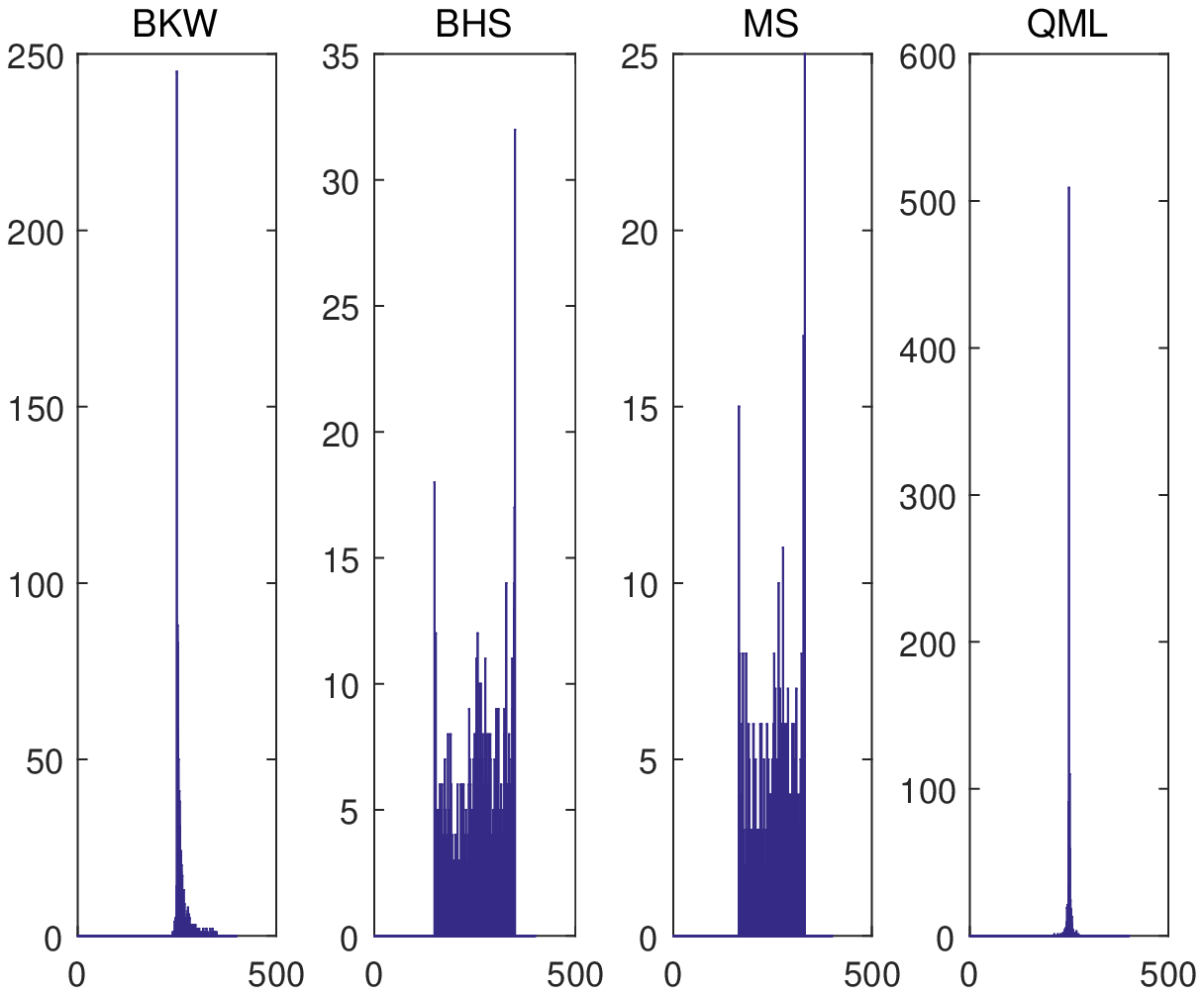}
%\caption{fig1}
}
\quad
\subfigure[$(\rho,\alpha,\beta)=(0,0.3,0)$]{
\includegraphics[width=7.0cm]{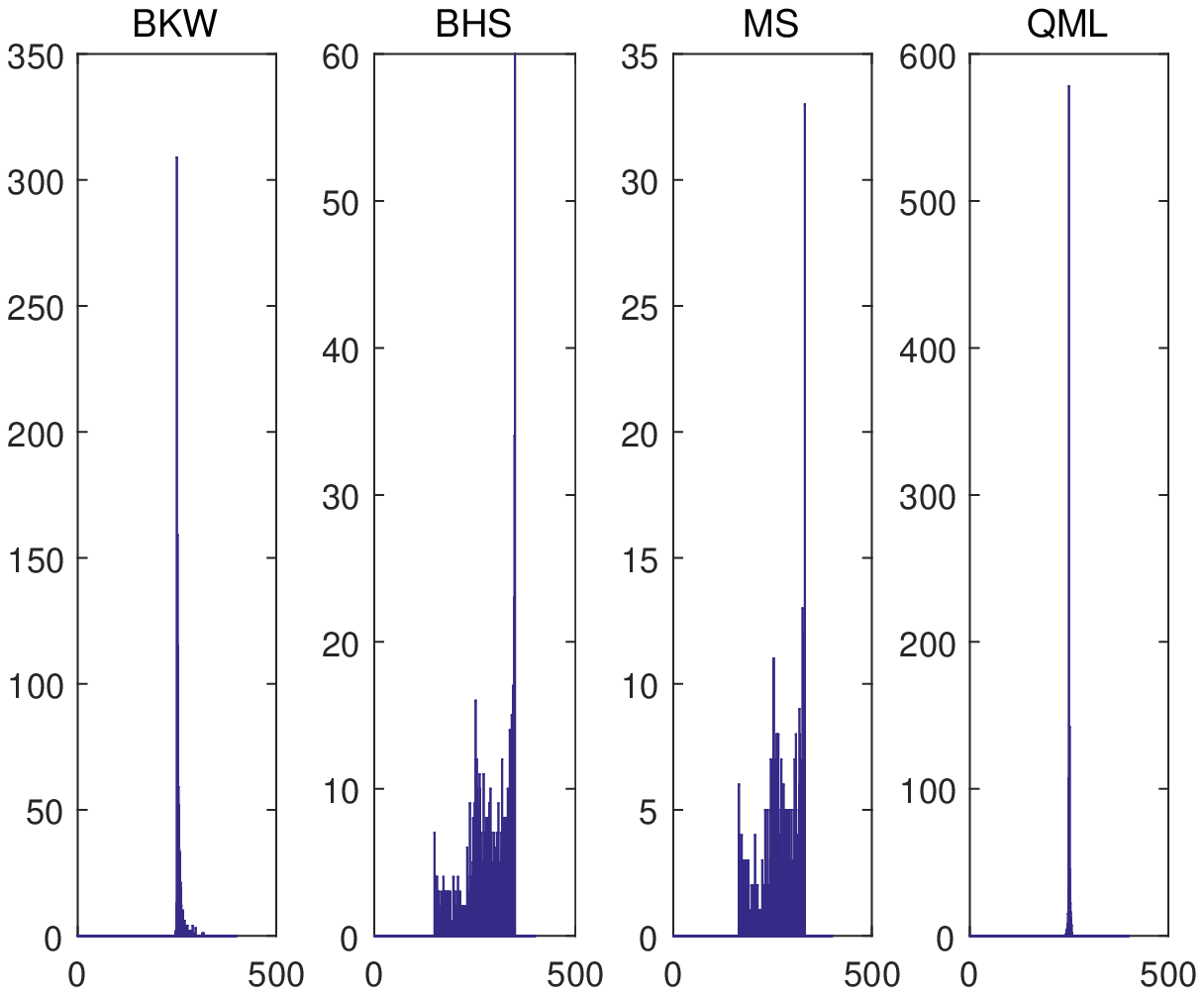}
}
\quad
\subfigure[$(\rho,\alpha,\beta)=(0,0,0.3)$]{
\includegraphics[width=7.0cm]{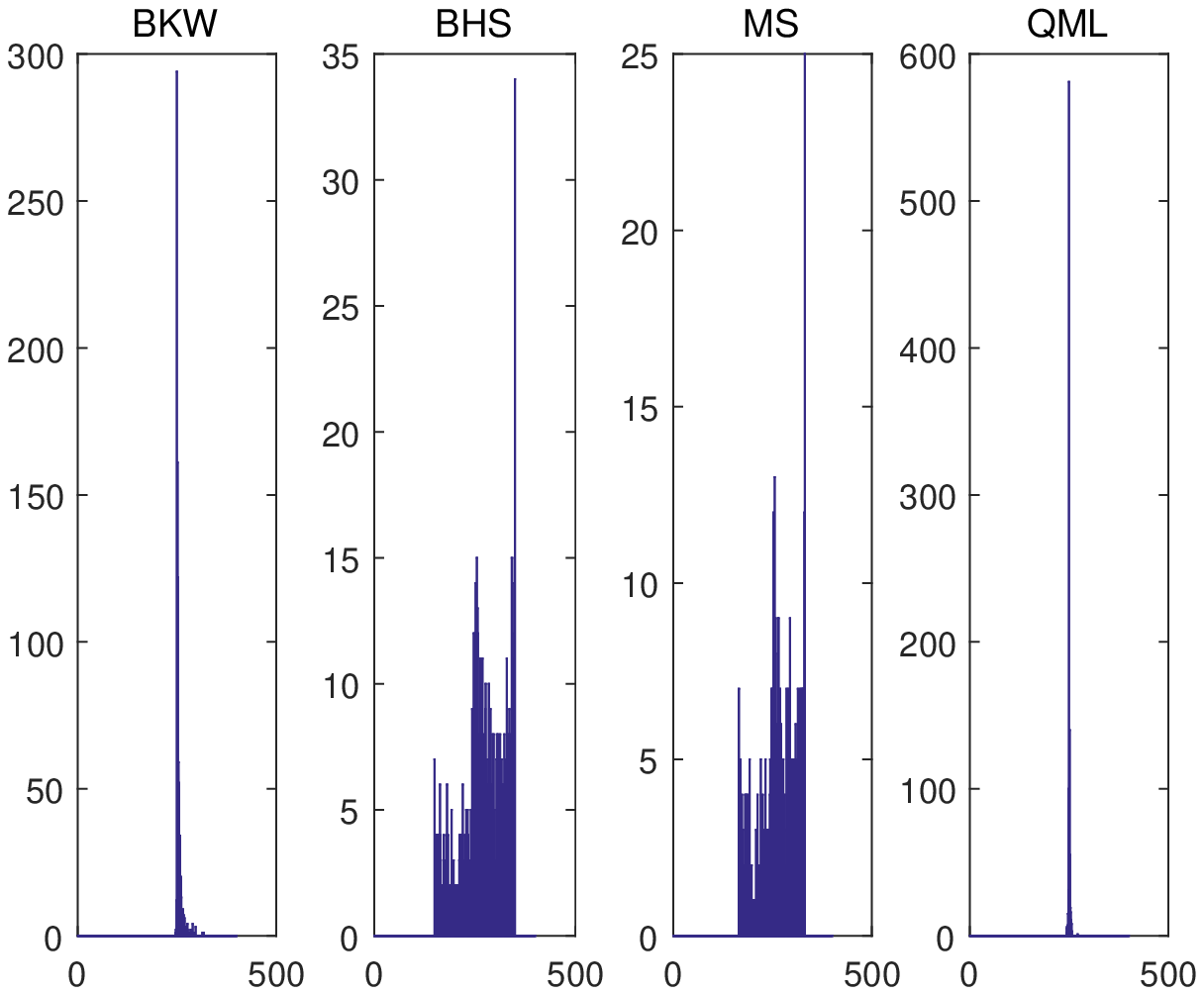}
}
\quad
\subfigure[$(\rho,\alpha,\beta)=(0.7,0.3,0.3)$]{
\includegraphics[width=7.0cm]{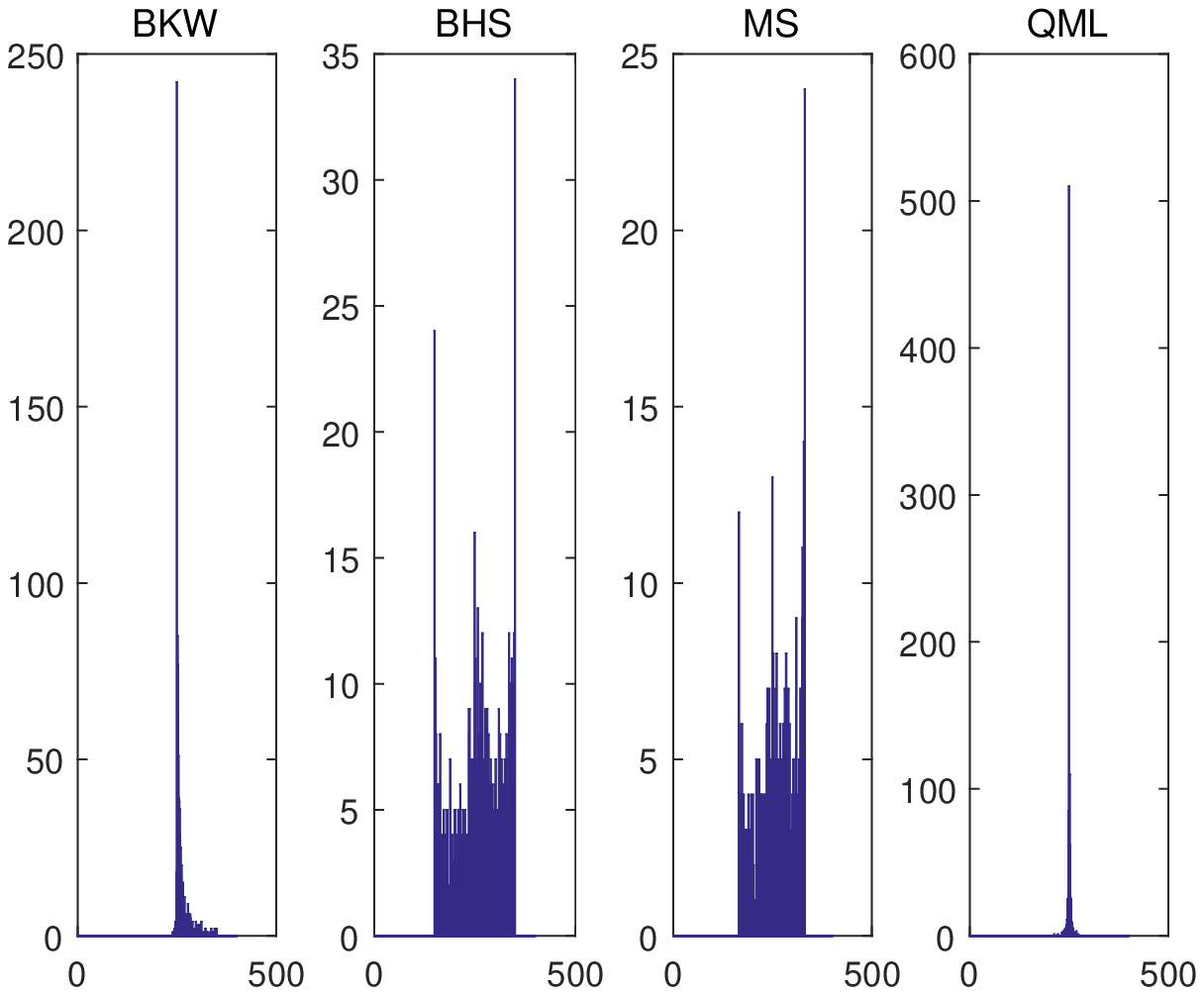}
}
\caption{Plots of the frequency of the estimated break points among 1000 replications for DGP 1.B and $N=500,T=500$.}
\label{1B_NT500}
\end{figure}

Recall that BKW and QML only have $O_p(1)$ estimation errors under DGP 1.B. However, Table \ref{rotation_full_rank_correct_Pro} shows that in all cases, the probabilities of correct estimation by the
QML estimator are much higher than those of correct estimation by the BKW estimator
Apparently, the BHS and MS methods cannot accurately estimate the true break point in this case.
Figures \ref{1B_NT100} and \ref{1B_NT500} show the distributions of the estimated change points under (1.B) for $N=100,T=100$ and $N=500,T=500$,
indicating that BHS and MS cannot handle rotational changes.
Although the estimation errors of BKW and QML are bounded under all settings, the QML estimators have a much tighter distribution around the true break point.

\begin{table}[H]
\renewcommand{\arraystretch}{0.8}
\caption{Probability of correct estimation under DGP 1.A.}
\centering
\label{rotation_singular_correct_Pro}
\begin{center}
\begin{tabular}{l l r r r r r r} \hline
$N,T$         &  & \multicolumn{1}{c}{$\hat{k}_{BKW}$} &  \multicolumn{1}{c}{$\hat{k}_{BHS}$} &  \multicolumn{1}{c}{$\hat{k}_{MS}$} & \multicolumn{1}{c}{$\hat{k}_{QML}$} \\
&  &  &  & & &  \\ \hline
  &               &$\rho=0$     &$\alpha=0$    &$\beta=0$  & \\\hline
100,100&  &0.1530    &0.1440  &0.1626   &0.4220         \\
100,200&  &0.1920    &0.1510  &0.1863   &0.4370         \\
200,200&  &0.2340    &0.1780  &0.1307   &0.5680         \\
200,500&  &0.2540    &0.2030  &0.2020   &0.5780 \\
500,500&  &0.2990    &0.2100  &0.2123   &0.7290    \\\hline
 &                  &$\rho=0.7$    &$\alpha=0$    &$\beta=0$  & \\\hline
100,100&  &0.1050    &0.2050  &0.2329   &0.5290         \\
100,200&  &0.1250    &0.1850  &0.1779   &0.5510         \\
200,200&  &0.1390    &0.1920  &0.1898   &0.6660         \\
200,500&  &0.1750    &0.1890  &0.2031   &0.6940        \\
500,500&  &0.2100    &0.2420  &0.2306   &0.7810 \\\hline
&                 &$\rho=0$       &$\alpha=0.3$  &$\beta=0$  & \\\hline
100,100&  &0.1790    &0.1300  &0.1072   &0.3280         \\
100,200&  &0.1850    &0.1380  &0.1897   &0.4090         \\
200,200&  &0.2260    &0.1650  &0.1931   &0.5320         \\
200,500&  &0.2530    &0.1730  &0.1845   &0.5650          \\
500,500&  &0.2750    &0.1920  &0.1964   &0.6880   \\\hline
 &                 &$\rho=0$     &$\alpha=0$      &$\beta=0.3$  & \\\hline
100,100&  &0.1480    &0.1700  &0.1956   &0.3840         \\
100,200&  &0.1730    &0.1810  &0.1847   &0.4210         \\
200,200&  &0.2240    &0.2110  &0.2069   &0.5700         \\
200,500&  &0.2770    &0.2250  &0.2370   &0.5930         \\
500,500&  &0.3220    &0.2790  &0.2790   &0.7500  \\\hline
  &                 &$\rho=0.7$   &$\alpha=0.3$    &$\beta=0.3$  & \\\hline
100,100&  &0.1070   &0.1510  &0.1739   &0.4670         \\
100,200&  &0.1210   &0.1860  &0.2157   &0.5030         \\
200,200&  &0.1370   &0.1820  &0.2072   &0.6360         \\
200,500&  &0.1670   &0.2180  &0.2149   &0.6520         \\
500,500&  &0.1900   &0.2510  &0.2427   &0.7640 \\\hline
\end{tabular}
\end{center}
\end{table}

\begin{table}[H]
\renewcommand{\arraystretch}{0.8}
\caption{Probability of correct estimation under DGP 1.B.}
\centering
\label{rotation_full_rank_correct_Pro}
\begin{center}
\begin{tabular}{l l r r r r r r} \hline
$N,T$         &  & \multicolumn{1}{c}{$\hat{k}_{BKW}$} &  \multicolumn{1}{c}{$\hat{k}_{BHS}$} &  \multicolumn{1}{c}{$\hat{k}_{MS}$} & \multicolumn{1}{c}{$\hat{k}_{QML}$} \\
&  &  &  & & &  \\ \hline
  &               &$\rho=0$     &$\alpha=0$    &$\beta=0$  & \\\hline
100,100&  &0.2760    &0.0690  &0.0769   &0.4790         \\
100,200&  &0.2920    &0.0540  &0.0362   &0.5180         \\
200,200&  &0.2720    &0.0320  &0.0655   &0.5270         \\
200,500&  &0.3110    &0.0140  &0.0091   &0.5340      \\
500,500&  &0.2960    &0.0100   &0.0123  &0.5580  \\\hline
 &                  &$\rho=0.7$    &$\alpha=0$    &$\beta=0$  & \\\hline
100,100&  &0.2710    &0.0640  &0.0909   &0.4540         \\
100,200&  &0.2500    &0.0270  &0.0398   &0.4530         \\
200,200&  &0.2180    &0.0160  &0.0200   &0.4790         \\
200,500&  &0.2370    &0.0120  &0.0144   &0.4970         \\
500,500&  &0.2450    &0.0080  &0.0080   &0.5090  \\\hline
&                 &$\rho=0$       &$\alpha=0.3$  &$\beta=0$  & \\\hline
100,100&  &0.3050    &0.1060  &0.1163   &0.5180         \\
100,200&  &0.2930    &0.0740  &0.0989   &0.5020         \\
200,200&  &0.2890    &0.0390  &0.0496   &0.5540         \\
200,500&  &0.3000    &0.0230  &0.0328   &0.5630         \\
500,500&  &0.3090    &0.0090  &0.0125   &0.5780  \\\hline
 &                 &$\rho=0$     &$\alpha=0$      &$\beta=0.3$  & \\\hline
100,100&  &0.2740    &0.0880  &0.1458   &0.5000         \\
100,200&  &0.2970    &0.0650  &0.0692   &0.5220         \\
200,200&  &0.2870    &0.0390  &0.0338   &0.5390         \\
200,500&  &0.3100    &0.0300  &0.0320   &0.5290         \\
500,500&  &0.2940    &0.0120  &0.0123   &0.5810   \\\hline
  &                 &$\rho=0.7$   &$\alpha=0.3$    &$\beta=0.3$  & \\\hline
100,100&  &0.2210    &0.1000  &0.1524   &0.4330         \\
100,200&  &0.2400    &0.0610  &0.0763   &0.4640         \\
200,200&  &0.2370    &0.0490  &0.0538   &0.4810         \\
200,500&  &0.2230    &0.0230  &0.0218   &0.4770         \\
500,500&  &0.2420    &0.0160   &0.0207  &0.5100  \\\hline
\end{tabular}
\end{center}
\end{table}

\begin{table}[H]
\renewcommand{\arraystretch}{0.7}
\caption{Probability of correct estimation under DGP 1.C with $N=100,T=100$.}
\centering
\label{full_rank_to_singular_correct_Pro}
\begin{center}
\begin{tabular}{l l r r r r r r} \hline
$m$         &  & \multicolumn{1}{c}{$\hat{k}_{BKW}$} &  \multicolumn{1}{c}{$\hat{k}_{BHS}$} &  \multicolumn{1}{c}{$\hat{k}_{MS}$} & \multicolumn{1}{c}{$\hat{k}_{QML}$} \\
&  &  &  & & &  \\ \hline
  &               &$\rho=0$     &$\alpha=0$    &$\beta=0$  & \\\hline
1  &   &0.3044    &0.1075  &0.1188   &0.6079         \\
0.8&   &0.3033    &0.1252  &0.1389   &0.6153         \\
0.5&   &0.3009    &0.1736  &0.1904   &0.6467         \\
0.1&   &0.2976    &0.1998  &0.2014   &0.6680         \\
0  &   &0.2977    &0.2031  &0.2192   &0.6705         \\\hline
  &               &$\rho=0.7$     &$\alpha=0$    &$\beta=0$  & \\\hline
1  &   &0.2841    &0.0781  &0.1040   &0.6051         \\
0.8&   &0.2871    &0.0975  &0.1135   &0.6254         \\
0.5&   &0.2896    &0.1532  &0.1620   &0.6641        \\
0.1&   &0.2876    &0.2073  &0.2369   &0.7131         \\
0  &   &0.2848    &0.2194  &0.2297   &0.7154         \\\hline
 &               &$\rho=0$     &$\alpha=0.3$    &$\beta=0$  & \\\hline
1  &   &0.2973    &0.1226  &0.1442   &0.6063         \\
0.8&   &0.2981    &0.1416  &0.1648   &0.6134         \\
0.5&   &0.2988    &0.1641  &0.1730   &0.6219         \\
0.1&   &0.2993    &0.1828  &0.1954   &0.6316         \\
0  &   &0.2988    &0.1860  &0.1995   &0.6342         \\\hline
 &               &$\rho=0$     &$\alpha=0$    &$\beta=0.3$  & \\\hline
1  &   &0.3018    &0.1344  &0.1399   &0.6075         \\
0.8&   &0.3016    &0.1549  &0.1679   &0.6164         \\
0.5&   &0.3044    &0.1927  &0.2078   &0.6383         \\
0.1&   &0.3009    &0.2141  &0.2093   &0.6461        \\
0  &   &0.3036    &0.2211  &0.2314   &0.6566         \\\hline
 &               &$\rho=0.7$     &$\alpha=0.3$    &$\beta=0.3$  & \\\hline
1  &   &0.2821    &0.1519  &0.1739   &0.5921         \\
0.8&   &0.2844    &0.1710  &0.1964   &0.6082         \\
0.5&   &0.2843    &0.2327  &0.2402   &0.6496         \\
0.1&   &0.2850    &0.2889  &0.2911   &0.6898         \\
0  &   &0.2868    &0.2951  &0.2966   &0.6951         \\\hline
\end{tabular}
\end{center}
\end{table}

\begin{table}[H]
\renewcommand{\arraystretch}{0.8}
\caption{Probability of correct estimation under DGP 1.D.}
\centering
\label{singular_pre_post_correct_Pro}
\begin{center}
\begin{tabular}{l l r r r r r r} \hline
$N,T$         &  & \multicolumn{1}{c}{$\hat{k}_{BKW}$} &  \multicolumn{1}{c}{$\hat{k}_{BHS}$} &  \multicolumn{1}{c}{$\hat{k}_{MS}$} & \multicolumn{1}{c}{$\hat{k}_{QML}$} \\
&  &  &  & & &  \\ \hline
  &               &$\rho=0$     &$\alpha=0$    &$\beta=0$  & \\\hline
100,100&  &0.7960    &0.9640  &0.9553   &0.9750         \\
100,200&  &0.8200    &0.9700  &0.9700   &0.9760         \\
200,200&  &0.8160    &0.9820  &0.9841   &0.9870         \\
200,500&  &0.8260    &0.9930  &0.9930   &0.9900         \\\hline
 &                  &$\rho=0.7$    &$\alpha=0$    &$\beta=0$  & \\\hline
100,100&  &0.7000    &0.9880  &0.9864   &0.9890         \\
100,200&  &0.7540    &0.9850  &0.9859   &0.9900         \\
200,200&  &0.7950    &0.9950  &0.9949   &0.9980         \\
200,500&  &0.8220    &0.9970  &0.9970   &0.9990         \\\hline
&                 &$\rho=0$       &$\alpha=0.3$  &$\beta=0$  & \\\hline
100,100&  &0.8020    &0.9580  &0.9563   &0.9630         \\
100,200&  &0.8170    &0.9570  &0.9589   &0.9670         \\
200,200&  &0.8140    &0.9850  &0.9842   &0.9870         \\
200,500&  &0.8510    &0.9840  &0.9840   &0.9880         \\\hline
 &                 &$\rho=0$     &$\alpha=0$      &$\beta=0.3$  & \\\hline
100,100&  &0.7910    &0.9620  &0.9671   &0.9640         \\
100,200&  &0.8090    &0.9670  &0.9674   &0.9780         \\
200,200&  &0.8150    &0.9900  &0.9904   &0.9890         \\
200,500&  &0.8330    &0.9890  &0.9890   &0.9950         \\\hline
  &                 &$\rho=0.7$   &$\alpha=0.3$    &$\beta=0.3$  & \\\hline
100,100&  &0.6670    &0.9740  &0.9766   &0.9830         \\
100,200&  &0.7670    &0.9790  &0.9801   &0.9910         \\
200,200&  &0.7910    &0.9940  &0.9940   &0.9920         \\
200,500&  &0.7900    &0.9920  &0.9920   &0.9950         \\\hline
\end{tabular}
\end{center}
\end{table}

\begin{figure}[htbp]
\centering
\subfigure[$(\rho,\alpha,\beta)=(0.7,0,0)$]{
\includegraphics[width=7.0cm]{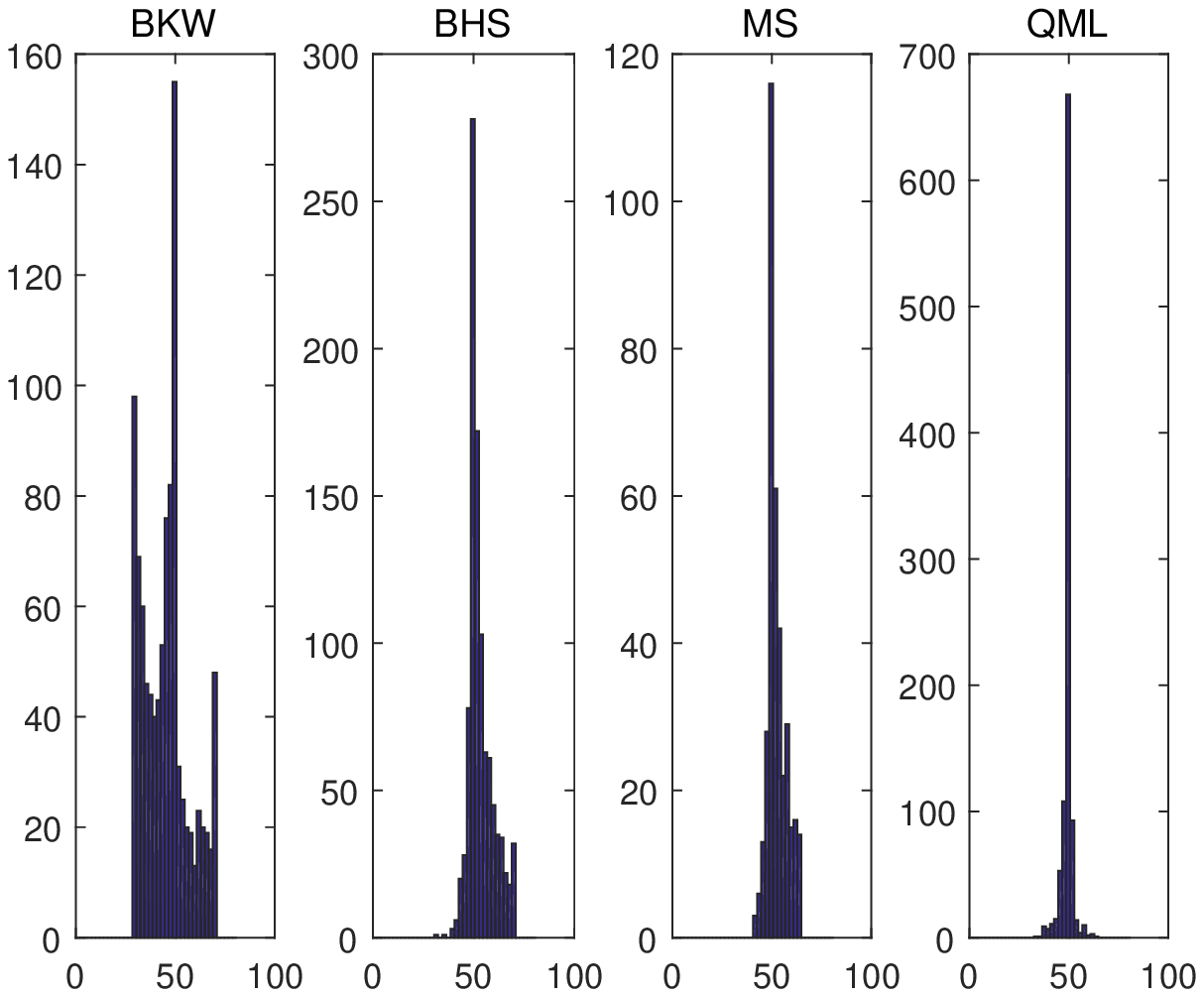}
%\caption{fig1}
}
\quad
\subfigure[$(\rho,\alpha,\beta)=(0,0.3,0)$]{
\includegraphics[width=7.0cm]{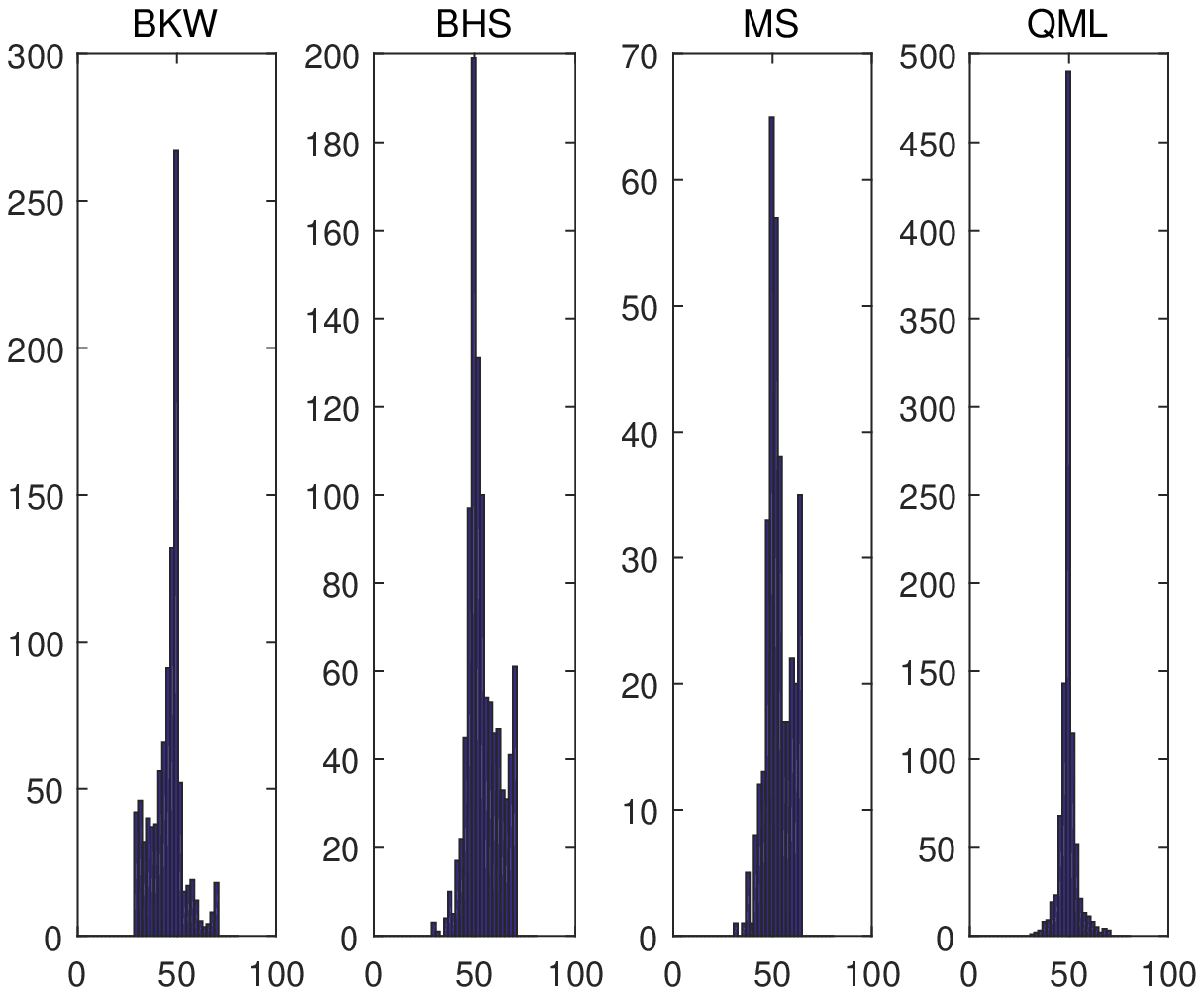}
}
\quad
\subfigure[$(\rho,\alpha,\beta)=(0,0,0.3)$]{
\includegraphics[width=7.0cm]{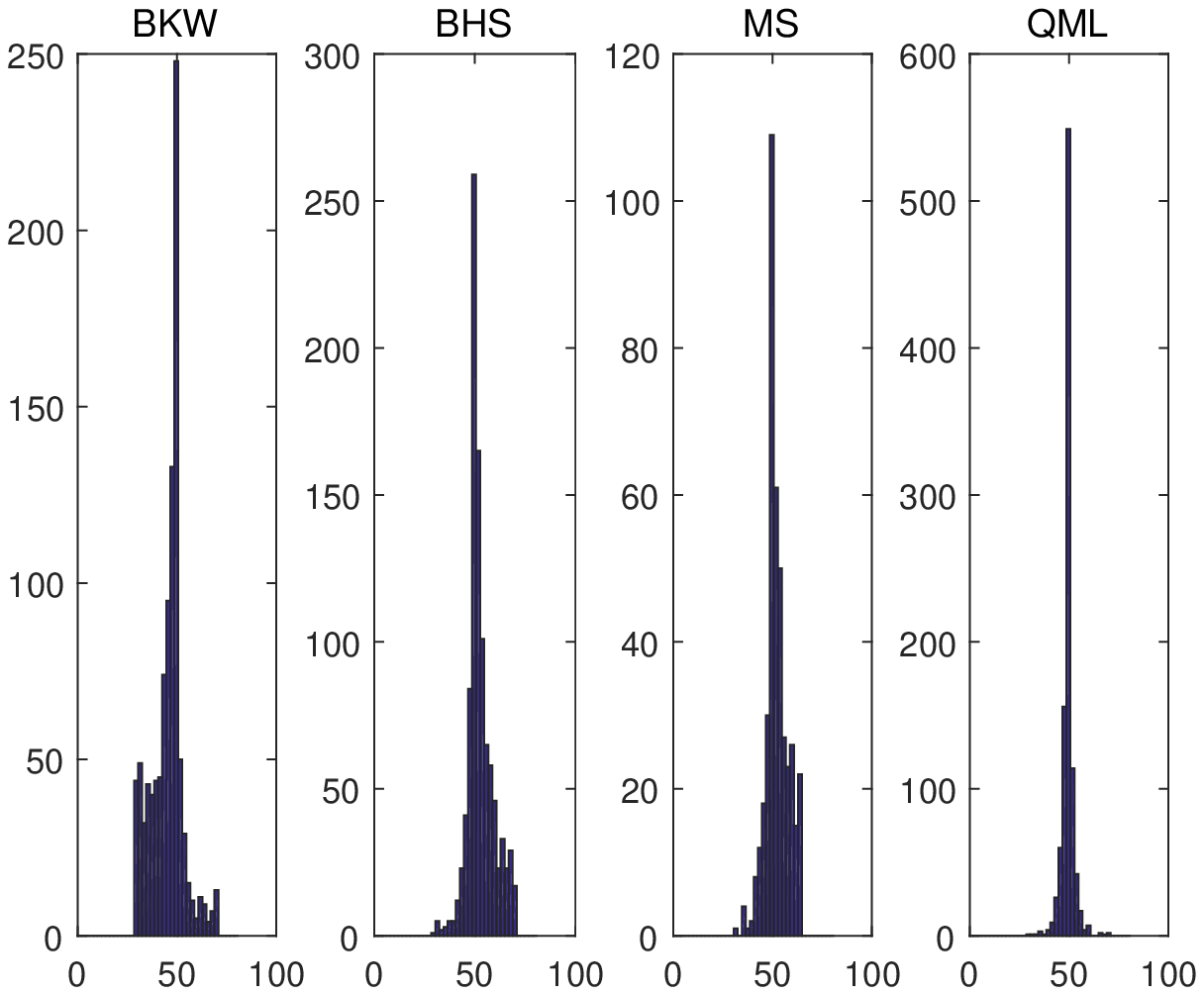}
}
\quad
\subfigure[$(\rho,\alpha,\beta)=(0.7,0.3,0.3)$]{
\includegraphics[width=7.0cm]{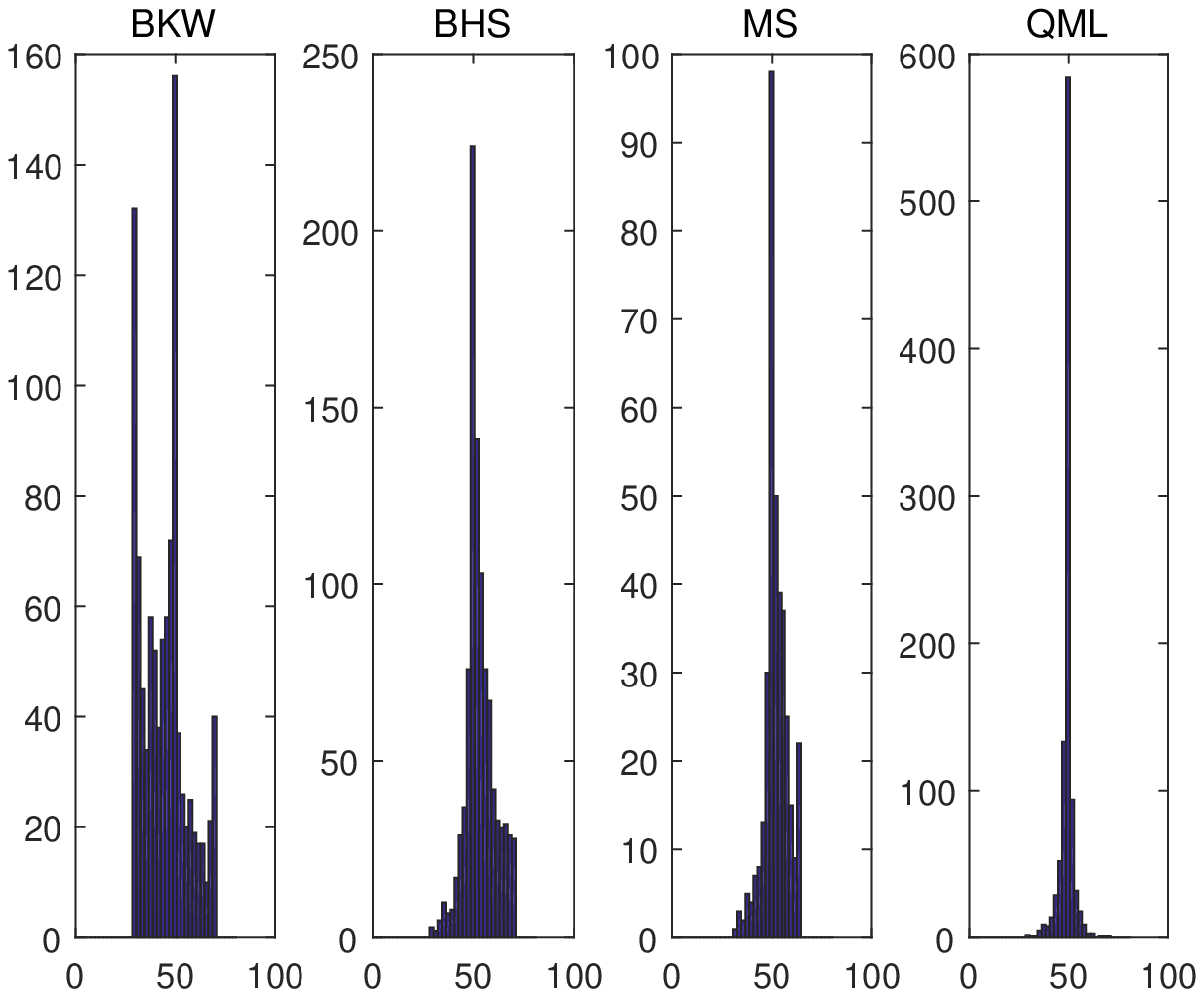}
}
\caption{Plots of the frequency of the estimated break points among 1000 replications for DGP 1.A and $N=100,T=100$.}
\label{1A_NT100}
\end{figure}

\begin{figure}[htbp]
\centering
\subfigure[$(\rho,\alpha,\beta)=(0.7,0,0)$]{
\includegraphics[width=7.0cm]{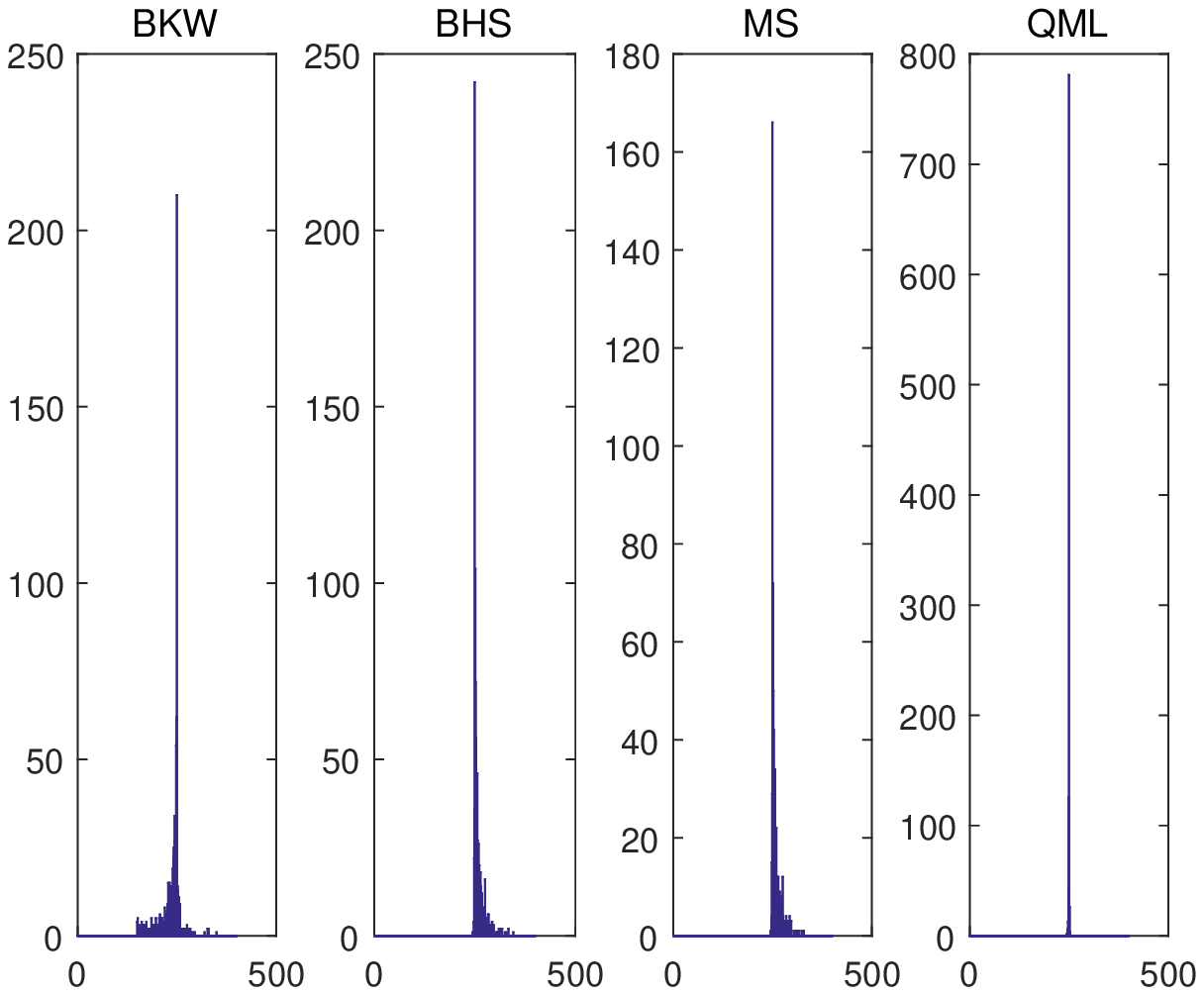}
%\caption{fig1}
}
\quad
\subfigure[$(\rho,\alpha,\beta)=(0,0.3,0)$]{
\includegraphics[width=7.0cm]{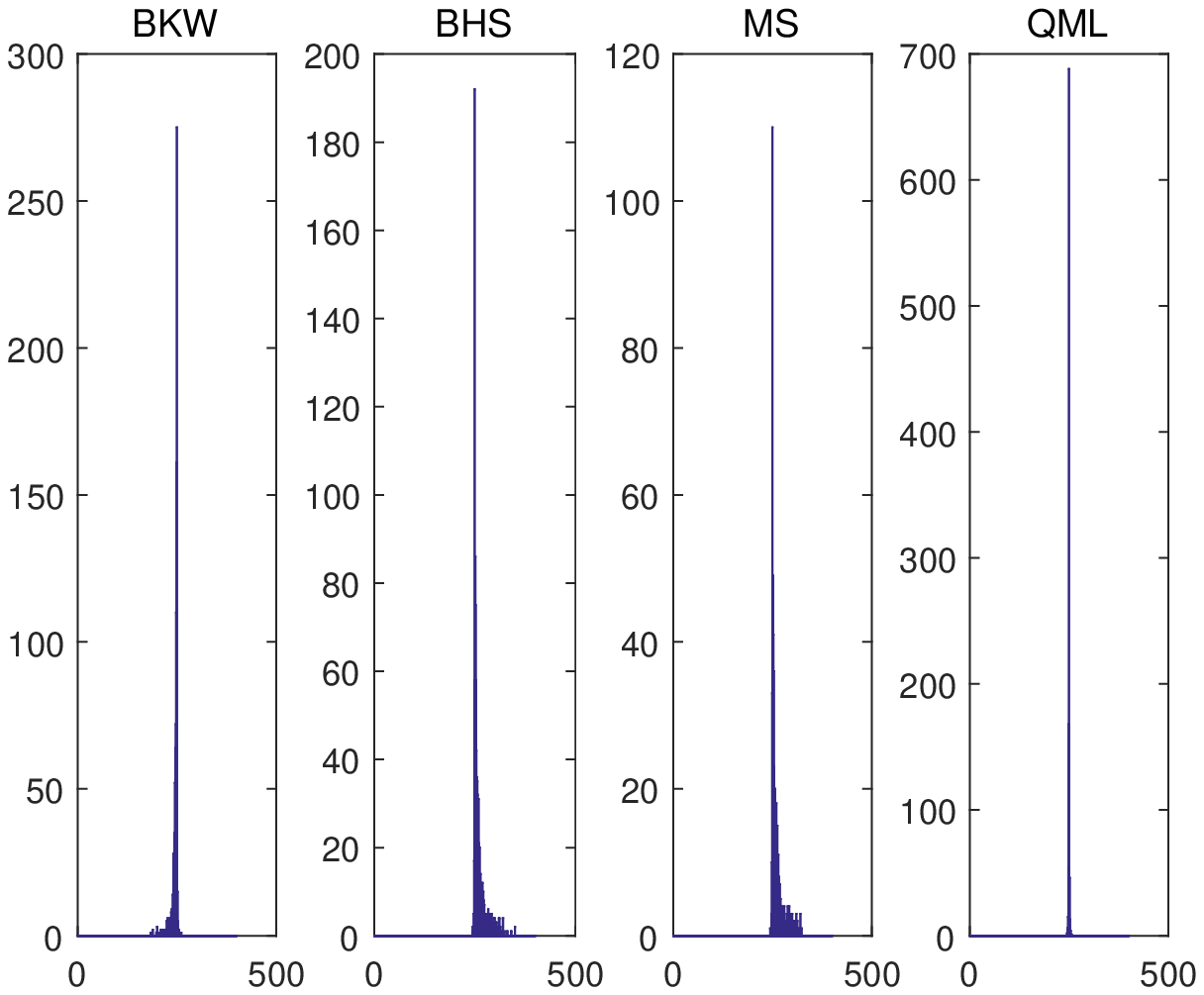}
}
\quad
\subfigure[$(\rho,\alpha,\beta)=(0,0,0.3)$]{
\includegraphics[width=7.0cm]{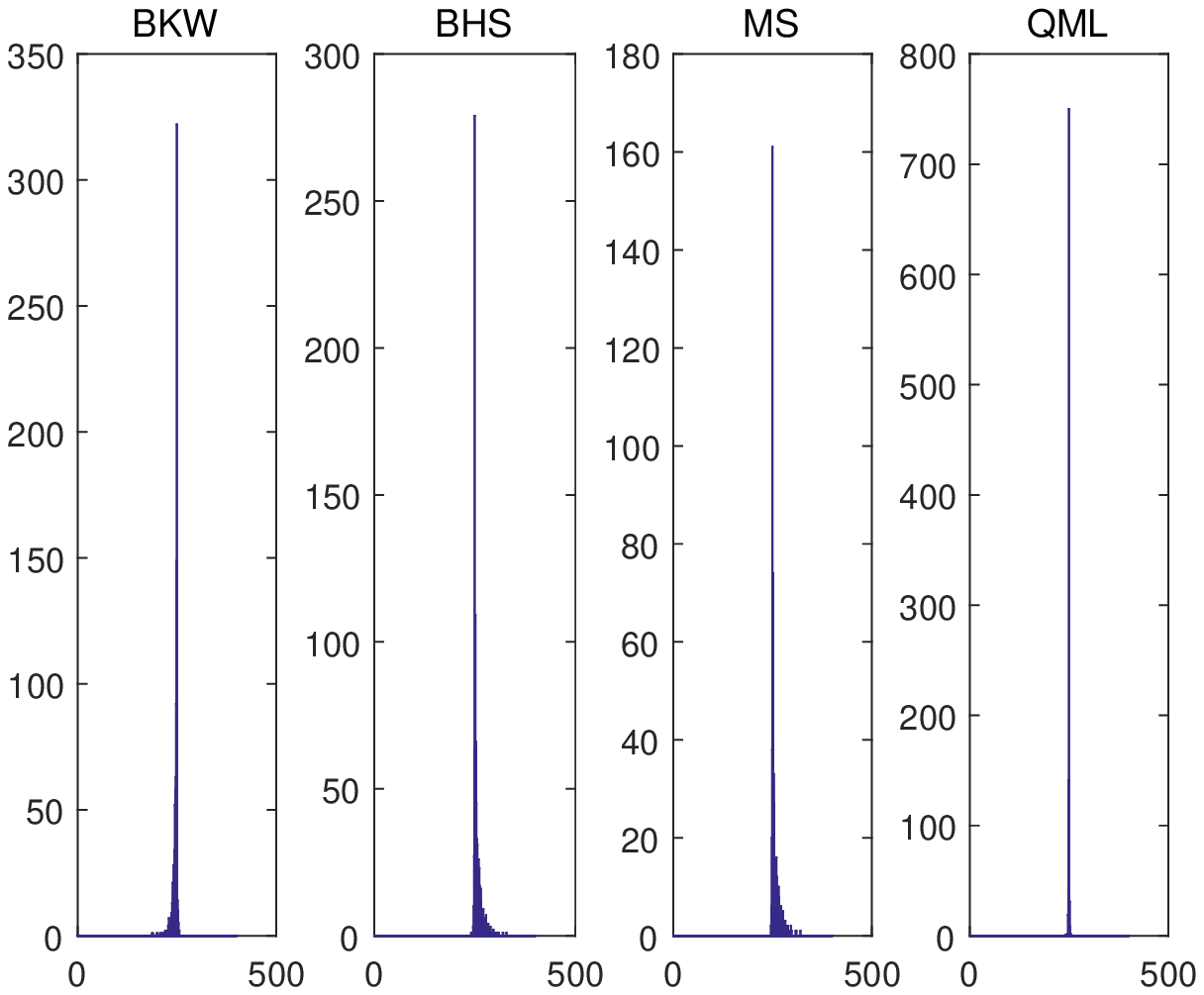}
}
\quad
\subfigure[$(\rho,\alpha,\beta)=(0.7,0.3,0.3)$]{
\includegraphics[width=7.0cm]{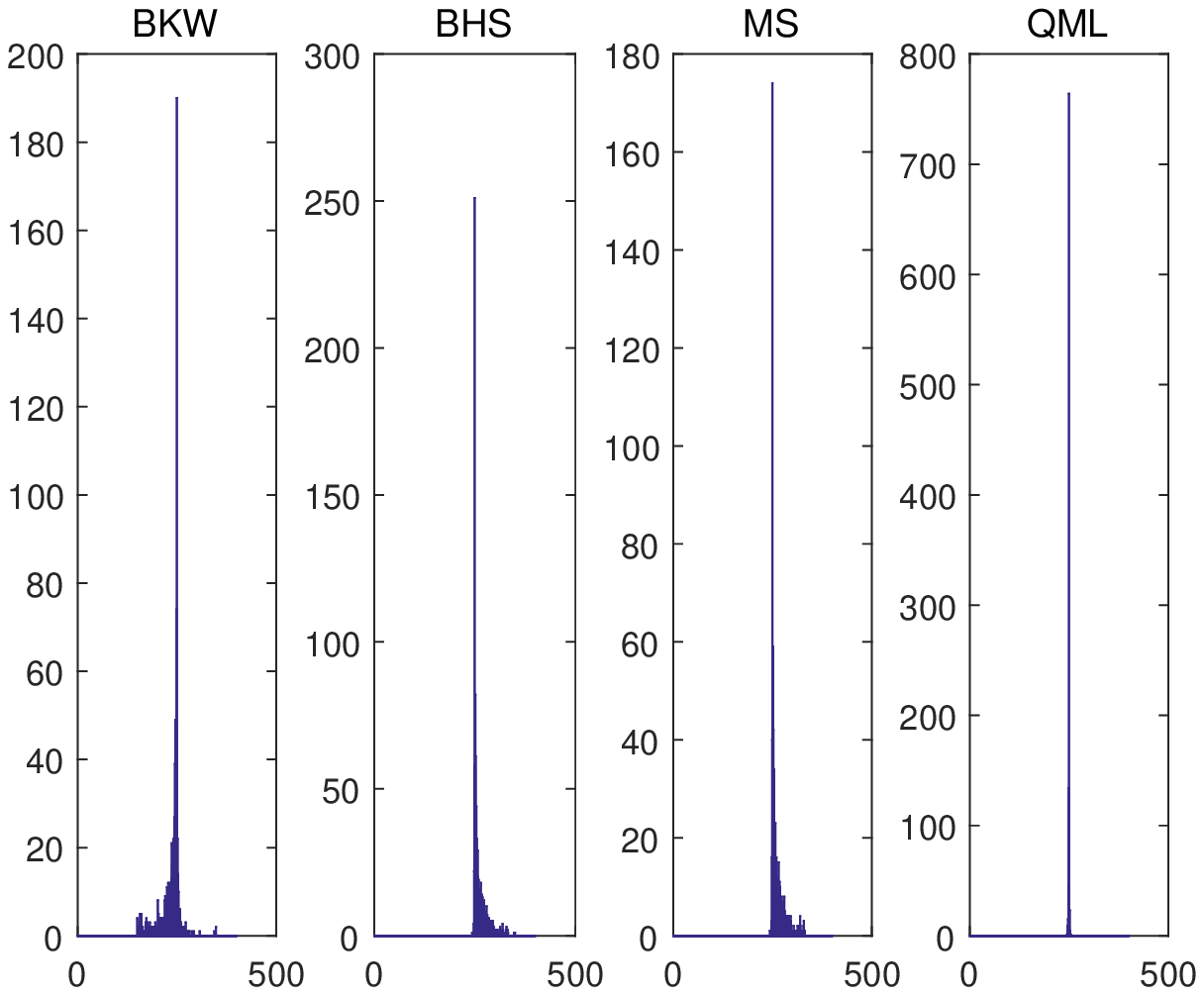}
}
\caption{Plots of the frequency of the estimated break points among 1000 replications for DGP 1.A and $N=500,T=500$.}
\label{1A_NT500}
\end{figure}

\vspace{-1em}
\section{Empirical Application}
\vspace{-1em}

\subsection{Macroeconomic data}
In the first empirical application, we apply our proposed method to a U.S. macroeconomic dataset (\cite{Stock2012}) to detect
the possible structural breaks in the underlying factor model. We use the dataset adopted by \cite{Cheng2016}, which comprises monthly observations of 102 U.S. macroeconomic variables. The sample begins
after the Great Moderation and ranges from 1985:01 to 2013:01 $(T = 337)$. Following \cite{Bai2017}, we focus on the subsample period between 2001:12 and 2013:01 $(T=134,N=102)$ because the complete
data may have multiple breaks.

\cite{Cheng2016} find that 2007:12 is a single-break date, and that the pre-break and post-break subsamples have one factor and two or three factors, respectively. Following \cite{Cheng2016},
\cite{Bai2017} also set the number of factors equal to one and two for the pre- and post-break subsamples, respectively. Then, they implement the LS estimation and obtain the estimated break point
$\hat{k}=2008:12$.
To implement our QML method, we first use Bai and Ng's information criterion IC1 and determine three pseudo-factors in the complete sample. Based on this result, we compute our QML estimator and obtain
2007:07 as the estimated break point, using which we split the sample into pre- and post-break subsamples. IC1 of \cite{Bai2002} detects two pre-break and three post-break factors. Based on the numbers of pre- and post-break factors and that of pseudo-factors, we can conclude that a new factor emerges after the break, so the QML estimator is consistent based on Theorem \ref{consistency}.

\subsection{Stock data}
The second empirical application uses the weekly rate of return for Nasdaq 100 Index from April 18, 2019, to October 1, 2020. As all companies have data starting from April 18, 2019, we choose that as
the start date. Traditionally, the index is limited to 100 common-stock issues, with only one issue allowed per issuer. Now, the index is limited to 100 issuers, some of which may have multiple issues as
index components. The current index has 103 components, representing 100 issuers, four of which are from China: Baidu, JD.com, Ctrip, and NetEase. Thus, the sample size is $T=76$ and $N=103$. As IC1 and
IC2 of \cite{Bai2002}, the methods proposed by \cite{Onatski2010}, \cite{Ahn_Horenstein2013}, and \cite{Fan2020} yield different numbers of pseudo-factors for the sample, we use different number of
factors $r=2,3,4,5,6,7$ to estimate the break date by using the QML method, and find that the estimated break date always falls in the week of February 20, 2020. This result agrees with that obtained
using the method developed by \cite{Baltagi2017}. In fact, the stock market began to fall sharply in the week of February 20, 2020, and two weeks later, the circuit breaker was triggered and U.S. stock market trading halted for a couple of times. Thus, the
factor loading matrix appears to have changed in the early days of the epidemic.

\vspace{-1em}
\section{Conclusions}
\vspace{-1em}

We study the QML method for estimating the break point in high-dimensional factor models with a single structural change. We consider three types of changes and develop an asymptotic theory for the
QML estimator.
We show that the QML estimator is consistent when the covariance matrices of the pre- or post-break factor loadings, or both, are singular.
In addition, the estimation error of the QML estimator is $O_p(1)$ when there is a rotation type of change in the factor loading matrix. We also derive the limiting distribution of the estimated break point in this case.
Moreover, our QML estimator is computationally easy and fast because the eigendecomposition is conducted only once.
The simulation results validate the suitable performance of the QML estimator.
We use the proposed method to estimate the break point for U.S. macroeconomic data and stocks data.

\newpage

\section*{Appendix}
\setcounter{equation}{0}
\setcounter{subsection}{0}
\setcounter{equation}{0}
\setcounter{proposition}{0}

In model (\ref{Baltagi}), \begin{eqnarray}
X & = & G\Lambda^{\prime}+e,\end{eqnarray}
 $G=(g_{1},...g_{T})^{\prime}$, $g_{t}=Bf_{t}$ for $t\le k_{0}$,
and $g_{t}=Cf_{t}$ for $t>k_{0}$. $\lambda_{i}$
and $f_{t}$ are always $r$-dimensional vectors and both $\Lambda_{1}$
and $\Lambda_{2}$ have dimension $N\times r$.
Let $\hat{G}=(\hat{g}_{1},...,\hat{g}_{T})^{\prime}$ denote the full-sample
PCA estimator for $G$. We define\begin{eqnarray*}
\hat{\Sigma}_{1}(k) & \equiv & k^{-1}\sum_{t=1}^{k}\hat{g}_{t}\hat{g}_{t}^{\prime},\\
\hat{\Sigma}_{2}(k) & \equiv & (T-k)^{-1}\sum_{t=k+1}^{T}\hat{g}_{t}\hat{g}_{t}^{\prime}.\end{eqnarray*}
These notations emphasize the dependence of $\hat{\Sigma}_1$ and $\hat{\Sigma}_2$ on $k$. In the proofs, we may use $\hat{\Sigma}_1$,  $\hat{\Sigma}_2$ and $\hat{\Sigma}_1(k)$, $\hat{\Sigma}_2(k)$ interchangeably as long as no confusion is caused. For notational simplicity, let $\hat{\Sigma}_{1}^{0}\equiv\hat{\Sigma}_{1}(k_{0})$ and $\hat{\Sigma}_{2}^{0}\equiv\hat{\Sigma}_{2}(k_{0})$.

The QML objective function can be expressed as \begin{eqnarray*}
U_{NT}(k) & = & k\log|\hat{\Sigma}_{1}|+(T-k)\log|\hat{\Sigma}_{2}|.\end{eqnarray*}
 If $k=k_{0}$, the objective function is \begin{eqnarray*}
U_{NT}(k_{0}) & = & k_{0}\log|\hat{\Sigma}_{1}^{0}|+(T-k_{0})\log|\hat{\Sigma}_{2}^{0}|,\end{eqnarray*}
 where $\hat{\Sigma}_{1}^{0}=k_{0}^{-1}\sum\limits _{t=1}^{k_{0}}\hat{g}_{t}\hat{g}_{t}^{'}$,
$\hat{\Sigma}_{2}^{0}=(T-k_{0})^{-1}\sum\limits _{t=k_{0}+1}^{T}\hat{g}_{t}\hat{g}_{t}^{'}$.

\vspace{2em}
 \textbf{{Representations of $\hat{g}_{t}$}.} \vspace{2em}

The full-sample PCA estimator $\hat{G}$ satisfies the following identity:

\begin{eqnarray}
\hat{G} & = & \frac{1}{NT}XX'\hat{G}V_{NT}^{-1}\nonumber \\
 & = & GH+\frac{1}{NT}e\Lambda G^{\prime}\hat{G}V_{NT}^{-1}+\frac{1}{NT}G\Lambda^{\prime}e^{\prime}\hat{G}V_{NT}^{-1}+\frac{1}{NT}ee^{\prime}\hat{G}V_{NT}^{-1},\label{eq:expan_G}\end{eqnarray}
 where $H=\Lambda^{\prime}\Lambda G^{\prime}\hat{G}V_{NT}^{-1}/NT$
and $V_{NT}$ is a diagonal matrix comprising the eigenvalues
of $XX'/NT$.

Hence, for each period $t$, we have \begin{equation}
\hat{g}_{t}-H^{\prime}g_{t}=V_{NT}^{-1}\left(\frac{\hat{G}^{\prime}G}{T}\frac{\Lambda^{\prime}e_{t}}{N}+\frac{\hat{G}^{\prime}e\Lambda}{NT}g_{t}+\frac{\hat{G}^{\prime}ee_{t}}{NT}\right)\label{eq:hat{g}}\end{equation}
 \citet{Bai2003} shows that\begin{equation}
\hat{g}_{k+1}-H^{\prime}g_{k+1}=O_{p}(\delta_{NT}^{-1})\label{eq:bai thm}\end{equation}
 \begin{equation}
T^{-1}\sum_{t=1}^{T}\|\hat{g}_{t}-H^{\prime}g_{t}\|^{2}=O_{p}(\delta_{NT}^{-2}),\ \mathrm{and}\ T^{-1}(\hat{G}'\hat{G}-H'G'GH)=O_{p}(\delta_{NT}^{-2})\label{eq:bai 2003}\end{equation}

From (A.1) and Lemma A.2 in \citet{Bai2003}, we have the
following lemma:

\begin{lemma} (i). Under Assumptions \ref{factors}--\ref{Central_Limit},
\label{Bai_implied}
\begin{eqnarray}
\max_{m}m^{-1}\sum_{t=k_{0}-m}^{k_{0}}\|\hat{g}_{t}-H^{\prime}g_{t}\|^{2} & = & O_{p}(\frac{1}{N}),\label{eq:Bai implied1}\\
\max_{m}m^{-1}\sum_{t=k_{0}+1}^{k_{0}+m}\|\hat{g}_{t}-H^{\prime}g_{t}\|^{2} & = & O_{p}(\frac{1}{N}).
\label{eq:Bai implied2}
\end{eqnarray}
(ii).  Under Assumptions \ref{factors}--\ref{as-.LeeL},
\begin{eqnarray}
\max_{m}m^{-1}\sum_{t=k_{0}-m}^{k_{0}}\|\tilde{g}_{t}-H^{\prime}g_{t}\|^{2} & \leq & \frac{\bar{c}}{N},\label{eq:Bai implied3}\\
\max_{m}m^{-1}\sum_{t=k_{0}+1}^{k_{0}+m}\|\tilde{g}_{t}-H^{\prime}g_{t}\|^{2} & \leq & \frac{\bar{c}}{N},
\label{eq:Bai implied4}
\end{eqnarray}
where $\bar{c}>0$ is a constant.
 \end{lemma} \textbf{{Proof.}} See the supplementary appendix. $\Box$

\subsection*{Both $\Sigma_{1}$ and $\Sigma_{2}$ are positive
definite matrices.}

We first consider the case in which both $\Sigma_{1}$
and $\Sigma_{2}$ are positive definite matrices.

Following \citet{Baltagi2017}, we define \begin{eqnarray*}
\zeta_{t} & = & \hat{g}_{t}\hat{g}_{t}^{'}-H_{0}^{'}g_{t}g_{t}^{'}H_{0},\text{ for }t=1,\cdots,T\end{eqnarray*}
 and \begin{eqnarray*}
\xi_{t} & = & H_{0}^{'}g_{t}g_{t}^{'}H_{0}-\Sigma_{1}\text{ for }t\leq k_{0},\\
\xi_{t} & = & H_{0}^{'}g_{t}g_{t}^{'}H_{0}-\Sigma_{2}\text{ for }t>k_{0},\end{eqnarray*}
 where $\Sigma_{1}=H_{0}^{'}\Sigma_{G,1}H_{0}$ and $\Sigma_{2}=H_{0}^{'}\Sigma_{G,2}H_{0}$
are the pre- and post-breaks of $H_{0}^{'}E(g_{t}g_{t}^{'})H_{0}$
and $H_{0}$ is the probability limit of $H$. Thus, we have \begin{eqnarray*}
\hat{g}_{t}\hat{g}_{t}^{'} & = & \Sigma_{1}+\xi_{t}+\zeta_{t}\text{ for }t\leq k_{0},\\
\hat{g}_{t}\hat{g}_{t}^{'} & = & \Sigma_{2}+\xi_{t}+\zeta_{t}\text{ for }t>k_{0}.\end{eqnarray*}
$H_{0}$ is nonsingular by Proposition 1 of Bai
(2003).

For $k\leq k_{0}$, \begin{eqnarray}
\hat{\Sigma}_{1} & = & \Sigma_{1}+\frac{1}{k}\sum_{t=1}^{k}\xi_{t}+\frac{1}{k}\sum_{t=1}^{k}\zeta_{t},\nonumber \\
\hat{\Sigma}_{2} & = & \frac{k_{0}-k}{T-k}[\Sigma_{1}-\Sigma_{2}]+\Sigma_{2}+\frac{1}{T-k}\sum_{t=k+1}^{T}\xi_{t}+\frac{1}{T-k}\sum_{t=k+1}^{T}\zeta_{t},\label{eq:Sigmahat rewrite}\end{eqnarray}
Thus, \begin{eqnarray}
\hat{\Sigma}_{1}-\hat{\Sigma}_{1}^{0} & = & \frac{k_{0}-k}{kk_{0}}\sum\limits _{t=1}^{k}(\xi_{t}+\zeta_{t})-\frac{1}{k_{0}}\sum\limits _{t=k+1}^{k_{0}}(\xi_{t}+\zeta_{t}),\nonumber \\
\hat{\Sigma}_{2}-\hat{\Sigma}_{2}^{0} & = & \frac{k-k_{0}}{T-k}(\Sigma_{2}-\Sigma_{1})+\frac{1}{T-k}\sum\limits _{t=k+1}^{k_{0}}(\xi_{t}+\zeta_{t})+\frac{k-k_{0}}{(T-k)(T-k_{0})}\sum\limits
_{t=k_{0}+1}^{T}(\xi_{t}+\zeta_{t}).\label{eq:expand Sigmahat}\end{eqnarray}

Before analyzing the consistency of the estimated fraction and the boundedness
of the estimation error, we need to prove the following lemmas. For any given $0<\eta\le\min(\tau_{0},1-\tau_{0})$
and $M>0$, define $D_{\eta}=\{k:(\tau_{0}-\eta)T\leq k\leq(\tau_{0}+\eta)T\}$,
$D_{\eta}^{c}$ as the complement of $D_{\eta}$, $\tau_{0}=\frac{k_{0}}{T}$,
and $D_{\eta,M}=\{k:(\tau_{0}-\eta)T\leq k\leq(\tau_{0}+\eta)T,\ |k_{0}-k|>M\}$.

\begin{lemma}\label{Baltagi_lemma} Under Assumptions \ref{factors}--\ref{Central_Limit},

\begin{eqnarray*}
(i) &  & \max_{[\tau_{1}T]\leq k\leq k_{0}}\left\Vert \frac{1}{k_{0}}\sum_{t=1}^{k}\xi_{t}\right\Vert =O_{p}(\frac{1}{\sqrt{T}}),\\
(ii) &  & \max_{k\in D_{\eta},k<k_{0}}\left\Vert \frac{1}{k_{0}-k}\sum_{t=k+1}^{k_{0}}\xi_{t}\right\Vert =O_{p}(1),\\
(iii) &  & \max_{k\in D_{\eta}^{c},k<k_{0}}\frac{1}{k_{0}-k}||\sum_{t=k+1}^{k_{0}}\xi_{t}||=O_{p}(\frac{1}{\sqrt{T}}),\\
(iv) &  & \max_{[\tau_{1}T]\leq k\leq k_{0}}\left\Vert \frac{1}{T-k}\sum_{t=k_{0}+1}^{T}\xi_{t}\right\Vert =O_{p}(\frac{1}{\sqrt{T}}),\\
(v) &  & \max\limits _{[\tau_{1}T]\leq k\leq k_{0}}\frac{1}{k_{0}}\left\Vert \sum_{t=1}^{k}\zeta_{t}\right\Vert =o_{p}(1),\\
(vi) &  & \max\limits _{[\tau_{1}T]\leq k\leq k_{0}}\frac{1}{k_{0}-k}\left\Vert \sum_{t=k+1}^{k_{0}}\zeta_{t}\right\Vert =o_{p}(1),\\
(vii) &  & \max\limits _{[\tau_{1}T]\leq k\leq k_{0}}\frac{1}{T-k}\left\Vert \sum_{t=k+1}^{T}\zeta_{t}\right\Vert =o_{p}(1),\end{eqnarray*}
 where $\tau_{1}\in(0,1)$ is the prior lower bound
for $\tau_{0}$, $[\tau_{1}T]$ denotes the prior lower bound
for the real break point $[\tau_{0}T]=k_{0}$, and $[\cdot]$ denotes
the integer part of a real number. \end{lemma}

\textbf{{Proof.}} See the supplementary appendix. $\Box$

\begin{lemma}\label{Sigma1_Dc_consistency} Under Assumptions
\ref{factors}--\ref{Central_Limit}, \begin{eqnarray*}
 &  & \max_{k\in D_{\eta}^{c},k<k_{0}}\|\hat{\Sigma}_{1}^{0}-\frac{1}{k_{0}-k}\sum_{t=k+1}^{k_{0}}\hat{g}_{t}\hat{g}_{t}^{\prime}\|=o_{p}(1)\end{eqnarray*}
 \end{lemma} \textbf{{Proof.}} See the supplementary appendix. $\Box$

\begin{lemma}\label{eq:term1_Lemma} Under Assumptions \ref{factors}--\ref{Central_Limit}, for
$k\in D_{\eta,M}$ and $k<k_{0}$, if both $\Sigma_{1}$
and $\Sigma_{2}$ are positive definite matrices, then
\begin{eqnarray*}
 &  & \frac{k}{k_{0}-k}\log|\hat{\Sigma}_{1}\hat{\Sigma}_{1}^{0-1}|=-\frac{k}{k_{0}}\frac{1}{k_{0}-k}\sum\limits _{t=k+1}^{k_{0}}tr(\xi_{t}\hat{\Sigma}_{1}^{0-1})+o_{p}(1),\end{eqnarray*}
where the $o_{p}(1)$ term is uniform in $k\in D_{\eta,M}$.

\end{lemma} \textbf{{Proof.}} See the supplementary appendix. $\Box$

\begin{lemma}\label{log_Sigm2_expansion} Under Assumptions \ref{factors}--\ref{Central_Limit},
for $|k-k_{0}|\leq M$ and $k<k_{0}$, if
both $\Sigma_{1}$ and $\Sigma_{2}$ are positive definite matrices,
then \[
(T-k)\log|\hat{\Sigma}_{2}\hat{\Sigma}_{2}^{0-1}|=(k-k_{0})tr(\Sigma_{2}-\Sigma_{1})\Sigma_{2}^{-1}+\sum\limits _{t=k+1}^{k_{0}}tr(\xi_{t}\Sigma_{2}^{-1})+o_{p}(1).\]

\end{lemma} \textbf{{Proof.}} See the supplementary appendix.  $\Box$

\subsection*{Proof of $\hat{\tau}-\tau_{0}=o_{p}(1)$}

By symmetry, it suffices to study the case of $k<k_{0}$.
Expanding $U_{NT}(k)-U_{NT}(k_{0})$ gives \begin{eqnarray}
U_{NT}(k)-U_{NT}(k_{0}) & = &
k\log|\hat{\Sigma}_{1}\hat{\Sigma}_{1}^{0-1}|+(T-k)\log|\hat{\Sigma}_{2}\hat{\Sigma}_{2}^{0-1}|-(k_{0}-k)\log|\hat{\Sigma}_{1}^{0}\hat{\Sigma}_{2}^{0-1}|.\label{eq:Uk-Uk0}\end{eqnarray}

To prove $\hat{\tau}-\tau_{0}=o_{p}(1)$, we need to show that
for any $\varepsilon>0$ and $\eta>0$, $P(|\hat{\tau}-\tau_{0}|>\eta)<\varepsilon$
as $(N,T)\rightarrow\infty$, and that $P(\hat{k}\in D_{\eta}^{c})<\varepsilon$.
For notational simplicity, we write $U_{NT}(k)$
as $U(k)$.

As $\hat{k}=\arg\min_{k}U(k)$, we
have $U(\hat{k})-U(k_{0})\leq0$. If $\hat{k}\in D_{\eta}^{c}$,
then $\min\limits _{k\in D_{\eta}^{c}}U(k)-U(k_{0})\leq0$. This implies
$P(\hat{k}\in D_{\eta}^{c})\leq P(\min\limits _{k\in D_{\eta}^{c}}U(k)-U(k_{0})\leq0)$;
thus, it suffices to show that for any given $\varepsilon>0$
and $\eta>0$, $P(\min\limits _{k\in D_{\eta}^{c}}U(k)-U(k_{0})\leq0)<\varepsilon$
as $N,T\rightarrow\infty$.

Suppose that $\min\limits _{k\in D_{\eta}^{c}}U(k)-U(k_{0})\leq0$
and $k^{*}=\arg\min\limits _{k\in D_{\eta}^{c}}U(k)-U(k_{0})$;
then, $U(k^{*})-U(k_{0})\leq0$ and $\frac{U(k^{*})-U(k_{0})}{|k^{*}-k_{0}|}\leq0$.
As $k^{*}\in D_{\eta}^{c}$, we have $\min\limits _{k\in D_{\eta}^{c}}\ \frac{U(k)-U(k_{0})}{|k-k_{0}|}\leq\frac{U(k^{*})-U(k_{0})}{|k^{*}-k_{0}|}\le0$.
Thus, $\min\limits _{k\in D_{\eta}^{c}}U(k)-U(k_{0})\leq0$ implies
$\min\limits _{k\in D_{\eta}^{c}}\frac{U(k)-U(k_{0})}{|k-k_{0}|}\leq0$.
Similarly, $\min\limits _{k\in D_{\eta}^{c}}\frac{U(k)-U(k_{0})}{|k-k_{0}|}\leq0$
implies $\min\limits _{k\in D_{\eta}^{c}}U(k)-U(k_{0})\leq0$. Therefore,
the following two events are equivalent:

\begin{equation}
\{w:\min\limits _{k\in D_{\eta}^{c}}U(k)-U(k_{0})\leq0\}=\{w:\min\limits _{k\in D_{\eta}^{c}}\frac{U(k)-U(k_{0})}{|k-k_{0}|}\leq0\}.\label{eq:equiv sets}\end{equation}
 Note that \begin{equation}
P(\min_{x\in\mathcal{X}}a(x)+b(x)\le0)\le P(\min_{x\in\mathcal{X}}a(x)+\min_{x\in\mathcal{X}}b(x)\le0)=P(\min_{x\in\mathcal{X}}a(x)+o_{p}(1)\le0)\label{eq:prob inequality}\end{equation}
 if $b(x)=o_{p}(1)$ uniformly for $x\in\mathcal{X}$.

Now, using \eqref{eq:Uk-Uk0} and \eqref{eq:equiv sets},
we have

\begin{eqnarray}
 &  & P(\min\limits _{k\in D_{\eta}^{c},k<k_{0}}U(k)-U(k_{0})\leq0)=P(\min\limits _{k\in D_{\eta}^{c},k<k_{0}}\frac{U(k)-U(k_{0})}{k_{0}-k}\leq0)\nonumber \\
 & = & P(\min\limits _{k\in
 D_{\eta}^{c},k<k_{0}}\frac{k}{k_{0}-k}\log|\hat{\Sigma}_{1}\hat{\Sigma}_{1}^{0-1}|+\frac{T-k}{k_{0}-k}\log|\hat{\Sigma}_{2}\hat{\Sigma}_{2}^{0-1}|-\log|\hat{\Sigma}_{1}^{0}\hat{\Sigma}_{2}^{0-1}|\leq0)\nonumber
 \\
 & \le & P(\min\limits _{k\in D_{\eta}^{c},k<k_{0}}\frac{k}{k_{0}-k}\log|\hat{\Sigma}_{1}\hat{\Sigma}_{1}^{0-1}|+\min\limits _{k\in
 D_{\eta}^{c},k<k_{0}}\frac{T-k}{k_{0}-k}\log|(\hat{\Sigma}_{2}-\hat{\Sigma}_{2}^{0})\hat{\Sigma}_{2}^{0-1}+I|-\log|\hat{\Sigma}_{1}^{0}\hat{\Sigma}_{2}^{0-1}|\leq0)\label{eq:P}\end{eqnarray}
 where $\min\limits _{k\in D_{\eta}^{c},k<k_{0}}\frac{k}{k_{0}-k}\log|\hat{\Sigma}_{1}\hat{\Sigma}_{1}^{0-1}|=o_{p}(1)$
because $\|\hat{\Sigma}_{1}-\hat{\Sigma}_{1}^{0}\|$ is uniformly
$o_{p}(1)$ for $[\tau_{1}T]\le k<k_{0}$ by \eqref{eq:expand Sigmahat}
and Lemmas \ref{Baltagi_lemma} $(i)$, $(iii)$, $(v)$ and $(vi)$.
Note that \begin{align}
\hat{\Sigma}_{2} &
=\frac{1}{T-k}\sum_{t=k+1}^{T}\hat{g}_{t}\hat{g}_{t}^{\prime}=\frac{1}{T-k}\sum_{t=k+1}^{k_{0}}\hat{g}_{t}\hat{g}_{t}^{\prime}+\frac{T-k_{0}}{T-k}\hat{\Sigma}_{2}^{0}=\frac{k_{0}-k}{T-k}(\hat{\Sigma}_{1}^{0}+o_{p}(1))+\frac{T-k_{0}}{T-k}\hat{\Sigma}_{2}^{0}\label{eq:rewrite
Sigmahat2}\end{align}
 because $\max_{k<k_{0}-\eta T}\|\hat{\Sigma}_{1}^{0}-\frac{1}{k_{0}-k}\sum_{t=k+1}^{k_{0}}\hat{g}_{t}\hat{g}_{t}^{\prime}\|=o_{p}(1)$
by Lemma \ref{Sigma1_Dc_consistency}. Thus, by \eqref{eq:prob inequality}
and \eqref{eq:rewrite Sigmahat2}, we can bound \eqref{eq:P} by \begin{align*}
 & P(\min\limits _{k\in
 D_{\eta}^{c},k<k_{0}}\frac{T-k}{k_{0}-k}\log|\frac{k-k_{0}}{T-k}(\hat{\Sigma}_{2}^{0}-\hat{\Sigma}_{1}^{0})\hat{\Sigma}_{2}^{0-1}+I|-\log|\hat{\Sigma}_{1}^{0}\hat{\Sigma}_{2}^{0-1}|+o_{p}(1)\leq0)\\
= & P(\min\limits _{k\in
D_{\eta}^{c},k<k_{0}}\frac{T-k}{k_{0}-k}\log|\frac{T-k_{0}}{T-k}I+\frac{k_{0}-k}{T-k}\hat{\Sigma}_{1}^{0}\hat{\Sigma}_{2}^{0-1}|-\log|\hat{\Sigma}_{1}^{0}\hat{\Sigma}_{2}^{0-1}|+o_{p}(1)\leq0).\end{align*}

Let $g({\bf {X}})=\frac{T-k}{k_{0}-k}\log|\frac{T-k_{0}}{T-k}I+\frac{k_{0}-k}{T-k}{\bf {X}}|-\log|{\bf {X}}|$
and $k\in D_{\eta}^{c},k<k_{0}$, where { ${\bf {X}}=\hat{\Sigma}_{1}^{0}\hat{\Sigma}_{2}^{0-1}$}.
By the property of a characteristic polynomial, we
have \begin{eqnarray}
 &  & g({\bf {X}})=\frac{T-k}{k_{0}-k}\sum\limits _{i=1}^{r}\log(\frac{T-k_{0}}{T-k}+\frac{k_{0}-k}{T-k}\rho_{i}({\bf {X}}))-\sum\limits _{i=1}^{r}\log\rho_{i}({\bf {X}}),\label{eq:g(X)}\end{eqnarray}
{where $\rho_{i}({\bf {X}})$ is the $i$-th
eigenvalue of ${\bf {X}}$ for $i=1,\cdots,r$.} On the partial derivative
with respect to $\rho_{i}({\bf {X}})$, we
have \begin{eqnarray}
 &  & \frac{\partial g({\bf {X}})}{\partial\rho_{i}({\bf {X}})}=\frac{(T-k_{0})(\rho_{i}({\bf {X}})-1)}{[(T-k_{0})+(k_{0}-k)\rho_{i}({\bf {X}})]\rho_{i}({\bf {X}})}.\label{eq:derivative}\end{eqnarray}
 From the derivative with respect to $\rho_{i}({\bf {X}})$,
$\frac{\partial g({\bf {X}})}{\partial\rho_{i}({\bf {X}})}<0$
for $0<\rho_{i}({\bf {X}})<1$ and $\frac{\partial g({\bf {X}})}{\partial\rho_{i}({\bf {X}})}>0$
for $\rho_{i}({\bf {X}})>1$. Thus, for $g({\bf {X}})$
to achieve its minimum value, all eigenvalues of ${\bf {X}}$ must
be one (i.e., all eigenvalues of the symmetric
matrix $\hat{\Sigma}_{2}^{0-1/2}\hat{\Sigma}_{1}^{0}\hat{\Sigma}_{2}^{0-1/2}$
should be equal to one); thus, $\hat{\Sigma}_{2}^{0-1/2}\hat{\Sigma}_{1}^{0}\hat{\Sigma}_{2}^{0-1/2}=I$
and $\hat{\Sigma}_{1}^{0}=\hat{\Sigma}_{2}^{0}$. This implies that
$g(I_{r})=0$ is a unique minimum of $g({\bf {X}})$.

Note that $\hat{\Sigma}_{1}^{0}-k_{0}^{-1}\sum\limits _{t=1}^{k_{0}}H^{\prime}g_{t}g_{t}^{\prime}H=o_{p}(1);$
thus, $\hat{\Sigma}_{1}^{0}-H^{\prime}B\Sigma_{F}B^{\prime}H=o_{p}(1)$
under Assumption \ref{factors} and the fact that $g_{t}=Bf_{t}$ for $t\le k_{0}$.
Similarly, $\hat{\Sigma}_{2}^{0}-H^{\prime}C\Sigma_{F}C^{\prime}H=o_{p}(1)$.
As $H\to_{p}H_{0}$ is a nonsingular matrix and $B$ and $C$ are
nonsingular as well, Assumption \ref{factors}(ii) implies \[
\hat{\Sigma}_{1}^{0}\hat{\Sigma}_{2}^{0-1}\to_{p}H_{0}^{\prime}B\Sigma_{F}B^{\prime}(C\Sigma_{F}C^{\prime})^{-1}H_{0}^{-1'}\ne I_{r},\]
 which has positive eigenvalues not equal to one. Thus, the sign of
\eqref{eq:derivative} implies that there exists a positive constant
$c_{\Delta}$ such that\begin{equation}
\min\limits _{k\in D^{c},k<k_{0}}\frac{T-k}{k_{0}-k}\log|\frac{T-k_{0}}{T-k}I+\frac{k_{0}-k}{T-k}\hat{\Sigma}_{1}^{0}\hat{\Sigma}_{2}^{0-1}|-\log|\hat{\Sigma}_{1}^{0}\hat{\Sigma}_{2}^{0-1}|\geq
c_{\Delta}>0=g(I_{r})\label{eq:g(X)>0}\end{equation}
 with w.p.a.1 as $N,T\to\infty$,
where $c_{\Delta}$ is a constant related to the difference $\Delta=B\Sigma_{F}B^{\prime}-C\Sigma_{F}C^{\prime}$.

Thus, we obtain the result $\hat{\tau} =\tau_0+o_p(1)$. $\Box$

\subsection*{Proof of Theorem \ref{bound_theorem}}

\label{sub:2.2}

To prove $\hat{k}-k_{0}=O_{p}(1)$, we need to show that
for any $\varepsilon>0$, there{ exists an} $M>0$
such that $P(|\hat{k}-k_{0}|>M)<\varepsilon$ as $(N,T)\rightarrow\infty$.
By the consistency of $\hat{\tau}$, for any $\varepsilon>0$
and $\min\{\tau_{0},1-\tau_{0}\}>\eta>0$, $P(\hat{k}\in D_{\eta}^{c})<\varepsilon$
as $(N,T)\rightarrow\infty$. For the given $\eta$ and $M$, we have
$D_{\eta,M}=\{k:(\tau_{0}-\eta)T\leq k\leq(\tau_{0}+\eta)T,\ |k_{0}-k|>M\}$;
thus, $P(|\hat{k}-k_{0}|>M)=P(\hat{k}\in D_{\eta}^{c})+P(\hat{k}\in D_{\eta,M})$.
Hence, it suffices to show that for any $\varepsilon>0$ and $\eta>0$,
there exists an $M>0$ such that $P(\hat{k}\in D_{\eta,M})<\varepsilon$
as $(N,T)\rightarrow\infty$. Again, by symmetry, it suffices to study
the case of $k<k_{0}$. Similar to the proof of
consistency of $\hat{\tau}$, we have \begin{eqnarray}
 &  & P(\min\limits _{k\in D_{\eta,M},k<k_{0}}U(k)-U(k_{0})\leq0),\nonumber \\
 & = & P(\min\limits _{k\in D_{\eta,M},k<k_{0}}\frac{U(k)-U(k_{0})}{k_{0}-k}\leq0),\nonumber \\
 & = & P(\min\limits _{k\in
 D_{\eta,M},k<k_{0}}\frac{k}{k_{0}-k}\log|\hat{\Sigma}_{1}\hat{\Sigma}_{1}^{0-1}|+\frac{T-k}{k_{0}-k}\log|\hat{\Sigma}_{2}\hat{\Sigma}_{2}^{0-1}|-\log|\hat{\Sigma}_{1}^{0}\hat{\Sigma}_{2}^{0-1}|\leq0)\nonumber
 \\
 & = & P(\min\limits _{k\in
 D_{\eta,M},k<k_{0}}\frac{k}{k_{0}-k}\log|\hat{\Sigma}_{1}\hat{\Sigma}_{1}^{0-1}|+\frac{T-k}{k_{0}-k}\log|(\hat{\Sigma}_{2}-\hat{\Sigma}_{2}^{0})\hat{\Sigma}_{2}^{0-1}+I|-\log|\hat{\Sigma}_{1}^{0}\hat{\Sigma}_{2}^{0-1}|\leq0)\nonumber
 \\
 & = & P(\min\limits _{k\in
 D_{\eta,M},k<k_{0}}\frac{k}{k_{0}-k}\log|\hat{\Sigma}_{1}\hat{\Sigma}_{1}^{0-1}|+\frac{T-k}{k_{0}-k}\log\Big|\frac{k_{0}-k}{T-k}\hat{\Sigma}_{1}^{0}\hat{\Sigma}_{2}^{0-1}+\frac{T-k_{0}}{T-k}I+\frac{k_{0}-k}{T-k}\Big(\frac{1}{k_{0}-k}\sum\limits
 _{t=k+1}^{k_{0}}\hat{g}_{t}\hat{g}_{t}^{'}-\hat{\Sigma}_{1}^{0}\Big)\hat{\Sigma}_{2}^{0-1}\Big|\nonumber \\
 &&- \log|\hat{\Sigma}_{1}^{0}\hat{\Sigma}_{2}^{0-1}|\leq0).\label{eq:p(Uk - Uk0 <0)}\end{eqnarray}
Note that \begin{eqnarray}
\frac{k}{k_{0}-k}\log|\hat{\Sigma}_{1}\hat{\Sigma}_{1}^{0-1}| & = & -\frac{k}{k_{0}}\frac{1}{k_{0}-k}\sum\limits _{t=k+1}^{k_{0}}tr(\xi_{t}\hat{\Sigma}_{1}^{0-1})+o_{p}(1),\label{eq:term1}\end{eqnarray}
where the $o_{p}(1)$ term is uniform in $k\in D_{\eta,M}$
and $k<k_{0}$ by Lemma \ref{eq:term1_Lemma}.

In addition, \begin{eqnarray}
\frac{1}{k_{0}-k}\sum\limits _{t=k+1}^{k_{0}}\hat{g}_{t}\hat{g}_{t}^{'}-\hat{\Sigma}_{1}^{0} & = & \frac{1}{k_{0}-k}\Big(\sum\limits _{t=1}^{k_{0}}\hat{g}_{t}\hat{g}_{t}^{'}-\sum\limits
_{t=1}^{k}\hat{g}_{t}\hat{g}_{t}^{'}\Big)-\hat{\Sigma}_{1}^{0}\nonumber \\
 & = & \frac{k_{0}}{k_{0}-k}\hat{\Sigma}_{1}^{0}-\frac{k}{k_{0}-k}\hat{\Sigma}_{1}-\hat{\Sigma}_{1}^{0}\nonumber \\
 & = & \frac{k}{k_{0}-k}\hat{\Sigma}_{1}^{0}-\frac{k}{k_{0}-k}\Big(\hat{\Sigma}_{1}^{0}+\frac{k_{0}-k}{kk_{0}}\sum\limits _{t=1}^{k}(\xi_{t}+\zeta_{t})-\frac{1}{k_{0}}\sum\limits
 _{t=k+1}^{k_{0}}(\xi_{t}+\zeta_{t})\Big)\nonumber \\
 & = & \frac{k}{(k_{0}-k)k_{0}}\sum\limits _{t=k+1}^{k_{0}}(\xi_{t}+\zeta_{t})-\frac{1}{k_{0}}\sum\limits _{t=1}^{k}(\xi_{t}+\zeta_{t})\nonumber \\
 & = & \frac{k}{(k_{0}-k)k_{0}}\sum\limits _{t=k+1}^{k_{0}}\xi_{t}+o_{p}(1),\label{eq:gg-Sigmahat1}\end{eqnarray}
where the third line uses \eqref{eq:expand Sigmahat}
and the $o_{p}(1)$ term in the last line is uniform in $k\in D_{\eta,M}$
according to Lemmas \ref{Baltagi_lemma} $(i)$, $(v)$, and
$(vi)$.

Let $\upsilon_{k}$ denote a uniform $o_{p}(1)$
term in \eqref{eq:gg-Sigmahat1}. For any given $\delta>0$, \eqref{eq:gg-Sigmahat1}
implies\begin{eqnarray}
 &  & P(\max\limits _{k\in D_{\eta,M},k<k_{0}}\Big\|\frac{1}{k_{0}-k}\sum\limits _{t=k+1}^{k_{0}}\hat{g}_{t}\hat{g}_{t}^{'}-\hat{\Sigma}_{1}^{0}\Big\|\geq\delta)\nonumber \\
 & \le & P(\max\limits _{k\in D_{\eta,M},k<k_{0}}\Big\|\frac{1}{k_{0}-k}\sum\limits _{t=k+1}^{k_{0}}\xi_{t}\Big\|+\|\upsilon_{k}\|\geq\delta)\nonumber \\
 & = & P(\max\limits _{k\in D_{\eta,M},k<k_{0}}\Big\|\frac{1}{k_{0}-k}\sum\limits _{t=k+1}^{k_{0}}\xi_{t}\Big\|+\|\upsilon_{k}\|\geq\delta,\ \max\limits _{k\in
 D_{\eta,M},k<k_{0}}\|\upsilon_{k}\|\le\delta/2)\nonumber \\
 &  & +P(\max\limits _{k\in D_{\eta,M},k<k_{0}}\Big\|\frac{1}{k_{0}-k}\sum\limits _{t=k+1}^{k_{0}}\xi_{t}\Big\|+\|\upsilon_{k}\|\geq\delta,\ \max\limits _{k\in
 D_{\eta,M},k<k_{0}}\|\upsilon_{k}\|>\delta/2)\nonumber \\
 & \le & P(\max\limits _{k\in D_{\eta,M},k<k_{0}}\Big\|\frac{1}{k_{0}-k}\sum\limits _{t=k+1}^{k_{0}}\xi_{t}\Big\|\geq\delta/2)+o(1)\nonumber \\
 & = & P(\max\limits _{(\tau_{0}-\eta)T\leq k<k_{0}-M}\Big\|\frac{1}{k_{0}-k}\sum\limits _{t=k+1}^{k_{0}}\xi_{t}\Big\|\geq\delta/2)+o(1).\label{eq:P_gg_Sigma}\end{eqnarray}
  Let $m=k_{0}-k$, \begin{eqnarray}
P(\max\limits _{(\tau_{0}-\eta)T\leq k<k_{0}-M}\Big\|\frac{1}{k_{0}-k}\sum\limits _{t=k+1}^{k_{0}}\xi_{t}\Big\|\geq\delta/2) & = & P(\max\limits _{M<m\leq\eta T}\Big\|\frac{1}{m}\sum\limits
_{t=1}^{m}\xi_{t}\Big\|\geq\delta/2)\nonumber \\
 & \leq & \frac{4}{\delta^{2}}(\frac{1}{M}+\sum\limits _{t=M+1}^{\eta T}\frac{1}{t^{2}})\nonumber \\
 & \le & \frac{4}{\delta^{2}}(\frac{2}{M}-\frac{1}{\eta T})\nonumber \\
 & = & \frac{C}{M\delta^{2}}+o(1)\rightarrow0,\quad as~~M\rightarrow\infty,\label{eq:gg HR}\end{eqnarray}
 where $0<C<\infty$ is a constant.

Similarly, \eqref{eq:term1} implies \begin{eqnarray}
 &  & P(\max\limits _{k\in D_{\eta,M},k<k_{0}}\Big|\frac{k}{k_{0}-k}\log|\hat{\Sigma}_{1}\hat{\Sigma}_{1}^{0-1}|\Big|\geq\delta)\nonumber \\
 & \leq & P(\max\limits _{k\in D_{\eta,M},k<k_{0}}\sqrt{r}\Big\|\frac{1}{k_{0}-k}\sum\limits _{t=k+1}^{k_{0}}\xi_{t}\hat{\Sigma}_{1}^{0-1}+o_{p}(1)\Big\|\geq\delta)\nonumber \\
 & = & P(\max\limits _{k\in D_{\eta,M},k<k_{0}}\Big\|\frac{1}{k_{0}-k}\sum\limits _{t=k+1}^{k_{0}}\xi_{t}(\hat{\Sigma}_{1}^{0-1}-\Sigma_{1}^{-1})+\frac{1}{k_{0}-k}\sum\limits
 _{t=k+1}^{k_{0}}\xi_{t}\Sigma_{1}^{-1}+o_{p}(1)\Big\|\geq\delta/\sqrt{r})\nonumber \\
 & \leq & P(\max\limits _{k\in D_{\eta,M},k<k_{0}}\Big\|\frac{1}{k_{0}-k}\sum\limits _{t=k+1}^{k_{0}}\xi_{t}\Sigma_{1}^{-1}\Big\|+o_{p}(1)\geq\delta/\sqrt{r})\nonumber \\
 & \leq & \frac{C}{M\delta^{2}}\rightarrow0,\quad as~~M\rightarrow\infty,\ \text{for some constant \ensuremath{C>0}},\label{eq:Hajek2}\end{eqnarray}
 where the fourth line follows from $\max\limits _{k\in D_{\eta,M},k<k_{0}}\Big\|\frac{1}{k_{0}-k}\sum\limits _{t=k+1}^{k_{0}}\xi_{t}(\hat{\Sigma}_{1}^{0-1}-\Sigma_{1}^{-1})\Big\|=o_{p}(1)$
by Lemma \ref{Baltagi_lemma} $(ii)$ and the fact that $||\hat{\Sigma}_{1}^{0-1}-\Sigma_{1}^{-1}||=o_{p}(1)$.
In addition, the last inequality holds through a similar derivation used in \eqref{eq:P_gg_Sigma}
and \eqref{eq:gg HR}.

By the continuity of $g$ defined in \eqref{eq:g(X)},
\eqref{eq:g(X)>0} indicates the presence of $\delta>0$ such that
$\|(k_{0}-k)^{-1}\sum\limits _{t=k+1}^{k_{0}}\hat{g}_{t}\hat{g}_{t}^{'}-\hat{\Sigma}_{1}^{0}\|<\delta$
holds for a sufficiently large $M$ by \eqref{eq:gg HR}
and \begin{equation}
\min\limits _{k\in D_{\eta,M},k<k_{0}}\frac{T-k}{k_{0}-k}\log\Big|\frac{k_{0}-k}{T-k}\hat{\Sigma}_{1}^{0}\hat{\Sigma}_{2}^{0-1}+\frac{T-k_{0}}{T-k}I+\frac{k_{0}-k}{T-k}\Big(\frac{1}{k_{0}-k}\sum\limits
_{t=k+1}^{k_{0}}\hat{g}_{t}\hat{g}_{t}^{'}-\hat{\Sigma}_{1}^{0}\Big)\hat{\Sigma}_{2}^{0-1}\Big|-\log|\hat{\Sigma}_{1}^{0}\hat{\Sigma}_{2}^{0-1}|\geq\frac{c_{\Delta}}{2}>0,\label{eq:bounded1}\end{equation}
 w.p.a.1 as $N,T\to\infty$. In addition, by \eqref{eq:Hajek2}, we have \begin{eqnarray}
  P(\Big|\min\limits _{k\in D_{\eta,M},k<k_{0}}|\frac{k}{k_{0}-k}\log|\hat{\Sigma}_{1}\hat{\Sigma}_{1}^{0-1}|\Big|\le\frac{c_{\Delta}}{4})&\ge & P(\max\limits _{k\in
  D_{\eta,M},k<k_{0}}\Big|\frac{k}{k_{0}-k}\log|\hat{\Sigma}_{1}\hat{\Sigma}_{1}^{0-1}|\Big|\le\frac{c_{\Delta}}{4})\nonumber\\
  &\ge&1-\frac{16C}{Mc_{\Delta}^{2}}\to1\label{eq:bounded2}\end{eqnarray}
 as $M\to\infty$. Using \eqref{eq:bounded1} and \eqref{eq:bounded2},
we can obtain \begin{eqnarray*}
 &  & \min\limits _{k\in
 D_{\eta,M},k<k_{0}}\frac{k}{k_{0}-k}\log|\hat{\Sigma}_{1}\hat{\Sigma}_{1}^{0-1}|+\frac{T-k}{k_{0}-k}\log\Big|\frac{k_{0}-k}{T-k}\hat{\Sigma}_{1}^{0}\hat{\Sigma}_{2}^{0-1}+\frac{T-k_{0}}{T-k}I+\frac{k_{0}-k}{T-k}\Big(\frac{1}{k_{0}-k}\sum\limits
 _{t=k+1}^{k_{0}}\hat{g}_{t}\hat{g}_{t}^{'}-\hat{\Sigma}_{1}^{0}\Big)\hat{\Sigma}_{2}^{0-1}\Big|\\
 &  &~~-\log|\hat{\Sigma}_{1}^{0}\hat{\Sigma}_{2}^{0-1}|\\
 &  &\ge \min\limits _{k\in
 D_{\eta,M},k<k_{0}}\frac{k}{k_{0}-k}\log|\hat{\Sigma}_{1}\hat{\Sigma}_{1}^{0-1}|+\frac{c_{\Delta}}{2}\ge-\frac{c_{\Delta}}{4}+\frac{c_{\Delta}}{2}\ge\frac{c_{\Delta}}{4}>0\end{eqnarray*}
 w.p.a.1 as $M\to\infty$. This shows that $P(\min\limits _{k\in D_{\eta,M},k<k_{0}}U(k)-U(k_{0})\leq0)<\varepsilon$
for a sufficiently large $M$.$\Box$

\subsection*{Proof of Theorem \ref{distribution_theorem}}

Let us recall \eqref{eq:Uk-Uk0}, \begin{eqnarray*}
U(k)-U(k_{0}) & = &
(k_{0}-k)\Big(\frac{k}{k_{0}-k}\log|\hat{\Sigma}_{1}\hat{\Sigma}_{1}^{0-1}|+\frac{T-k}{k_{0}-k}\log|\hat{\Sigma}_{2}\hat{\Sigma}_{2}^{0-1}|-\log|\hat{\Sigma}_{1}^{0}\hat{\Sigma}_{2}^{0-1}|\Big).\end{eqnarray*}
 For the second term in the above equation, we have \begin{eqnarray*}
(T-k)\log|\hat{\Sigma}_{2}\hat{\Sigma}_{2}^{0-1}| & = & (k-k_{0})tr(\Sigma_{2}-\Sigma_{1})\Sigma_{2}^{-1}+\sum\limits _{t=k+1}^{k_{0}}tr(\xi_{t}\Sigma_{2}^{-1})+o_{p}(1),\end{eqnarray*}
 by Lemma \ref{log_Sigm2_expansion}. Similarly,
by \eqref{eq:expand Sigmahat} and Lemma \ref{Baltagi_lemma}, we
have \begin{eqnarray*}
k\log|\hat{\Sigma}_{1}\hat{\Sigma}_{1}^{0-1}| & = & k\log|(\hat{\Sigma}_{1}-\hat{\Sigma}_{1}^{0})\hat{\Sigma}_{1}^{0-1}+I|,\\
 & = & k\log|\Big[\frac{k_{0}-k}{kk_{0}}\sum\limits _{t=1}^{k}(\xi_{t}+\zeta_{t})-\frac{1}{k_{0}}\sum\limits _{t=k+1}^{k_{0}}(\xi_{t}+\zeta_{t})\Big]\hat{\Sigma}_{1}^{0-1}+I|,\\
 & = & k\cdot tr(\frac{k_{0}-k}{kk_{0}}\sum\limits _{t=1}^{k}(\xi_{t}+\zeta_{t})\hat{\Sigma}_{1}^{0-1})-k\cdot tr(\frac{1}{k_{0}}\sum\limits
 _{t=k+1}^{k_{0}}(\xi_{t}+\zeta_{t})\hat{\Sigma}_{1}^{0-1})+o_{p}(1),\\
 & = & -\sum\limits _{t=k+1}^{k_{0}}tr(\xi_{t}\Sigma_{1}^{-1})+o_{p}(1).\end{eqnarray*}
 Thus, \begin{eqnarray*}
U(k)-U(k_{0})\xrightarrow{d}\sum\limits _{t=k+1}^{k_{0}}tr(\xi_{t}(\Sigma_{2}^{-1}-\Sigma_{1}^{-1}))+(k_{0}-k)tr\left(\Sigma_{1}\Sigma_{2}^{-1}-r-\log|\Sigma_{1}\Sigma_{2}^{-1}|\right)\end{eqnarray*}

Similarly, for the case of $k>k_{0}$, the limit can be written as
$\sum\limits _{t=k_{0}+1}^{k}tr(\xi_{t}(\Sigma_{1}^{-1}-\Sigma_{2}^{-1}))+(k-k_{0})tr\left(\Sigma_{1}^{-1}\Sigma_{2}-r-\log|\Sigma_{1}^{-1}\Sigma_{2}|\right)$.
$\Box$

\section*{$\Sigma_{1}$ or $\Sigma_{2}$, or both, is a singular matrix.}

Before proving the theorem, we need to prove the following lemmas, where $A^{-}$ denotes the MP inverse of $A$, $\rho_{i}(A)$ represents
the $i$-th eigenvalue of matrix $A$, and $\sigma_{i}(A)$ represents
the $i$-th singular value of matrix $A$. \vspace{2em}

\begin{lemma} \label{eef} Under Assumptions \ref{factors}--\ref{Central_Limit},\begin{align*}
\max_{k\in[[\tau_{1}T],k_{0}]}\frac{1}{k}\sum_{s=1}^{k}\|\sum_{t=1}^{T}\hat{g}_{t}e_{t}^{\prime}e_{s}/NT\|^{2} & =O_{p}(\delta_{NT}^{-4})\\
\max_{k\in[k_{0},[\tau_{2}T]]}\frac{1}{T-k}\sum_{s=k+1}^{T}\|\sum_{t=1}^{T}\hat{g}_{t}e_{t}^{\prime}e_{s}/NT\|^{2} & =O_{p}(\delta_{NT}^{-4}).\end{align*}

\end{lemma}

\textbf{Proof}: By symmetry, it is sufficient to
focus on the case of $k\in[k_{0},[\tau_{2}T]]$. \begin{align*}
 & \frac{1}{N^{2}T^{2}(T-k)}\sum_{s=k+1}^{T}\|\sum_{t=1}^{T}e_{s}^{\prime}e_{t}\hat{g}_{t}^{\prime}\|^{2}\\
\le &
\frac{2}{N^{2}T^{2}(T-k)}\sum_{s=k+1}^{T}\|\sum_{t=1}^{T}(\hat{g}_{t}-H^{\prime}g_{t})e_{t}^{\prime}e_{s}\|^{2}+\frac{2}{N^{2}T^{2}(T-k)}\sum_{s=k+1}^{T}\|\sum_{t=1}^{T}H^{\prime}g_{t}e_{t}^{\prime}e_{s}\|^{2}\end{align*}
 Recall that $E(e_{t}^{\prime}e_{s})/N=\gamma_{N}(s,t)$. Consider
the equation\begin{align*}
\max_{k\in[k_{0},[\tau_{2}T]]}\frac{2}{N^{2}T^{2}(T-k)}\sum_{s=k+1}^{T}\|\sum_{t=1}^{T}H^{\prime}g_{t}e_{t}^{\prime}e_{s}\|^{2} &
\le\frac{2}{T^{3}(1-\tau_{2})}\sum_{s=1}^{T}\|\sum_{t=1}^{T}H^{\prime}g_{t}\frac{e_{t}^{\prime}e_{s}-E(e_{t}^{\prime}e_{s})}{N}\|^{2}\\
 &~~~ +\frac{2}{T^{3}(1-\tau_{2})}\sum_{s=1}^{T}\|\sum_{t=1}^{T}H^{\prime}g_{t}\gamma_{N}(s,t)\|^{2},\end{align*}
 where the first term can be written as\begin{equation}
\frac{2}{T^{2}(1-\tau_{2})N}\sum_{s=1}^{T}\|\frac{1}{\sqrt{NT}}\sum_{t=1}^{T}\sum_{i=1}^{N}H^{\prime}g_{t}[e_{it}e_{is}-E(e_{it}e_{is})]\|^{2}=O_{p}\left(\frac{1}{NT}\right)\label{eq:eef1}\end{equation}
 under Assumption \ref{Central_Limit}(i) and the second term is $O_{p}(T^{-2})$ because the
expectation can be bounded by \begin{align}
\frac{2}{T^{3}}\sum_{s=1}^{T}E\|\sum_{t=1}^{T}g_{t}\gamma_{N}(s,t)\|^{2} & \le\frac{2}{T^{3}}\sum_{s=1}^{T}\sum_{t=1}^{T}\sum_{u=1}^{T}E(\|g_{t}\|\|g_{u}\|)|\gamma_{N}(s,t)||\gamma_{N}(s,u)|\nonumber \\
 & \le\frac{2}{T^{3}}\sum_{s=1}^{T}\sum_{t=1}^{T}\sum_{u=1}^{T}\max_{t}E\|g_{t}\|^{2}|\gamma_{N}(s,t)||\gamma_{N}(s,u)|\nonumber \\
 & \le\frac{2}{T^{3}}\sum_{s=1}^{T}\max_{t}E\|g_{t}\|^{2}\left(\sum_{t=1}^{T}|\gamma_{N}(s,t)|\right)^{2}=O(T^{-2}),\label{eq:eef2}\end{align}
 where we use the facts that $E(\|g_{t}\|\|g_{u}\|)\le[E\|g_{t}\|^{2}E\|g_{u}\|^{2}]^{1/2}\le\max_{t}E\|g_{t}\|^{2}$
under Assumption \ref{factors} and $\sum_{t=1}^{T}|\gamma_{N}(s,t)|\le M$ by Assumption
\ref{Depen_and_Hetero}(ii).

Next, consider the term \begin{align*}
\max_{k\in[k_{0},[\tau_{2}T]]}\frac{1}{N^{2}T^{2}(T-k)}\sum_{s=k+1}^{T}\|\sum_{t=1}^{T}(\hat{g}_{t}-H^{\prime}g_{t})e_{t}^{\prime}e_{s}\|^{2} &
\le\frac{2}{T^{3}(1-\tau_{2})}\sum_{s=1}^{T}\|\sum_{t=1}^{T}(\hat{g}_{t}-H^{\prime}g_{t})\frac{e_{t}^{\prime}e_{s}-E(e_{t}^{\prime}e_{s})}{N}\|^{2}\\
 &~~~ +\frac{2}{T^{3}(1-\tau_{2})}\sum_{s=1}^{T}\|\sum_{t=1}^{T}(\hat{g}_{t}-H^{\prime}g_{t})\gamma_{N}(s,t)\|^{2},\end{align*}
 where the first term can be bounded by\begin{align}
\frac{2}{T^{3}(1-\tau_{2})}\sum_{s=1}^{T}\|\sum_{t=1}^{T}(\hat{g}_{t}-H^{\prime}g_{t})\frac{e_{t}^{\prime}e_{s}-E(e_{t}^{\prime}e_{s})}{N}\|^{2} &
\le\frac{2}{T^{3}(1-\tau_{2})N}\sum_{s=1}^{T}\sum_{t=1}^{T}\|\hat{g}_{t}-H^{\prime}g_{t}\|^{2}\sum_{t=1}^{T}\left[\frac{1}{\sqrt{N}}\sum_{i=1}^{N}e_{it}e_{is}-E(e_{it}e_{is})\right]^{2}\nonumber \\
 &
 =\frac{2}{NT(1-\tau_{2})}\sum_{t=1}^{T}\|\hat{g}_{t}-H^{\prime}g_{t}\|^{2}\cdot\frac{1}{T^{2}}\sum_{s=1}^{T}\sum_{t=1}^{T}\left[\frac{1}{\sqrt{N}}\sum_{i=1}^{N}e_{it}e_{is}-E(e_{it}e_{is})\right]^{2}\nonumber
 \\
 & =O_{p}\left(\frac{1}{N\delta_{NT}^{2}}\right),\label{eq:eef3}\end{align}
 by Assumption \ref{Depen_and_Hetero}(v) and the second term is bounded by \begin{equation}
\frac{2}{T^{3}(1-\tau_{2})}\sum_{s=1}^{T}\|\sum_{t=1}^{T}(\hat{g}_{t}-H^{\prime}g_{t})\gamma_{N}(s,t)\|^{2}\le\frac{2}{T}\sum_{t=1}^{T}\|\hat{g}_{t}-H^{\prime}g_{t}\|^{2}\frac{1}{T^{2}(1-\tau_{2})}\sum_{s=1}^{T}\sum_{t=1}^{T}|\gamma_{N}(s,t)|^{2}=O_{p}\left(\frac{1}{T\delta_{NT}^{2}}\right)\label{eq:eef4}\end{equation}
 because $\sum_{t=1}^{T}|\gamma_{N}(s,t)|^{2}\le(\sum_{t=1}^{T}|\gamma_{N}(s,t)|)^{2}\le M^{2}$
under Assumption \ref{Depen_and_Hetero}(ii). Combining the results obtained in \eqref{eq:eef1}--\eqref{eq:eef4},
we obtain the desired result. $\Box$ \vspace{2em}

When $C$ is singular, $\hat{\Sigma}_{2}(k)$ converges
in probability to a singular matrix for $k\ge k_{0}$. In finite samples,
however, the smallest eigenvalue of $\hat{\Sigma}_{2}(k)$ is not
zero. The following proposition establishes a lower bound for the smallest
eigenvalue of $\hat{\Sigma}_{2}(k)$, which ensures that it is meaningful
to compute the logarithm of $|\hat{\Sigma}_{2}(k)|$ in the objective
function for any given sample size. Symmetrically, a similar lower bound can be established for the smallest eigenvalue of $\hat{\Sigma}_{1}(k)$
for $k\le k_{0}$ when $B$ is singular. Because of space restrictions, Proposition \ref{low_bound} here only
states the result for the case of $\hat{\Sigma}_{2}(k)$.
\vspace{1em}

\begin{proposition}
\label{low_bound} Under Assumptions \ref{factors}--\ref{invariance}, for $k\ge k_{0}$
and $k\le[\tau_{2}T]$, if $C$ is singular and $\sqrt{N}/T\to0$
as $N,T\to\infty$, there exists a constant
$c_{U}\ge c_{L}>0$ such that \begin{align*}
P\left(\min_{k\in[k_{0},[\tau_{2}T]]}\rho_{j}(\hat{\Sigma}_{2}(k))\ge\frac{c_{L}}{N}\right) & \to1,\\
P\left(\max_{k\in[k_{0},[\tau_{2}T]]}\rho_{j}(\hat{\Sigma}_{2}(k)\le\frac{c_{U}}{N}\right) & \to1,\end{align*}
 for $j=r_{2}+1,...,r$.

\end{proposition}

\textbf{Proof}:

\textbf{{Part 1.}} For $k\ge k_{0}$, $\hat{\Sigma}_{2}(k)=(T-k)^{-1}\hat{G}_{2}^{k'}\hat{G}_{2}^{k}$,
where $\hat{G}_{2}^{k}=[\hat{g}_{k+1},...,\hat{g}_{T}]'$. Let $X_{2}^{k}=[X_{k+1},...,X_{T}]'$,
$e_{2}^{k}=[e_{k+1},...,e_{T}]'$, and $G_{2}^{k}=[G_{k+1},...,G_{T}]'$.

From $XX^{\prime}\hat{G}/NT=\hat{G}V_{NT}$, eq. \eqref{eq:expan_G}
implies \begin{align}
\hat{G}_{2}^{k} & =X_{2}^{k}X^{\prime}\hat{G}V_{NT}^{-1}/NT=\frac{1}{NT}(G_{2}^{k}\Lambda^{\prime}+e_{2}^{k})(\Lambda G^{\prime}+e^{\prime})\hat{G}V_{NT}^{-1}\nonumber \\
\hat{G}_{2}^{k}-G_{2}^{k}H & =\frac{1}{NT}e_{2}^{k}\Lambda
G^{\prime}\hat{G}V_{NT}^{-1}+\frac{1}{NT}e_{2}^{k}e^{\prime}\hat{G}V_{NT}^{-1}+\frac{1}{NT}G_{2}^{k}\Lambda^{\prime}e^{\prime}\hat{G}V_{NT}^{-1}.\label{eq:G2hat - G2H}\end{align}
In addition, note that \[
\hat{\Sigma}_{2}(k)-\frac{1}{T-k}\hat{G}_{2}^{k'}M_{F_{2}^{k}}\hat{G}_{2}^{k}=\frac{1}{T-k}\hat{G}_{2}^{k'}P_{F_{2}^{k}}\hat{G}_{2}^{k}\ge0,\]
 where $P_{F_{2}^{k}}=F_{2}^{k}(F_{2}^{k'}F_{2}^{k})^{-1}F_{2}^{k'}$,
$M_{F_{2}^{k}}=I_{T-k}-P_{F_{2}^{k}}$, and $F_{2}^{k}=[f_{k+1},...,f_{T}]'$.
Thus, Weyl's inequality for eigenvalues implies \begin{align}
\min_{k\in[k_{0},[\tau_{2}T]]}\rho_{j}(\hat{\Sigma}_{2}(k)) &
\ge\min_{k\in[k_{0},[\tau_{2}T]]}\left[\rho_{j}\left(\frac{1}{T-k}\hat{G}_{2}^{k'}M_{F_{2}^{k}}\hat{G}_{2}^{k}\right)+\rho_{r}\left(\frac{1}{T-k}\hat{G}_{2}^{k'}P_{F_{2}^{k}}\hat{G}_{2}^{k}\right)\right]\nonumber
\\
 & \ge\min_{k\in[k_{0},[\tau_{2}T]]}\rho_{j}\left(\frac{1}{T-k}\hat{G}_{2}^{k'}M_{F_{2}^{k}}\hat{G}_{2}^{k}\right),\ \mathrm{for\ }j=r_{2}+1,...,r\label{eq:rhor Sigma2k}\end{align}
Thus, it suffices to find the lower bound for $\min_{k\in[k_{0},[\tau_{2}T]]}\rho_{j}\left(\hat{G}_{2}^{k'}M_{F_{2}^{k}}\hat{G}_{2}^{k}/(T-k)\right)$.
As $F_{2}^{k}C^{\prime}=G_{2}^{k}$ for $k\ge k_{0}$, we have
$M_{F_{2}^{k}}G_{2}^{k}=0$ and \begin{equation}
\frac{1}{T-k}\hat{G}_{2}^{k'}M_{F_{2}^{k}}\hat{G}_{2}^{k}=\frac{1}{T-k}(\hat{G}_{2}^{k}-G_{2}^{k}H)^{\prime}M_{F_{2}^{k}}(\hat{G}_{2}^{k}-G_{2}^{k}H).\label{eq:GMG}\end{equation}
 Now, using \eqref{eq:G2hat - G2H}, we can obtain \begin{align}
\frac{1}{\sqrt{T-k}}M_{F_{2}^{k}}(\hat{G}_{2}^{k}-G_{2}^{k}H) & =\frac{1}{\sqrt{T-k}}M_{F_{2}^{k}}\left(\frac{1}{NT}e_{2}^{k}\Lambda
G^{\prime}\hat{G}V_{NT}^{-1}+\frac{1}{NT}G_{2}^{k}\Lambda^{\prime}e^{\prime}\hat{G}V_{NT}^{-1}+\frac{1}{NT}e_{2}^{k}e^{\prime}\hat{G}V_{NT}^{-1}\right)\nonumber \\
 & =a_{1k}+a_{2k}+a_{3k}.\label{eq:a1k a2k a3k}\end{align}
 Let us consider the term $a_{1k}$ in \eqref{eq:a1k a2k a3k}. As $\sigma_{i}(\mathbb{A}+\mathbb{B})\le\sigma_{i}(\mathbb{A})+\sigma_{1}(\mathbb{B})$,
we have \begin{align*}
\sigma_{j}\left(\frac{e_{2}^{k}\Lambda}{N\sqrt{T-k}}\frac{G^{\prime}\hat{G}}{T}V_{NT}^{-1}\right) &
\le\sigma_{j}\left(M_{F_{2}^{k}}\frac{e_{2}^{k}\Lambda}{N\sqrt{T-k}}\frac{G^{\prime}\hat{G}}{T}V_{NT}^{-1}\right)+\sigma_{1}\left(P_{F_{2}^{k}}\frac{e_{2}^{k}\Lambda}{N\sqrt{T-k}}\frac{G^{\prime}\hat{G}}{T}V_{NT}^{-1}\right),\end{align*}
 which implies \begin{align}
 & \min_{k\in[k_{0},[\tau_{2}T]]}\sigma_{j}\left(M_{F_{2}^{k}}\frac{e_{2}^{k}\Lambda}{N\sqrt{T-k}}\frac{G^{\prime}\hat{G}}{T}V_{NT}^{-1}\right)\nonumber \\
\ge &
\min_{k\in[k_{0},[\tau_{2}T]]}\sigma_{j}\left(\frac{e_{2}^{k}\Lambda}{N\sqrt{T-k}}\frac{G^{\prime}\hat{G}}{T}V_{NT}^{-1}\right)-\max_{k\in[k_{0},[\tau_{2}T]]}\sigma_{1}\left(P_{F_{2}^{k}}\frac{e_{2}^{k}\Lambda}{N\sqrt{T-k}}\frac{G^{\prime}\hat{G}}{T}V_{NT}^{-1}\right)\nonumber
\\
\ge &
\frac{1}{\sqrt{N}}\sigma_{r}\left(\frac{G^{\prime}\hat{G}}{T}V_{NT}^{-1}\right)\sqrt{\min_{k\in[k_{0},[\tau_{2}T]]}\rho_{j}\left(\frac{\Lambda^{\prime}e_{2}^{k'}e_{2}^{k}\Lambda}{N(T-k)}\right)}\nonumber
\\
 & -
 \max_{k\in[k_{0},[\tau_{2}T]]}\sqrt{\frac{1}{N(T-k)}\rho_{1}\left(V_{NT}^{-1}\frac{\hat{G}^{\prime}G}{T}\left(\frac{\Lambda^{\prime}e_{2}^{k'}F_{2}^{k}}{\sqrt{NT}}\right)\left(\frac{F_{2}^{k'}F_{2}^{k}}{T}\right)^{-1}\left(\frac{F_{2}^{k'}e_{2}^{k}\Lambda}{\sqrt{NT}}\right)\frac{G^{\prime}\hat{G}}{T}V_{NT}^{-1}\right)}\nonumber
 \\
\ge & \frac{1}{\sqrt{N}}\underline{c}\cdot\sigma_{r}\left(\frac{G^{\prime}\hat{G}}{T}V_{NT}^{-1}\right)+O_{p}\left(\frac{1}{\sqrt{NT}}\right)\ge\frac{1}{\sqrt{N}}c,\ w.p.a.1\ \mathrm{for\ some\
}c>0\label{eq:sigma a1k}\end{align}
 where the third and fourth lines use the inequality $\sigma_{j}(\mathbb{A}\mathbb{B})\ge\sigma_{j}(\mathbb{A})\sigma_{r}(\mathbb{B})$
and the relation $\rho_{r}(\mathbb{A}^{\prime}\mathbb{A})^{1/2}=\sigma_{r}(\mathbb{A})$,
and the fifth line uses Assumptions \ref{as-.LeeL} and \ref{invariance} %\ref{as-.LeeL} and \ref{invariance}
and the fact that $\frac{G^{\prime}\hat{G}}{T}V_{NT}^{-1}$ is nonsingular
as $N,T\to\infty$ by Proposition 1 and Lemma A.3 of Bai (2003).

The term $a_{2k}$ in \eqref{eq:a1k a2k a3k} is zero because $M_{F_{2}^{k}}G_{2}^{k}=0$.
For term $a_{3k}$ in \eqref{eq:a1k a2k a3k}, we can obtain its upper
bound as \begin{align}
\|\frac{1}{NT\sqrt{T-k}}M_{F_{2}^{k}}e_{2}^{k}e^{\prime}\hat{G}V_{NT}^{-1}\|^{2} &
\le2\|\frac{1}{NT\sqrt{T-k}}e_{2}^{k}e^{\prime}\hat{G}V_{NT}^{-1}\|^{2}+2\|\frac{1}{NT\sqrt{T-k}}P_{F_{2}^{k}}e_{2}^{k}e^{\prime}\hat{G}V_{NT}^{-1}\|^{2}\nonumber \\
 & =\frac{1}{N^{2}T^{2}(T-k)}2tr\left(V_{NT}^{-1}\hat{G}^{\prime}ee_{2}^{k'}e_{2}^{k}e^{\prime}\hat{G}V_{NT}^{-1}\right)\nonumber \\
 &~~~ +\frac{1}{N^{2}T^{2}(T-k)}2tr\left(V_{NT}^{-1}\hat{G}^{\prime}ee_{2}^{k'}P_{F_{2}^{k}}e_{2}^{k}e^{\prime}\hat{G}V_{NT}^{-1}\right).\label{eq:a3k}\end{align}
 For the first term in \eqref{eq:a3k}, we have \begin{align*}
\frac{1}{N^{2}T^{2}(T-k)}\|V_{NT}^{-1}\hat{G}^{\prime}ee_{2}^{k'}e_{2}^{k}e^{\prime}\hat{G}V_{NT}^{-1}\| &
=\frac{1}{N^{2}T^{2}(T-k)}\|V_{NT}^{-1}\sum_{s=k+1}^{T}\sum_{t=1}^{T}\hat{g}_{t}e_{t}^{\prime}e_{s}e_{s}^{\prime}\sum_{u=1}^{T}e_{u}\hat{g}_{u}^{\prime}V_{NT}^{-1}\|\\
 & \le\frac{1}{T-k}\|V_{NT}^{-1}\|\sum_{s=k+1}^{T}\|\sum_{t=1}^{T}\hat{g}_{t}e_{t}^{\prime}e_{s}/NT\|^{2}.\end{align*}
 Note that \begin{equation}
\max_{k\in[k_{0},[\tau_{2}T]]}(T-k)^{-1}\sum_{s=k+1}^{T}\|\sum_{t=1}^{T}\hat{g}_{t}e_{t}^{\prime}e_{s}/NT\|^{2}=O_{p}(\delta_{NT}^{-4})\label{eq:a3k1}\end{equation}
 by Lemma \ref{eef}.

For the second term in \eqref{eq:a3k}, we can obtain \begin{equation}
\max_{k\in[k_{0},[\tau_{2}T]]}\left\Vert
V_{NT}^{-1}\frac{\hat{G}^{\prime}ee_{2}^{k'}F_{2}^{k}}{NT(T-k)}\left(\frac{F_{2}^{k'}F_{2}^{k}}{T-k}\right)^{-1}\frac{F_{2}^{k'}e_{2}^{k}e^{\prime}\hat{G}}{NT(T-k)}V_{NT}^{-1}\right\Vert
=O_{p}(\delta_{NT}^{-4}).\label{eq:a3k2}\end{equation}
 To observe this, note that $[F_{2}^{k'}F_{2}^{k}/(T-k)]^{-1}=O_{p}(1)$
uniformly over $k\in[k_{0},[\tau_{2}T]]$ and \begin{align*}
\frac{1}{NT(T-k)}\hat{G}^{\prime}ee_{2}^{k'}F_{2}^{k} & =\frac{1}{NT(T-k)}\sum_{s=k+1}^{T}\sum_{t=1}^{T}\hat{g}_{t}e_{t}^{\prime}e_{s}f_{s}^{\prime}\\
 &
 =\frac{1}{T(T-k)}\sum_{s=k+1}^{T}\sum_{t=1}^{T}\hat{g}_{t}\left[\frac{e_{t}^{\prime}e_{s}}{N}-\gamma_{N}(s,t)\right]f_{s}^{\prime}+\frac{1}{T(T-k)}\sum_{s=k+1}^{T}\sum_{t=1}^{T}\hat{g}_{t}\gamma_{N}(s,t)f_{s}^{\prime}=O_{p}(\delta_{NT}^{-2})\end{align*}
 uniformly over $k\in[k_{0},[\tau_{2}T]]$ following the derivation
of terms I and II in Lemma B.2 of Bai (2003).

Thus, combining the results in \eqref{eq:a3k}--\eqref{eq:a3k2},
we have \begin{equation}
\max_{k\in[k_{0},[\tau_{2}T]]}\sigma_{1}(\frac{1}{NT\sqrt{T-k}}M_{F_{2}^{k}}e_{2}^{k}e^{\prime}\hat{G}V_{NT}^{-1})\le\max_{k\in[k_{0},[\tau_{2}T]]}\|\frac{1}{NT\sqrt{T-k}}M_{F_{2}^{k}}e_{2}^{k}e^{\prime}\hat{G}V_{NT}^{-1}\|=O_{p}(\delta_{NT}^{-2}).\label{eq:sigma
a3k}\end{equation}
 Next, rearranging the terms in \eqref{eq:a1k a2k a3k} yields $a_{1k}=\frac{1}{\sqrt{T-k}}M_{F_{2}^{k}}(\hat{G}_{2}^{k}-G_{2}^{k}H)-a_{3k}$,
which implies that $\sigma_{j}(a_{1k})\le\sigma_{j}\left(\frac{1}{\sqrt{T-k}}M_{F_{2}^{k}}(\hat{G}_{2}^{k}-G_{2}^{k}H)\right)+\sigma_{1}(-a_{3k})$
and \begin{align}
 & \min_{k\in[k_{0},[\tau_{2}T]]}\sigma_{j}\left(\frac{1}{\sqrt{T-k}}M_{F_{2}^{k}}(\hat{G}_{2}^{k}-G_{2}^{k}H)\right)\nonumber \\
 &\ge \min_{k\in[k_{0},[\tau_{2}T]]}\sigma_{j}(a_{1k})-\max_{k\in[k_{0},[\tau_{2}T]]}\sigma_{1}(a_{3k})\ge\frac{1}{\sqrt{N}}c,\ w.p.a.1\ \mathrm{for\ some\ }c>0\label{eq:MG}\end{align}
 because $\sigma_{1}(a_{3k})$ is uniformly $O_{p}(\delta_{NT}^{-2})$
by \eqref{eq:sigma a3k} and dominated by $\sigma_{j}(a_{1k})$ in
\eqref{eq:sigma a1k} under the condition that $\sqrt{N}/T\to0$ as
$N,T\to\infty$. Hence, combining \eqref{eq:rhor Sigma2k}, \eqref{eq:GMG}, and
\eqref{eq:MG} yields \begin{align*}
\min_{k\in[k_{0},[\tau_{2}T]]}\rho_{j}(\hat{\Sigma}_{2}(k)) &
\ge\min_{k\in[k_{0},[\tau_{2}T]]}\rho_{j}\left(\frac{1}{T-k}(\hat{G}_{2}^{k}-G_{2}^{k}H)^{\prime}M_{F_{2}^{k}}(\hat{G}_{2}^{k}-G_{2}^{k}H)\right)\\
 & =\min_{k\in[k_{0},[\tau_{2}T]]}\sigma_{j}\left(\frac{1}{\sqrt{T-k}}M_{F_{2}^{k}}(\hat{G}_{2}^{k}-G_{2}^{k}H)\right)^{2}\ge\frac{1}{N}c^{2}=\frac{1}{N}c_{L},\end{align*}
 w.p.a.1 if $\sqrt{N}/T\to0$ as $N,T\to\infty$. \vspace{1em}

\textbf{{Part 2.}} Note that \begin{eqnarray*}
 &  & \sigma_{j}(\hat{G}_{2}^{k}/\sqrt{T-k})\leq\sigma_{j}(G_{2}^{k}H/\sqrt{T-k})+\sigma_{1}((\hat{G}_{2}^{k}-G_{2}^{k}H)/\sqrt{T-k}).\end{eqnarray*}
 In addition, $\sigma_{j}(G_{2}^{k}H/\sqrt{T-k})=0$ for $r_{2}<j\le r$
and \[
\max_{k\in[k_{0},[\tau_{2}T]]}\sigma_{1}((\hat{G}_{2}^{k}-G_{2}^{k}H)/\sqrt{T-k})\le\max_{k\in[k_{0},[\tau_{2}T]]}\frac{1}{\sqrt{T-k}}\|\hat{G}_{2}^{k}-G_{2}^{k}H\|\le\frac{1}{\sqrt{(1-\tau_{2})T}}\|\hat{G}-GH\|\le\frac{c}{\sqrt{N}}\]
w.p.a.1 for some $0<c<\infty$ by Lemma \ref{Bai_implied} (ii)
if $\sqrt{N}/T\to0$ as $N,T\to\infty$. Thus,
\begin{eqnarray*}
\max_{k\in[k_{0},[\tau_{2}T]]}\rho_{j}(\hat{\Sigma}_{2}(k))=\max_{k\in[k_{0},[\tau_{2}T]]}\sigma_{j}(\hat{G}_{2}^{k}/\sqrt{T-k})^{2}\le c_{U}/N,~~w.p.a.1.
\end{eqnarray*}
as $N,T\to\infty$ for $j=r_{2}+1,...,r$ and some positive
constant $c_{U}$.$\Box$\vspace{2em}

The following lemma yields a bound on the difference
between $|\hat{\Sigma}_{2}^{0}|$ and $|\hat{\Sigma}_{2}(k)|$ for
$k_{0}<k\le\tau_{2}T$ when $C$ is singular. The same result applies
to the difference between $|\hat{\Sigma}_{1}^{0}|$ and $|\hat{\Sigma}_{1}(k)|$
for $\tau_{1}T\le k<k_{0}$ when $B$ is singular. \vspace{1em}

\begin{lemma}\label{differ} Under Assumptions \ref{factors}--\ref{invariance}, for $k>k_{0}$
and $k\le[\tau_{2}T]$, if $C$ is singular and $T/N\to\kappa$ as
$N,T\to\infty$ for $0<\kappa<\infty$, then \[
\max_{k_{0}<k\le\tau_{2}T}\frac{1}{k-k_{0}}\left||\hat{\Sigma}_{2}^{0}|-|\hat{\Sigma}_{2}(k)|\right|=O_{p}(T^{-(r-r_{2})-1}).\]
 \end{lemma} \textbf{Proof}:

First, note that \begin{eqnarray*}
\hat{\Sigma}_{2}^{0} & = &
\frac{1}{T-k_{0}}\sum_{t=k_{0}+1}^{T}\hat{g}_{t}\hat{g}_{t}^{\prime}=\frac{1}{T-k_{0}}\sum_{t=k+1}^{T}\hat{g}_{t}\hat{g}_{t}^{\prime}+\frac{1}{T-k_{0}}\sum_{t=k_{0}+1}^{k}\hat{g}_{t}\hat{g}_{t}^{\prime}\\
 & = & A_{k}+\frac{1}{T-k_{0}}\mathcal{\hat{G}}^{\prime}\mathcal{\hat{G}},\end{eqnarray*}
 where $A_{k}\equiv(T-k_{0})^{-1}\sum_{t=k+1}^{T}\hat{g}_{t}\hat{g}_{t}^{\prime}$
and $\mathcal{\hat{G}}\equiv[\hat{g}_{k_{0}+1},...,\hat{g}_{k}]^{\prime}$.
By (7.10) of \citet{Lange2010}, we have\begin{equation}
|\hat{\Sigma}_{2}^{0}|=|A_{k}|\cdot|I_{k-k_{0}}+\frac{1}{T-k_{0}}\mathcal{\hat{G}}A_{k}^{-1}\mathcal{\hat{G}}^{\prime}|,\label{eq:det(Sig20)}\end{equation}
 where $A_{k}^{-1}$ is reasonable because the smallest eigenvalue of
$N\cdot A_{k}$ is bounded away from zero by proposition \ref{low_bound}.

We now analyze the term $(T-k_{0})^{-1}\mathcal{\hat{G}}A_{k}^{-1}\mathcal{\hat{G}}^{\prime}$,
which can be written as \begin{eqnarray}
\frac{1}{T-k_{0}}\mathcal{\hat{G}}A_{k}^{-1}\mathcal{\hat{G}}^{\prime} & = & \frac{1}{|A_{k}|}\left(\frac{1}{T-k_{0}}\mathcal{\hat{G}}A_{k}^{\#}\mathcal{\hat{G}}^{\prime}\right)\nonumber \\
 & = &
 \frac{1}{|A_{k}|(T-k_{0})}\mathcal{\hat{G}}\left(A_{k}^{\#}-\left[\frac{1}{T-k_{0}}\sum_{t=k+1}^{T}H^{\prime}g_{t}g_{t}^{\prime}H\right]^{\#}\right)\mathcal{\hat{G}}^{\prime}+\frac{1}{|A_{k}|(T-k_{0})}\mathcal{\hat{G}}\left[\frac{1}{T-k_{0}}\sum_{t=k+1}^{T}H^{\prime}g_{t}g_{t}^{\prime}H\right]^{\#}\mathcal{\hat{G}}^{\prime}\nonumber
 \\
 & \equiv & \mathbb{S}_{1}+\mathbb{S}_{2}.\label{eq:lemma A1 eq1}\end{eqnarray}
 For $\mathbb{S}_{1}$, \begin{equation}
\max_{k_{0}<k\le\tau_{2}T}\|A_{k}-\frac{1}{T-k_{0}}\sum_{t=k+1}^{T}H^{\prime}g_{t}g_{t}^{\prime}H\|=O_{p}(N^{-1})\label{eq:bai2003lemmab2b3}\end{equation}
by a uniform version of Lemmas B2 and B3 of \cite{Bai2003}. When $r-r_{2}=1$, \eqref{eq:bai2003lemmab2b3}
implies \[
\max_{k_{0}<k\le\tau_{2}T}\left\Vert A_{k}^{\#}-\left[(T-k_{0})^{-1}\sum_{t=k+1}^{T}H^{\prime}g_{t}g_{t}^{\prime}H\right]^{\#}\right\Vert =O_{p}(N^{-1}).\]
Then, it follows that \begin{equation}
\max_{k_{0}<k\le\tau_{2}T}\|\frac{1}{k-k_{0}}\mathbb{S}_{1}\|\le\max_{k_{0}<k\le\tau_{2}T}\left(\frac{1}{|A_{k}|(T-k_{0})}\frac{\|\mathcal{\hat{G}}\|^{2}}{k-k_{0}}\right)O_{p}(N^{-1})=O_{p}\left(\frac{1}{T}\right)\
\ \mathrm{for}\ r_{2}=r-1\label{eq:S1}\end{equation}
given the fact that $1/|A_{k}|$ is uniformly $O_{p}(T)$ when $r-r_{2}=1$
by proposition \ref{low_bound} under the condition $N\propto T$ and the
fact that \[
\max_{k_{0}<k\le\tau_{2}T}\frac{1}{k-k_{0}}\|\mathcal{\hat{G}}\|^{2}\le\max_{k_{0}<k\le\tau_{2}T}\frac{2}{k-k_{0}}\sum_{t=k_{0}+1}^{k}\|\hat{g}_{t}-H^{\prime}g_{t}\|^{2}+\frac{2}{k-k_{0}}\sum_{t=k_{0}+1}^{k}\|H^{\prime}g_{t}\|^{2}=O_{p}(1)\]
 by Lemma \ref{Bai_implied}.

When $r-r_{2}\ge2$, \[
\rho_{1}(A_{k}^{\#})=|A_{k}|/\rho_{r}(A_{k})=O_{p}(N^{-(r-r_{2}-1)})\]
 uniformly over $k_{0}<k\le\tau_{2}T$ by proposition \ref{low_bound};
thus, we have \begin{equation}
\|A_{k}^{\#}\|\le\sqrt{r}\sigma_{1}(A_{k}^{\#})=\sqrt{r}\rho_{1}(A_{k}^{\#})=O_{p}(N^{-(r-r_{2}-1)})\ \ \mathrm{for}\ r-r_{2}\ge2.\label{eq:Taylor}\end{equation}
Now, let $f_{t}=\Sigma_{f}^{1/2}\varepsilon_{t}$ with $E\varepsilon_{t}\varepsilon_{t}^{\prime}=I_{r}$.
Hence, for $k\ge k_{0}$, we have $g_{k+1}=C\Sigma_{f}^{1/2}\varepsilon_{k+1}$.
By \eqref{eq:bai2003lemmab2b3} and\\
$\max_{k_{0}<k\le\tau_{2}T}\|\frac{1}{T-k}\sum_{t=k+1}^{T}f_{t}f_{t}^{\prime}-\Sigma_{F}\|=\max_{k_{0}<k\le\tau_{2}T}\|\frac{1}{T-k}\sum_{t=k+1}^{T}\epsilon_t\|=O_{p}(T^{-1/2})$
by H\'{a}jek-R\'{e}nyi inequality, we have $\max_{k_{0}<k\le\tau_{2}T}\|\frac{1}{T-k_{0}}\sum_{t=k+1}^{T}H^{\prime}g_{t}g_{t}^{\prime}H-H^{\prime}C\Sigma_{f}C^{\prime}H\|=O_{p}(T^{-1/2})$
and \[
\max_{k_{0}<k\le\tau_{2}T}\|A_{k}-H^{\prime}C\Sigma_{f}C^{\prime}H\|=O_{p}(\delta_{NT}^{-1});\]
thus, \begin{equation}
\max_{k_{0}<k\le\tau_{2}T}\|A_{k}^{1/2}U_{k}-H^{\prime}C\Sigma_{f}^{1/2}\|=O_{p}(\delta_{NT}^{-1}),\label{eq:Bk^(1/2) U}\end{equation}
 where $U_{k}=A_{k}^{1/2}(\Sigma_{f}^{1/2}C^{\prime}H)^{-}$. Therefore,
for $t>k_{0}$, we have\begin{eqnarray*}
\hat{g}_{t} & = & H^{\prime}Cf_{t}+(\hat{g}_{t}-H^{\prime}g_{t})=H^{\prime}C\Sigma_{f}^{1/2}\varepsilon_{t}+(\hat{g}_{t}-H^{\prime}g_{t})\\
 & = & A_{k}^{1/2}U_{k}\varepsilon_{t}+(H^{\prime}C\Sigma_{f}^{1/2}-A_{k}^{1/2}U_{k})\varepsilon_{t}+(\hat{g}_{t}-H^{\prime}g_{t})\\
 & = & A_{k}^{1/2}U_{k}\varepsilon_{t}+O_{p}(\delta_{NT}^{-1})\varepsilon_{t}+(\hat{g}_{t}-H^{\prime}g_{t})\end{eqnarray*}
 by \eqref{eq:Bk^(1/2) U} and\begin{eqnarray}
\mathcal{\hat{G}}^{\prime}-A_{k}^{1/2}U_{k}\mathcal{E}^{\prime} & = & O_{p}(\delta_{NT}^{-1})\cdot\mathcal{E}^{\prime}+\mathcal{\hat{G}}^{\prime}-H^{\prime}\mathcal{G}^{\prime}\label{eq:gk+1
expansion}\end{eqnarray}
where $\mathcal{E}^{\prime}\equiv[\varepsilon_{k_{0}+1},...,\varepsilon_{k}]$,
$\mathcal{G}^{\prime}\equiv[g_{k_{0}+1},...,g_{k}]$, and $O_{p}(\delta_{NT}^{-1})$
term is uniform in $k_{0}<k\le\tau_{2}T$. In addition, \[
\left[\frac{1}{T-k_{0}}\sum_{t=k+1}^{T}H^{\prime}g_{t}g_{t}^{\prime}H\right]^{\#}=0,\ \ \mathrm{for}\ r-r_{2}\ge2.\]
 Thus, we have\begin{eqnarray}
 &  & \frac{1}{T-k_{0}}\max_{k_{0}<k\le\tau_{2}T}\left\Vert
 \frac{1}{k-k_{0}}\mathcal{\hat{G}}\left(A_{k}^{\#}-\left[\frac{1}{T-k_{0}}\sum_{t=k+1}^{T}H^{\prime}g_{t}g_{t}^{\prime}H\right]^{\#}\right)\mathcal{\hat{G}}^{\prime}\right\Vert \nonumber \\
 & = & \frac{1}{T-k_{0}}\max_{k_{0}<k\le\tau_{2}T}\left\Vert
 \frac{1}{k-k_{0}}\mathcal{E}U_{k}^{\prime}A_{k}^{1/2}A_{k}^{\#}A_{k}^{1/2}U_{k}\mathcal{E}^{\prime}+\frac{1}{k-k_{0}}(\mathcal{\hat{G}}-\mathcal{E}U_{k}^{\prime}A_{k}^{1/2})A_{k}^{\#}A_{k}^{1/2}U_{k}\mathcal{E}^{\prime}\right.\nonumber
 \\
 &  &
 +\left.\frac{1}{k-k_{0}}\mathcal{E}U_{k}^{\prime}A_{k}^{1/2}A_{k}^{\#}(\mathcal{\hat{G}}^{\prime}-A_{k}^{1/2}U_{k}\mathcal{E}^{\prime})+\frac{1}{k-k_{0}}(\mathcal{\hat{G}}-\mathcal{E}U_{k}^{\prime}A_{k}^{1/2})A_{k}^{\#}(\mathcal{\hat{G}}^{\prime}-A_{k}^{1/2}U_{k}\mathcal{E}^{\prime})\right\Vert
 \nonumber \\
 & \le &
 \frac{1}{T-k_{0}}\max_{k_{0}<k\le\tau_{2}T}\frac{1}{k-k_{0}}|A_{k}|\sum_{t=k_{0}+1}^{k}\|\varepsilon_{t}\|^{2}\|U_{k}\|^{2}+\frac{2}{T-k_{0}}\max_{k_{0}<k\le\tau_{2}T}\frac{1}{k-k_{0}}\|\mathcal{\hat{G}}-\mathcal{E}U_{k}^{\prime}A_{k}^{1/2}\|\|(A_{k}^{1/2})^{\#}\|\|U_{k}\mathcal{E}^{\prime}\||A_{k}^{1/2}|\nonumber
 \\
 &  & +\frac{1}{T-k_{0}}\max_{k_{0}<k\le\tau_{2}T}\frac{1}{k-k_{0}}\|\mathcal{\hat{G}}-\mathcal{E}U_{k}^{\prime}A_{k}^{1/2}\|^{2}\|A_{k}^{\#}\|\label{eq:term1 in LemmaA1 eq1}\end{eqnarray}
where we use the fact that $A_{k}^{\#}=(A_{k}^{1/2})^{\#}(A_{k}^{1/2})^{\#}$
and $(A_{k}^{1/2})^{\#}A_{k}^{1/2}=|A_{k}|^{1/2}I_{r}$. The definition
of $U_{k}$ implies that $U_{k}=O_{p}(1)$ uniformly over $k_{0}<k\le\tau_{2}T$.
The first term in \eqref{eq:term1 in LemmaA1 eq1} is $O_{p}(N^{-(r-r_{2})}T^{-1})$
because $|A_{k}|=O_{p}(N^{-(r-r_{2})})$ by proposition \ref{low_bound},
the second term is $O_{p}(N^{-(r-r_{2})}\sqrt{N}\delta_{NT}^{-1}T^{-1})$
because $|A_{k}^{1/2}|=O_{p}(N^{-(r-r_{2})/2})$, $(k-k_{0})^{-1/2}\|\mathcal{\hat{G}}^{\prime}-B_{k}^{1/2}U\mathcal{E}^{\prime}\|$
is uniformly $O_{p}(\delta_{NT}^{-1})$ by \eqref{eq:gk+1 expansion}
and Lemma \ref{Bai_implied}, $(A_{k}^{1/2})^{\#}$ is uniformly
$O_{p}(N^{-(r-r_{2}-1)/2})$ by \eqref{eq:Taylor}, and the last term
in \eqref{eq:term1 in LemmaA1 eq1} is $O_{p}(\delta_{NT}^{-2}N^{-(r-r_{2}-1)}T^{-1})$
by \eqref{eq:Taylor}, \eqref{eq:gk+1 expansion}, and Lemma \ref{Bai_implied}.
 The result in \eqref{eq:term1 in LemmaA1 eq1} indicates that \begin{equation}
\max_{k_{0}<k\le\tau_{2}T}\|\frac{1}{k-k_{0}}\mathbb{S}_{1}\|=O_{p}(T^{-1})\label{eq:S1 case2}\end{equation}
under the condition $N\propto T$. Recalling \eqref{eq:S1}, we obtain
the same rate of $\mathbb{S}_{1}$ for both $r_{2}=r-1$ and $r_{2}\le r-2$.

Term $\mathbb{S}_{2}$ in \eqref{eq:lemma A1 eq1} is zero if $r_{2}\le r-2$
because the adjoint matrix of an $r\times r$ matrix $\mathbb{A}$
is zero when rank$(\mathbb{A})\le r-2$.

When $r_{2}=r-1$, we have \begin{eqnarray*}
 &  & \max_{k_{0}<k\le\tau_{2}T}\frac{1}{|A_{k}|(k-k_{0})}\left\Vert
 \mathcal{\hat{G}}\left[\frac{1}{T-k_{0}}\sum_{t=k+1}^{T}H^{\prime}g_{t}g_{t}^{\prime}H\right]^{\#}\mathcal{\hat{G}}^{\prime}\right\Vert \\
 & = & \max_{k_{0}<k\le\tau_{2}T}\frac{1}{|A_{k}|(k-k_{0})}\left\Vert
 \mathcal{\hat{G}}\left[\frac{1}{T-k_{0}}\sum_{t=k+1}^{T}H^{\prime}g_{t}g_{t}^{\prime}H\right]^{\#}(\mathcal{\hat{G}}^{\prime}-H^{\prime}\mathcal{G}^{\prime})+\frac{1}{|A_{k}|(k-k_{0})}\mathcal{\hat{G}}\left[\frac{1}{T-k_{0}}\sum_{t=k+1}^{T}H^{\prime}g_{t}g_{t}^{\prime}H\right]^{\#}H^{\prime}\mathcal{G}^{\prime}\right\Vert
 \\
 & = & \max_{k_{0}<k\le\tau_{2}T}\frac{1}{|A_{k}|(k-k_{0})}\left\Vert
 \mathcal{\hat{G}}\left[\frac{1}{T-k_{0}}\sum_{t=k+1}^{T}H^{\prime}g_{t}g_{t}^{\prime}H\right]^{\#}(\mathcal{\hat{G}}^{\prime}-H^{\prime}\mathcal{G}^{\prime})+\frac{1}{|A_{k}|(k-k_{0})}\mathcal{\hat{G}}\left[\frac{1}{T-k_{0}}\sum_{t=k+1}^{T}f_{t}g_{t}^{\prime}H\right]^{\#}(H^{\prime}C)^{\#}H^{\prime}\mathcal{G}^{\prime}\right\Vert
 \\
 & = & \max_{k_{0}<k\le\tau_{2}T}\frac{1}{|A_{k}|(k-k_{0})}\left\Vert
 \mathcal{\hat{G}}\left[\frac{1}{T-k_{0}}\sum_{t=k+1}^{T}H^{\prime}g_{t}g_{t}^{\prime}H\right]^{\#}(\mathcal{\hat{G}}^{\prime}-H^{\prime}\mathcal{G}^{\prime})\right\Vert \\
 & = & \max_{k_{0}<k\le\tau_{2}T}\frac{1}{|A_{k}|(k-k_{0})}\left\Vert
 (\mathcal{\hat{G}}-\mathcal{G}H)\left[\frac{1}{T-k_{0}}\sum_{t=k+1}^{T}H^{\prime}g_{t}g_{t}^{\prime}H\right]^{\#}(\mathcal{\hat{G}}^{\prime}-H^{\prime}\mathcal{G}^{\prime})\right\Vert
 =O_{p}(1),\end{eqnarray*}
 where the second term in the third line is zero because $(H^{\prime}C^{\prime})^{\#}H^{\prime}g_{t}=(H^{\prime}C)^{\#}H^{\prime}Cf_{t}=|H^{\prime}C^{\prime}|I_{r}=0$
for $t>k_{0}$, and the last equality follows from Lemma \ref{Bai_implied}
and the result that $1/|A_{k}|=O_{p}(N)$ by proposition \ref{low_bound}
for $r=r_{2}+1$. Hence, under the condition $T\propto N$, we have
\begin{equation}
\max_{k_{0}<k\le\tau_{2}T}\|\frac{1}{k-k_{0}}\mathbb{S}_{2}\|=O_{p}(T^{-1}).\label{eq:S2 final}\end{equation}
Thus, combining the results in \eqref{eq:lemma A1 eq1}, \eqref{eq:S1},
\eqref{eq:S1 case2}, and \eqref{eq:S2 final}, we obtain
\begin{equation}
\frac{1}{T-k_{0}}\max_{k_{0}<k\le\tau_{2}T}\|\frac{1}{k-k_{0}}\mathcal{\hat{G}}A_{k}^{-1}\mathcal{\hat{G}}^{\prime}\|=O_{p}(T^{-1}),\label{eq:GBG}\end{equation}
Thus, \eqref{eq:det(Sig20)} can be written as\begin{eqnarray*}
|\hat{\Sigma}_{2}^{0}| & = & |A_{k}|\prod_{j=1}^{r}\left[1+\rho_{j}\left(\frac{1}{T-k_{0}}\mathcal{\hat{G}}A_{k}^{-1}\mathcal{\hat{G}}^{\prime}\right)\right]\\
 & \le &
 |A_{k}|\left[1+\rho_{1}\left(\frac{1}{T-k_{0}}\mathcal{\hat{G}}A_{k}^{-1}\mathcal{\hat{G}}^{\prime}\right)\right]^{r}\le|A_{k}|\left[1+O_{p}\left(\frac{k-k_{0}}{T}\right)\right]^{r},\end{eqnarray*}
 where we use \eqref{eq:GBG} and the fact that
 $\rho_{1}(\mathcal{\hat{G}}A_{k}^{-1}\mathcal{\hat{G}}^{\prime})=\sigma_{1}(\mathcal{\hat{G}}A_{k}^{-1}\mathcal{\hat{G}}^{\prime})\le\|\mathcal{\hat{G}}A_{k}^{-1}\mathcal{\hat{G}}^{\prime}\|$.
Thus, by proposition \ref{low_bound}, we have \begin{eqnarray}
0<\frac{1}{k-k_{0}}(|\hat{\Sigma}_{2}^{0}|-|A_{k}|) & \le & O_{p}\left(\frac{1}{T}\right)|A_{k}|=O_{p}\left(T^{-1}N^{-(r-r_{2})}\right),\label{eq:lemma A1 eq1 bound}\end{eqnarray}
where the $O_{p}(T^{-1}N^{-(r-r_{2})})$ term is uniform over $k_{0}<k\le\tau_{2}T$.

Next, comparing $|A_{k}|$ and $|\hat{\Sigma}_{2}(k)|$, we have \begin{eqnarray}
 &  & \max_{k_{0}<k<\tau_{2}T}\frac{1}{k-k_{0}}\left||A_{k}|-|\hat{\Sigma}_{2}(k)|\right|\nonumber \\
 & = & \max_{k_{0}<k<\tau_{2}T}\frac{1}{k-k_{0}}\left||\frac{1}{T-k_{0}}\sum_{t=k+1}^{T}\hat{g}_{t}\hat{g}_{t}^{\prime}|-|\frac{1}{T-k}\sum_{t=k+1}^{T}\hat{g}_{t}\hat{g}_{t}^{\prime}|\right|\nonumber \\
 & = & \max_{k_{0}<k<\tau_{2}T}\frac{1}{k-k_{0}}\left|\left(1-\frac{k-k_{0}}{T-k_{0}}\right)^{r}-1\right||\hat{\Sigma}_{2}(k)|\nonumber \\
 &=&O_{p}\left(T^{-1}N^{-(r-r_{2})}\right),\label{eq:lemma A1 eq2}\end{eqnarray}
where we use the fact that $\max_{k_{0}<k<\tau_{2}T}|\hat{\Sigma}_{2}(k)|=O_{p}(N^{-(r-r_{2})})$
by proposition \ref{low_bound}. As both \eqref{eq:lemma A1 eq1 bound}
and \eqref{eq:lemma A1 eq2} are shown to be $O_{p}(T^{-1}N^{-(r-r_{2})})$,
we obtain the desired result for this lemma under the condition $T\propto N$.
$\Box$\vspace{2em}

The following lemma yields a lower bound on the difference
between $|\hat{\Sigma}_{2}(k)|$ and $|\hat{\Sigma}_{2}^{0}|$ for
$\tau_{1}T\le k<k_{0}$ when $C$ is singular and $B$ is either
singular or nonsingular. The same result applies to the difference
between $|\hat{\Sigma}_{1}(k)|$ and $|\hat{\Sigma}_{1}^{0}|$ for
$k_{0}<k\le\tau_{2}T$ when $B$ is singular.\vspace{1em}

\begin{lemma}\label{differ2} Under Assumptions \ref{factors}--\ref{B_C_full_rank_project}, for
$\tau_{1}T\le k<k_{0}$, if $C$ is singular and $T/N\to\kappa$ as
$N,T\to\infty$ for $0<\kappa<\infty$, then \[
\frac{|\hat{\Sigma}_{2}(k)|-|\hat{\Sigma}_{2}^{0}|}{|\hat{\Sigma}_{2}^{0}|}\ge c\cdot(k_{0}-k)\ \mathrm{w.p.a.1}\]
 for a constant $c>0$ as $N,T\to\infty$. \end{lemma}

\noindent \textbf{Proof}:

Let us rewrite $\hat{\Sigma}_{2}(k)$ as \begin{eqnarray*}
\hat{\Sigma}_{2}(k) & = &
\frac{1}{T-k}\sum_{t=k+1}^{T}\hat{g}_{t}\hat{g}_{t}^{\prime}=\frac{1}{T-k}\sum_{t=k_{0}+1}^{T}\hat{g}_{t}\hat{g}_{t}^{\prime}+\frac{1}{T-k}\sum_{t=k+1}^{k_{0}}\hat{g}_{t}\hat{g}_{t}^{\prime}\\
 & = & D_{k}+\frac{1}{T-k}\mathcal{\hat{G}}^{\prime}\mathcal{\hat{G}},\end{eqnarray*}
 where $D_{k}\equiv(T-k)^{-1}\sum_{t=k_{0}+1}^{T}\hat{g}_{t}\hat{g}_{t}^{\prime}$
and $\mathcal{\hat{G}}\equiv[\hat{g}_{k+1},...,\hat{g}_{k_{0}}]^{\prime}$.
By (7.10) of \citet{Lange2010}, we have \begin{eqnarray}
|\hat{\Sigma}_{2}(k)| & = & |D_{k}|\cdot|I_{k_{0}-k}+\frac{1}{T-k}\mathcal{\hat{G}}D_{k}^{-1}\mathcal{\hat{G}}^{\prime}|\nonumber \\
 & \ge & |D_{k}|\left[1+\rho_{1}\left(\frac{1}{T-k}\mathcal{\hat{G}}D_{k}^{-1}\mathcal{\hat{G}}^{\prime}\right)\right]\label{eq:det(Sig2k)}\end{eqnarray}
 We would like to find the lower bound of the largest eigenvalue of
matrix $\frac{1}{T-k}\mathcal{\hat{G}}D_{k}^{-1}\mathcal{\hat{G}}^{\prime}$,
which can be written as \begin{eqnarray}
\frac{1}{T-k}\mathcal{\hat{G}}D_{k}^{-1}\mathcal{\hat{G}}^{\prime} & = & \frac{1}{|D_{k}|}\left(\frac{1}{T-k}\mathcal{\hat{G}}D_{k}^{\#}\mathcal{\hat{G}}^{\prime}\right)\nonumber \\
 & = &
 \frac{1}{|D_{k}|(T-k)}\mathcal{\hat{G}}\left(D_{k}^{\#}-\left[\frac{1}{T-k}\sum_{t=k_{0}+1}^{T}H^{\prime}g_{t}g_{t}^{\prime}H\right]^{\#}\right)\mathcal{\hat{G}}^{\prime}+\frac{1}{|D_{k}|(T-k)}\mathcal{\hat{G}}\left[\frac{1}{T-k}\sum_{t=k_{0}+1}^{T}H^{\prime}g_{t}g_{t}^{\prime}H\right]^{\#}\mathcal{\hat{G}}^{\prime}\nonumber
 \\
 & \equiv & \frac{1}{|D_{k}|}(\mathbb{P}_{1}+\mathbb{P}_{2}).\label{eq:GD^(-1)G}\end{eqnarray}
 The subsequent proof will be performed in two steps. \vspace{1em}

\textbf{Step 1}. When $r-1=r_{2}>0$, we have \begin{equation}
\max_{\tau_{1}T\le k<k_{0}}\frac{1}{k_{0}-k}\|\mathbb{P}_{1}\|=O_{p}(T^{-1}N^{-1})\label{eq:P1}\end{equation}
because $\max_{\tau_{1}T\le k<k_{0}}\frac{1}{k_{0}-k}\|\mathcal{\hat{G}}\|^{2}=O_{p}(1)$,
\[
\max_{\tau_{1}T\le k<k_{0}}\frac{1}{T-k}\left\Vert D_{k}^{\#}-\left[\frac{1}{T-k}\sum_{t=k_{0}+1}^{T}H^{\prime}g_{t}g_{t}^{\prime}H\right]^{\#}\right\Vert =O_{p}(N^{-1})\]
by \eqref{eq:bai2003lemmab2b3}. Hence, we have\begin{equation}
\frac{\rho_{r}(\mathbb{P}_{1})}{k_{0}-k}\ge-\frac{\max_{j}|\rho_{j}(\mathbb{P}_{1})|}{k_{0}-k}=-\frac{\sigma_{r}(\mathbb{P}_{1})}{k_{0}-k}\ge-\frac{\|\mathbb{P}_{1}\|}{k_{0}-k}=O_{p}(T^{-1}N^{-1})\label{eq:rhomin
P1}\end{equation}
given the fact that the singular values are absolute values of the eigenvalues
of a symmetric matrix.

Next, for the term $\mathbb{P}_{2}$ in \eqref{eq:GD^(-1)G}, we have
\begin{eqnarray}
\mathcal{\hat{G}}\left[\frac{1}{T-k}\sum_{t=k_{0}+1}^{T}H^{\prime}g_{t}g_{t}^{\prime}H\right]^{\#}\mathcal{\hat{G}}^{\prime} & = &
\mathcal{\hat{G}}H^{\#}C^{\#'}\left[\frac{1}{T-k}\sum_{t=k_{0}+1}^{T}f_{t}f_{t}^{\prime}\right]^{\#}C^{\#}H^{\#'}\mathcal{\hat{G}}^{\prime}\nonumber \\
 & = &
 [(\mathcal{\hat{G}}-\mathcal{G}H)+\mathcal{G}H]H^{\#}C^{\#'}\underbrace{\left[\frac{1}{T-k}\sum_{t=k_{0}+1}^{T}f_{t}f_{t}^{\prime}\right]^{\#}}_{\equiv\mathbb{Q}_{k}}C^{\#}H^{\#'}[H^{\prime}\mathcal{G}^{\prime}+(\mathcal{\hat{G}}^{\prime}-H^{\prime}\mathcal{G}^{\prime})],\label{eq:P2
 expand}\end{eqnarray}
where $\mathcal{G}\equiv[g_{k+1},...,g_{k_{0}}]^{\prime}$, and the first
line uses the fact that $g_{t}=Cf_{t}$ for $t\ge k_{0}$.
As $\sigma_{1}(\mathbb{A}+\mathbb{B})\le\sigma_{1}(\mathbb{A})+\sigma_{1}(\mathbb{B})$,
we have \begin{equation}
\sigma_{1}(\mathbb{Q}_{k}^{1/2}C^{\#}H^{\#'}H^{\prime}\mathcal{G}^{\prime})\le\sigma_{1}(\mathbb{Q}_{k}^{1/2}C^{\#}H^{\#'}\mathcal{\hat{G}}^{\prime})+\sigma_{1}(-\mathbb{Q}_{k}^{1/2}C^{\#}H^{\#'}(\mathcal{\hat{G}}^{\prime}-H^{\prime}\mathcal{G}^{\prime}))\label{eq:sigma_max
inequality}\end{equation}
by setting $\mathbb{A}=\mathbb{Q}_{k}^{1/2}C^{\#}H^{\#'}\mathcal{\hat{G}}^{\prime}$
and $\mathbb{B}=-\mathbb{Q}_{k}^{1/2}C^{\#}H^{\#'}(\mathcal{\hat{G}}^{\prime}-H^{\prime}\mathcal{G}^{\prime})$.
Rearranging the inequality in \eqref{eq:sigma_max inequality} and
using \eqref{eq:P2 expand}, we obtain \begin{eqnarray}
\frac{1}{\sqrt{k_{0}-k}}\sigma_{1}(\mathbb{Q}_{k}^{1/2}C^{\#}H^{\#'}\mathcal{\hat{G}}^{\prime}) & \ge &
\frac{1}{\sqrt{k_{0}-k}}\sigma_{1}(\mathbb{Q}_{k}^{1/2}C^{\#}\mathcal{G}^{\prime})|H|-\frac{1}{\sqrt{k_{0}-k}}\sigma_{1}(\mathbb{Q}_{k}^{1/2}C^{\#}H^{\#'}(\mathcal{\hat{G}}^{\prime}-H^{\prime}\mathcal{G}^{\prime}))\nonumber
\\
\sqrt{\frac{1}{k_{0}-k}\rho_{1}(\mathcal{\hat{G}}H^{\#}C^{\#'}\mathbb{Q}_{k}C^{\#}H^{\#'}\mathcal{\hat{G}}^{\prime})} & \ge &
|H|\sqrt{\frac{1}{k_{0}-k}\rho_{1}(\mathcal{G}C^{'\#}\mathbb{Q}_{k}C^{\#}\mathcal{G}^{\prime})}-\frac{1}{\sqrt{k_{0}-k}}\|\mathbb{Q}_{k}^{1/2}C^{\#}H^{\#'}(\mathcal{\hat{G}}^{\prime}-H^{\prime}\mathcal{G}^{\prime})\|\nonumber
\\
 & \ge & |H|\sqrt{\frac{\rho_{1}(\mathcal{G}C^{'\#}\mathbb{Q}_{k}C^{\#}\mathcal{G}^{\prime})}{k_{0}-k}}-O_{p}(N^{-1/2})\label{eq:sqrt_rhomax}\end{eqnarray}
where the first line is based on the fact that $H^{\#}H=I_{r}|H|$, the second
line uses the inequality that the maximum singular value is bounded
by the Frobenius norm, and the third line follows from the derivation
below: \begin{eqnarray*}
\frac{1}{k_{0}-k}\|\mathbb{Q}_{k}^{1/2}C^{\#}H^{\#'}(\mathcal{\hat{G}}^{\prime}-H^{\prime}\mathcal{G}^{\prime})\|^{2} & = &
\frac{1}{k_{0}-k}tr[(\mathcal{\hat{G}}-\mathcal{G}H)H^{\#}C^{\#'}\mathbb{Q}_{k}C^{\#}H^{\#'}(\mathcal{\hat{G}}^{\prime}-H^{\prime}\mathcal{G}^{\prime})]\\
 & \le & \rho_{1}(\mathbb{Q}_{k})\frac{1}{k_{0}-k}tr[C^{\#}H^{\#'}(\mathcal{\hat{G}}^{\prime}-H^{\prime}\mathcal{G}^{\prime})(\mathcal{\hat{G}}-\mathcal{G}H)H^{\#}C^{\#'}]\\
 & = & O_{p}(N^{-1})\ \mathrm{uniformly\ over\ }\tau_{1}T\le k<k_{0}\end{eqnarray*}
 given the fact that $\max_{\tau_{1}T\le k<k_{0}}(k_{0}-k)^{-1}\|(\mathcal{\hat{G}}^{\prime}-H^{\prime}\mathcal{G}^{\prime})(\mathcal{\hat{G}}-\mathcal{G}H)\|=O_{p}(N^{-1})$
by Lemma \ref{Bai_implied}.

Now, it suffices to find the lower bound of $\rho_{1}(\mathcal{G}C^{'\#}\mathbb{Q}_{k}C^{\#}\mathcal{G}^{\prime})$.
Using the definition of $\mathbb{Q}_{k}$ and inequality $\rho_{1}(\mathbb{AB})\ge\rho_{r}(\mathbb{A})\rho_{1}(\mathbb{B})$
for $r\times r$ positive semidefinite matrices $\mathbb{A}$ and
$\mathbb{B}$, we have \begin{equation}
\rho_{1}(\mathcal{G}C^{'\#}\mathbb{Q}_{k}C^{\#}\mathcal{G}^{\prime})\ge\rho_{r}\left(\left[\frac{T-k_{0}}{T-k}\hat{\Sigma}_{f,2}^{0}\right]^{\#}\right)\rho_{1}\left(C^{\#}\mathcal{G}^{\prime}\mathcal{G}C^{'\#}\right)=\frac{T-k_{0}}{T-k}\left(\frac{|\Sigma_{f}|}{\rho_{1}(\Sigma_{f})}+o_{p}(1)\right)\rho_{1}\left(C^{\#}\mathcal{G}^{\prime}\mathcal{G}C^{'\#}\right)\label{eq:GCQCG}\end{equation}
where we use the facts that $\hat{\Sigma}_{f,2}^{0}\equiv\frac{1}{T-k_{0}}\sum_{t=k_{0}+1}^{T}f_{t}f_{t}^{\prime}\to_{p}\Sigma_{f}$
and $\rho_{r}(\Sigma_{f}^{\#})=\rho_{r}(|\Sigma_{f}|\Sigma_{f}^{-1})=|\Sigma_{f}|/\rho_{1}(\Sigma_{f})$.

For $k_{0}-k\to\infty$ as $N,T\to\infty$, \begin{equation}
\rho_{1}\left(C^{\#}\frac{\mathcal{G}^{\prime}\mathcal{G}}{k_{0}-k}C^{'\#}\right)=\rho_{1}\left(C^{\#}B\Sigma_{f}B^{\prime}C^{'\#}+o_{p}(1)\right)>c_{1}\ \mathrm{w.p.a.}1\label{eq:k0-k
unbounded}\end{equation}
 for a constant $c_{1}>0$ as $N,T\to\infty$, because $C^{\#}B\ne0$
and $\Sigma_{f}$ is positive definite according to Assumption \ref{B_C_full_rank_project} (i).
As $|\Sigma_{f}|/\rho_{1}(\Sigma_{f})>0$ in \eqref{eq:GCQCG},
we have $\frac{1}{k_{0}-k}\rho_{1}\left(\mathcal{G}C^{'\#}\mathbb{Q}_{k}C^{\#}\mathcal{G}^{\prime}\right)\ge c_{2}$
w.p.a.1 for a constant $c_{2}>0$ as $N,T\to\infty$.

For $k_{0}-k$ being bounded, \begin{equation}
\rho_{1}\left(C^{\#}\mathcal{G}^{\prime}\mathcal{G}C^{'\#}\right)\ge\rho_{1}\left(C^{\#}g_{k_{0}}g_{k_{0}}^{\prime}C^{'\#}\right)=\rho_{1}\left(C^{\#}Bf_{k_{0}}f_{k_{0}}^{\prime}B^{\prime}C^{'\#}\right)=f_{k_{0}}^{\prime}B^{\prime}C^{'\#}C^{\#}Bf_{k_{0}}>c\label{eq:k0-k
bounded}\end{equation}
for a constant $c>0$ according to Assumption \ref{B_C_full_rank_project} (ii), where $C^{\#}Bf_{k_{0}}\ne0$.
 Combining \eqref{eq:GCQCG},
\eqref{eq:k0-k unbounded}, and \eqref{eq:k0-k bounded}, we have $\rho_{1}(\mathcal{G}C^{'\#}\mathbb{Q}_{k}C^{\#}\mathcal{G}^{\prime})/(k_{0}-k)>c_{2}$
w.p.a.1 for a constant $c_{2}>0$. Thus, we can obtain the lower
bound for the RHS of \eqref{eq:sqrt_rhomax} as \begin{eqnarray}
\frac{1}{\sqrt{k_{0}-k}}\sigma_{1}(\mathbb{Q}_{k}^{1/2}C^{\#}H^{\#'}\mathcal{\hat{G}}^{\prime}) & \ge & |H|c_{2}^{1/2}-O_{p}(N^{-1/2})\nonumber \\
\frac{1}{k_{0}-k}\rho_{1}(\mathcal{\hat{G}}H^{\#}C^{\#'}\mathbb{Q}_{k}C^{\#}H^{\#'}\mathcal{\hat{G}}^{\prime}) & \ge & c_{3}\ \mathrm{w.p.a.}1\label{eq:P2 bound}\end{eqnarray}
 as $N,T\to\infty$ for a constant $c_{3}>0$, because $H$ has
a nonsingular limit.

Based on \eqref{eq:GD^(-1)G} and Weyl's inequality, we have \begin{eqnarray}
\frac{1}{k_{0}-k}\rho_{1}\left(\frac{1}{T-k}\mathcal{\hat{G}}D_{k}^{-1}\mathcal{\hat{G}}^{\prime}\right) & \ge &
\frac{1}{|D_{k}|(k_{0}-k)}\rho_{r}(\mathbb{P}_{1})+\frac{1}{|D_{k}|(k_{0}-k)}\rho_{1}(\mathbb{P}_{2})\nonumber \\
 & = & \frac{1}{|D_{k}|(k_{0}-k)}\rho_{r}(\mathbb{P}_{1})+\frac{1}{|D_{k}|(T-k)}\frac{\rho_{1}(\mathcal{\hat{G}}H^{\#}C^{\#'}\mathbb{Q}_{k}C^{\#}H^{\#'}\mathcal{\hat{G}}^{\prime})}{k_{0}-k}\nonumber \\
 & \ge & O_{p}(T^{-1})+\underbrace{\frac{c_{4}N}{T-k}}_{\mathrm{dominating\ term}}\ \ \mathrm{w.p.a.}1\label{eq:case_1_rank=r-1}\end{eqnarray}
 for a constant $c_{4}>0$ as $N,T\to\infty$ by \eqref{eq:rhomin P1}
and \eqref{eq:P2 bound}, where the last line follows from proposition \ref{low_bound}
and the $O_{p}(T^{-1})$ term is uniform over $\tau_{1}T\le k<k_{0}$.
\vspace{1em}

\textbf{Step 2}. When $r-r_{2}\ge2$ or $r_{2}=0$,
the term $\mathbb{P}_{2}$ in \eqref{eq:GD^(-1)G} is zero. For the
term $\mathbb{P}_{1}$ in \eqref{eq:GD^(-1)G}, we have\[
\mathcal{\hat{G}}D_{k}^{\#}\mathcal{\hat{G}}^{\prime}=[\mathcal{G}H+(\mathcal{\hat{G}}-\mathcal{G}H)]D_{k}^{\#}[H^{\prime}\mathcal{G}^{\prime}+(\mathcal{\hat{G}}^{\prime}-H^{\prime}\mathcal{G}^{\prime})].\]
 Using similar techniques to those in \eqref{eq:sigma_max inequality} and \eqref{eq:sqrt_rhomax},
we have \begin{eqnarray}
\frac{1}{\sqrt{k_{0}-k}}\sigma_{1}(D_{k}^{\#^{1/2}}\mathcal{\hat{G}}^{\prime}) & \ge &
\frac{1}{\sqrt{k_{0}-k}}\sigma_{1}(D_{k}^{\#^{1/2}}H^{\prime}\mathcal{G}^{\prime})-\frac{1}{\sqrt{k_{0}-k}}\sigma_{1}(D_{k}^{\#^{1/2}}(\mathcal{\hat{G}}^{\prime}-H^{\prime}\mathcal{G}^{\prime}))\nonumber
\\
 & \ge &
 \frac{1}{\sqrt{k_{0}-k}}\sigma_{1}(D_{k}^{\#^{1/2}}H^{\prime}\mathcal{G}^{\prime})-\frac{\|D_{k}^{\#^{1/2}}(\mathcal{\hat{G}}^{\prime}-H^{\prime}\mathcal{G}^{\prime})\|}{\sqrt{k_{0}-k}}\nonumber \\
 & = &  \frac{1}{\sqrt{k_{0}-k}}\sigma_{1}(D_{k}^{\#^{1/2}}H^{\prime}\mathcal{G}^{\prime})-O_{p}(N^{-(r-r_{2})/2}),\label{eq:sigma DGhat}\end{eqnarray}
 where the last line is based on the fact that\begin{eqnarray*}
\frac{1}{k_{0}-k}\|D_{k}^{\#^{1/2}}(\mathcal{\hat{G}}^{\prime}-H^{\prime}\mathcal{G}^{\prime})\|^{2} & = &
\frac{1}{k_{0}-k}tr[(\mathcal{\hat{G}}-\mathcal{G}H)D_{k}^{\#}(\mathcal{\hat{G}}^{\prime}-H^{\prime}\mathcal{G}^{\prime})]\\
 & \le & \rho_{1}(D_{k}^{\#})tr\left[\frac{(\mathcal{\hat{G}}^{\prime}-H^{\prime}\mathcal{G}^{\prime})(\mathcal{\hat{G}}-\mathcal{G}H)}{k_{0}-k}\right]\\
 & = & O_{p}(N^{-(r-r_{2})})\end{eqnarray*}
uniformly over $\tau_{1}T\le k<k_{0}$ under the condition $N\propto T$,
because of Lemma \ref{Bai_implied}, and the fact
that \begin{equation}
\rho_{1}(D_{k}^{\#})=|D_{k}|/\rho_{r}(D_{k})=O_{p}(N^{-(r-r_{2})})O_{p}(N)\label{eq:rhomax_Dkadj}\end{equation}
by proposition \ref{low_bound}.

Now, it suffices to determine the lower bound of $\rho_{1}(\mathcal{G}HD_{k}^{\#}H^{\prime}\mathcal{G}^{\prime})$.
Similar to \eqref{eq:Bk^(1/2) U}, we have \begin{equation}
D_{k}^{1/2}U_{1}-H^{\prime}C\Sigma_{f}^{1/2}=O_{p}(\delta_{NT}^{-1}),\label{eq:Dk^(1/2)U1}\end{equation}
uniformly over $\tau_{1}T\le k<k_{0}$, where $U_{1}=D_{k}^{1/2}(\Sigma_{f}^{1/2}C^{\prime}H)^{-}$.
For $t\le k_{0}$, we have $g_{t}=Bf_{t}$;
thus, \begin{eqnarray}
H^{\prime}g_{t} & = & H^{\prime}Bf_{t}=D_{k}^{1/2}U_{1}\varepsilon_{t}+(H^{\prime}Bf_{t}-D_{k}^{1/2}U_{1}\varepsilon_{t})\nonumber \\
\mathrm{and}\ H^{\prime}\mathcal{G}^{\prime} & = & D_{k}^{1/2}U_{1}\mathcal{E}^{\prime}+(H^{\prime}\mathcal{G}^{\prime}-D_{k}^{1/2}U_{1}\mathcal{E}^{\prime}),\label{eq:HG decompose}\end{eqnarray}
 where we set $f_{t}=\Sigma_{f}^{1/2}\varepsilon_{t}$ with $E(\varepsilon_{t}\varepsilon_{t}^{\prime})=I_{r}$
and $\mathcal{E}\equiv[\varepsilon_{k+1},...,\varepsilon_{k_{0}}]^{\prime}$.

First, we consider the case in which $k_{0}-k$ is bounded.
Note that $D_{k}^{\#}$ has $r_{2}$ eigenvalues of order $O_{p}(N^{-(r-r_{2})})$
and $r-r_{2}$ eigenvalues of order $O_{p}(N^{-(r-r_{2}-1)})$ by
\eqref{eq:rhomax_Dkadj}, and let $v_{1}(r\times r_{2})$ and $v_{2}(r\times(r-r_{2}))$
denote the corresponding eigenvectors. By proposition \ref{low_bound},
for $t\le k_{0}$, \begin{equation}
\varepsilon_{t}^{\prime}U_{1}D_{k}^{1/2}D_{k}^{\#}D_{k}^{1/2}U_{1}\varepsilon_{t}=|D_{k}|(\varepsilon_{t}^{\prime}U_{1}U_{1}\varepsilon_{t})=O_{p}(N^{-(r-r_{2})}),\label{eq:eUDDDUe}\end{equation}
 which implies that $D_{k}^{1/2}U_{1}\varepsilon_{t}$ lies in the space
spanned by $v_{1}$. Thus, for $t\le k_{0}$, \begin{align}
f_{t}C^{\prime}HD_{k}^{\#}H^{\prime}Cf_{t} & =\|D_{k}^{\#^{1/2}}[(H^{\prime}C\Sigma_{f}^{1/2}\varepsilon_{t}-D_{k}^{1/2}U_{1}\varepsilon_{t})+D_{k}^{1/2}U_{1}\varepsilon_{t}]\|^{2}\nonumber \\
 & \le2\|D_{k}^{\#^{1/2}}(H^{\prime}C\Sigma_{f}^{1/2}-D_{k}^{1/2}U_{1})\varepsilon_{t}\|^{2}+O_{p}(N^{-(r-r_{2})})\nonumber \\
 & \le2\rho_{1}(D_{k}^{\#})\|(H^{\prime}C\Sigma_{f}^{1/2}-D_{k}^{1/2}U_{1})\varepsilon_{t}\|^{2}+O_{p}(N^{-(r-r_{2})})=O_{p}(N^{-(r-r_{2})}),\label{eq:fCHDHCf}\end{align}
where the second line follows from \eqref{eq:eUDDDUe} and the last
line follows from \eqref{eq:rhomax_Dkadj} and \eqref{eq:Dk^(1/2)U1}
under the condition $N\propto T$.

To bound $\rho_{1}(\mathcal{G}HD_{k}^{\#}H^{\prime}\mathcal{G}^{\prime})$,
we consider \begin{align}
D_{k}^{\#^{1/2}}H^{\prime}g_{k_{0}} &
=D_{k}^{\#^{1/2}}[H^{\prime}Bf_{k_{0}}-\mathrm{Proj}(H^{\prime}Bf_{k_{0}}|H^{\prime}C\Sigma_{f}^{1/2})]+D_{k}^{\#^{1/2}}\mathrm{Proj}(H^{\prime}Bf_{k_{0}}|H^{\prime}C\Sigma_{f}^{1/2})\nonumber \\
 &
 =D_{k}^{\#^{1/2}}[H^{\prime}Bf_{k_{0}}-\mathrm{Proj}(H^{\prime}Bf_{k_{0}}|D_{k}^{1/2}U_{1})+O_{p}(\delta_{NT}^{-1})]+D_{k}^{\#^{1/2}}\mathrm{Proj}(H^{\prime}Bf_{k_{0}}|H^{\prime}C\Sigma_{f}^{1/2})\nonumber
 \\
 & =D_{k}^{\#^{1/2}}[H^{\prime}Bf_{k_{0}}-\mathrm{Proj}(H^{\prime}Bf_{k_{0}}|D_{k}^{1/2}U_{1})]+O_{p}(N^{-(r-r_{2})/2}),\label{eq:DHgk0}\end{align}
where $\mathrm{Proj}(\mathbb{A}|\mathbb{Z})$ denotes the projection
of $\mathbb{A}$ onto the columns of $\mathbb{Z}$, the second line
follows from \eqref{eq:Dk^(1/2)U1}, and the $O_{p}(N^{-(r-r_{2})/2})$
term in the third line follows from \eqref{eq:rhomax_Dkadj} and the fact
that $D_{k}^{\#^{1/2}}\mathrm{Proj}(H^{\prime}Bf_{k_{0}}|H^{\prime}C\Sigma_{f}^{1/2})=O_{p}(N^{-(r-r_{2})/2})$
by \eqref{eq:fCHDHCf}, because $\mathrm{Proj}(H^{\prime}Bf_{k_{0}}|H^{\prime}C\Sigma_{f}^{1/2})$
is a linear combination of $H^{\prime}C\Sigma_{f}^{1/2}$ columns.
Under Assumption \ref{B_C_full_rank_project} (iii), according to which $\|Bf_{k_{0}}-\mathrm{Proj}(Bf_{k_{0}}|C)\|\ge d>0$,
we have $H^{\prime}Bf_{k_{0}}-\mathrm{Proj}(H^{\prime}Bf_{k_{0}}|H^{\prime}C\Sigma_{f}^{1/2})$
bounded away from zero. This implies that the term $H^{\prime}Bf_{k_{0}}-\mathrm{Proj}(H^{\prime}Bf_{k_{0}}|D_{k}^{1/2}U_{1})$
in the last line of \eqref{eq:DHgk0} is also bounded away from zero
and lies in the space spanned by $v_{2}$, because it is, by design,
orthogonal to $D_{k}^{1/2}U_{1}$, which lies in the space of $v_{1}$
by \eqref{eq:eUDDDUe}. As $v_{2}$ corresponds to the $O_{p}(N^{-(r-r_{2}-1)})$
eigenvalues of $D_{k}^{\#}$, we have \[
\rho_{1}(\mathcal{G}HD_{k}^{\#}H^{\prime}\mathcal{G}^{\prime})\ge
g_{k_{0}}^{\prime}HD_{k}^{\#}H^{\prime}g_{k_{0}}\ge\rho_{r-r_{2}}(D_{k}^{\#})\|H^{\prime}g_{k_{0}}\|^{2}=\frac{|D_{k}|}{\rho_{r_{2}+1}(D_{k})}\|H^{\prime}g_{k_{0}}\|^{2}\ge\frac{N}{c_{U}}\|H^{\prime}g_{k_{0}}\|^{2}|D_{k}|\]
w.p.a.1 as $N,T\to\infty$ by proposition \ref{low_bound}. Thus, for $k_{0}-k$
being bounded, \begin{equation}
\rho_{1}(\mathcal{G}HD_{k}^{\#}H^{\prime}\mathcal{G}^{\prime})\ge g_{k_{0}}^{\prime}HD_{k}^{\#}H^{\prime}g_{k_{0}}\ge c_{1}N\cdot|D_{k}|\label{eq:step2 k0-k bounded}\end{equation}
w.p.a.1. for a constant $c_{1}>0$ as $N,T\to\infty$ under the
condition $N\propto T$.

Second, we consider the case in which $k_{0}-k\to\infty$ as $N,T\to\infty$.
Using \eqref{eq:HG decompose}, we rewrite $\mathcal{G}HD_{k}^{\#}H^{\prime}\mathcal{G}^{\prime}$
as \begin{eqnarray*}
\mathcal{G}HD_{k}^{\#}H^{\prime}\mathcal{G}^{\prime} & = &
[(\mathcal{G}H-\mathcal{E}U_{1}^{\prime}D_{k}^{1/2})+\mathcal{E}U_{1}^{\prime}D_{k}^{1/2}]D_{k}^{\#}[D_{k}^{1/2}U_{1}\mathcal{E}^{\prime}+(H^{\prime}\mathcal{G}^{\prime}-D_{k}^{1/2}U_{1}\mathcal{E}^{\prime})].\end{eqnarray*}
 Based on the same techniques as in \eqref{eq:sigma_max inequality} and \eqref{eq:sqrt_rhomax},
the decomposition in \eqref{eq:HG decompose} implies\[
\sigma_{1}(D_{k}^{\#^{1/2}}(H^{\prime}\mathcal{G}^{\prime}-D_{k}^{1/2}U_{1}\mathcal{E}^{\prime}))\le\sigma_{1}(D_{k}^{\#^{1/2}}H^{\prime}\mathcal{G}^{\prime})+\sigma_{1}(-D_{k}^{\#^{1/2}}D_{k}^{1/2}U_{1}\mathcal{E}^{\prime}),\]
 so we have \begin{eqnarray}
\frac{1}{\sqrt{k_{0}-k}}\sigma_{1}(D_{k}^{\#^{1/2}}H^{\prime}\mathcal{G}^{\prime}) & \ge &
\frac{1}{\sqrt{k_{0}-k}}\sigma_{1}(D_{k}^{\#^{1/2}}(H^{\prime}\mathcal{G}^{\prime}-D_{k}^{1/2}U_{1}\mathcal{E}^{\prime}))-\frac{1}{\sqrt{k_{0}-k}}\sigma_{1}(D_{k}^{\#^{1/2}}D_{k}^{1/2}U_{1}\mathcal{E}^{\prime})\nonumber
\\
 & \ge &
 \frac{1}{\sqrt{k_{0}-k}}\sigma_{1}(D_{k}^{\#^{1/2}}(H^{\prime}\mathcal{G}^{\prime}-D_{k}^{1/2}U_{1}\mathcal{E}^{\prime}))-|D_{k}^{1/2}|\sigma_{1}(U_{1})\sigma_{1}\left(\frac{\mathcal{E}^{\prime}}{\sqrt{k_{0}-k}}\right)\nonumber
 \\
 & = & \frac{1}{\sqrt{k_{0}-k}}\sigma_{1}(D_{k}^{\#^{1/2}}(H^{\prime}\mathcal{G}^{\prime}-D_{k}^{1/2}U_{1}\mathcal{E}^{\prime}))-O_{p}(N^{-(r-r_{2})/2})O_{p}(1)\label{eq:sigma_DHG}\end{eqnarray}
 where the second inequality is based on the fact that $\sigma_{1}(\mathbb{AB})\le\sigma_{1}(\mathbb{A})\sigma_{1}(\mathbb{B})$
and the last line follows from the facts that $|D_{k}|$ is uniformly $O_{p}(N^{-(r-r_{2})})$
by proposition \ref{low_bound}, $\sigma_{1}(U_{1})\le\|U_{1}\|=O_{p}(1)$
by the structure of $U_{1}$, and $\frac{1}{\sqrt{k_{0}-k}}\sigma_{1}(\mathcal{E}^{\prime})\le\sqrt{\|\mathcal{E}\|^{2}/(k_{0}-k)}=O_{p}(1)$
uniformly over $\tau_{1}T\le k<k_{0}$.

Next, we need to determine the lower bound of $\rho_{1}((\mathcal{G}H-\mathcal{E}U_{1}D_{k}^{1/2})D_{k}^{\#}(H^{\prime}\mathcal{G}^{\prime}-D_{k}^{1/2}U_{1}\mathcal{E}^{\prime}))/(k_{0}-k)$.
From \eqref{eq:Dk^(1/2)U1}, we have \begin{align}
H^{\prime}\mathcal{G}^{\prime}-D_{k}^{1/2}U_{1}\mathcal{E}^{\prime} &
=H^{\prime}\mathcal{G}^{\prime}-H^{\prime}C\Sigma_{f}^{1/2}\mathcal{E}^{\prime}+(H^{\prime}C\Sigma_{f}^{1/2}\mathcal{E}^{\prime}-D_{k}^{1/2}U_{1}\mathcal{E}^{\prime})\nonumber \\
 & =H^{\prime}\mathcal{G}^{\prime}-H^{\prime}C\mathcal{F}^{\prime}+O_{p}(\delta_{NT}^{-1})\mathcal{E}^{\prime}\nonumber \\
 & =H^{\prime}(B-C)\mathcal{F}^{\prime}+O_{p}(\delta_{NT}^{-1})\mathcal{E}^{\prime}\label{eq:HG - DUE}\end{align}
where $\mathcal{F}^{\prime}\equiv[f_{k+1},...,f_{k_{0}}]=\Sigma_{f}^{1/2}\mathcal{E}^{\prime}$
in the second line. Again, using the inequality in \eqref{eq:sigma_max inequality},
we have\[
\sigma_{1}(D_{k}^{\#^{1/2}}(H^{\prime}\mathcal{G}^{\prime}-H^{\prime}C\Sigma_{f}^{1/2}\mathcal{E}^{\prime}))\le\sigma_{1}(D_{k}^{\#^{1/2}}(H^{\prime}\mathcal{G}^{\prime}-D_{k}^{1/2}U_{1}\mathcal{E}^{\prime}))+\sigma_{1}(-D_{k}^{\#^{1/2}}(H^{\prime}C\Sigma_{f}^{1/2}\mathcal{E}^{\prime}-D_{k}^{1/2}U_{1}\mathcal{E}^{\prime})).\]
Thus, in combination with \eqref{eq:HG - DUE}, we obtain \begin{align}
\sigma_{1}(D_{k}^{\#^{1/2}}(H^{\prime}\mathcal{G}^{\prime}-D_{k}^{1/2}U_{1}\mathcal{E}^{\prime})) &
\ge\sigma_{1}(D_{k}^{\#^{1/2}}(H^{\prime}\mathcal{G}^{\prime}-H^{\prime}C\Sigma_{f}^{1/2}\mathcal{E}^{\prime}))-\sigma_{1}(D_{k}^{\#^{1/2}}(H^{\prime}C\Sigma_{f}^{1/2}\mathcal{E}^{\prime}-D_{k}^{1/2}U_{1}\mathcal{E}^{\prime}))\nonumber
\\
\frac{1}{\sqrt{k_{0}-k}}\sigma_{1}(D_{k}^{\#^{1/2}}(H^{\prime}\mathcal{G}^{\prime}-D_{k}^{1/2}U_{1}\mathcal{E}^{\prime})) &
\ge\frac{1}{\sqrt{k_{0}-k}}\sigma_{1}(D_{k}^{\#^{1/2}}(H^{\prime}(B-C)\mathcal{F}^{\prime})-\frac{1}{\sqrt{k_{0}-k}}\sigma_{1}(D_{k}^{\#^{1/2}}O_{p}(\delta_{NT}^{-1})\mathcal{E}^{\prime})\nonumber \\
 & \ge\sqrt{\frac{1}{k_{0}-k}\rho_{1}(D_{k}^{\#}(H^{\prime}(B-C)\mathcal{F}^{\prime}\mathcal{F}(B-C)^{\prime}H)}+O_{p}(N^{-(r-r_{2})/2})\nonumber \\
 & \ge\left[\rho_{1}(D_{k}^{\#})\rho_{r}\left(H^{\prime}(B-C)\frac{1}{k_{0}-k}\sum_{t=k+1}^{k_{0}}f_{t}f_{t}^{\prime}(B-C)^{\prime}H\right)\right]^{1/2}+O_{p}(N^{-(r-r_{2})/2})\nonumber \\
 & =\underbrace{\left(\rho_{1}(D_{k}^{\#})[\rho_{r}(H^{\prime}(B-C)\Sigma_{f}(B-C)^{\prime}H)+o_{p}(1)]\right)^{1/2}}_{leading\ term}+O_{p}(N^{-(r-r_{2})/2})\nonumber \\
 &\ge\rho_{1}(D_{k}^{\#})^{1/2}c_{2}\label{eq:DHG-DUE}\end{align}
w.p.a.1 for a constant $c_{2}>0$ as $N,T\to\infty$,
where the $O_{p}(N^{-(r-r_{2})/2})$ term in the third line follows
from \eqref{eq:rhomax_Dkadj} and the condition $N\propto T$, the
last line follows from the fact that $[\rho_{r}(H^{\prime}(B-C)\Sigma_{f}(B-C)^{\prime}H)]^{1/2}\ge c_{2}>0$
because $B-C\ne0$ and $\Sigma_{f}$ are positive definite by Assumption
\ref{factors}, and the leading term in the last line is $O_{p}(N^{-(r-r_{2}-1)/2})$
by \eqref{eq:rhomax_Dkadj}. Hence, combining \eqref{eq:sigma_DHG}
and \eqref{eq:DHG-DUE} yields\begin{align}
\frac{1}{\sqrt{k_{0}-k}}\sigma_{1}(D_{k}^{\#^{1/2}}H^{\prime}\mathcal{G}^{\prime}) & \ge\rho_{1}(D_{k}^{\#})^{1/2}c_{2}\ \mathrm{and}\nonumber \\
\frac{1}{k_{0}-k}\rho_{1}(\mathcal{G}HD_{k}^{\#}H^{\prime}\mathcal{G}^{\prime}) & \ge\frac{|D_{k}|}{\rho_{r}(D_{k})}c_{2}^{2}\ge\frac{c_{2}^{2}}{c_{U}}N|D_{k}|=c_{3}N\cdot|D_{k}|\label{eq:step2 k0-k
unbounded}\end{align}
w.p.a.1 for a constant $c_{3}>0$ as $N,T\to\infty$.

According to \eqref{eq:sigma DGhat}, \eqref{eq:step2 k0-k bounded}, and
\eqref{eq:step2 k0-k unbounded}, there exists a constant
$c_{4}>0$ such that \begin{eqnarray*}
\frac{1}{\sqrt{k_{0}-k}}\sigma_{1}(D_{k}^{\#^{1/2}}\mathcal{\hat{G}}^{\prime}) & \ge & \underbrace{\sqrt{c_{4}N\cdot|D_{k}|}}_{leading\ term}-O_{p}(N^{-(r-r_{2})/2}),\end{eqnarray*}
w.p.a.1 as $N,T\to\infty$; thus, we have \[
\frac{1}{k_{0}-k}\rho_{1}(\mathcal{\hat{G}}D_{k}^{\#}\mathcal{\hat{G}}^{\prime})\ge c_{4}N\cdot|D_{k}|\ \ \mathrm{w.p.a.}1\]
 as $N,T\to\infty$. Hence, \begin{equation}
\frac{1}{k_{0}-k}\rho_{1}\left(\frac{1}{T-k}\mathcal{\hat{G}}D_{k}^{-1}\mathcal{\hat{G}}^{\prime}\right)=\frac{1}{|D_{k}|}\frac{1}{T-k}\frac{\rho_{1}\left(\mathcal{\hat{G}}D_{k}^{\#}\mathcal{\hat{G}}^{\prime}\right)}{k_{0}-k}\ge
c_{4}\frac{N}{T-k}\ \ \mathrm{w.p.a.}1\label{eq:case 2 rank < r-1}\end{equation}
 as $N,T\to\infty$. Summarizing the results in \eqref{eq:case_1_rank=r-1}
and \eqref{eq:case 2 rank < r-1}, we obtain the lower bound of $\rho_{1}\left(\frac{1}{T-k}\mathcal{\hat{G}}D_{k}^{-1}\mathcal{\hat{G}}^{\prime}\right)$.
Thus, steps 1 and 2 are completed. \vspace{1em}

Finally, using the lower bound of the largest eigenvalue of matrix
$\frac{1}{T-k}\mathcal{\hat{G}}D_{k}^{-1}\mathcal{\hat{G}}^{\prime}$,
we can rewrite \eqref{eq:det(Sig2k)} as\[
|\hat{\Sigma}_{2}(k)|=|D_{k}|\cdot|I_{k_{0}-k}+\frac{1}{T-k}\mathcal{\hat{G}}D_{k}^{-1}\mathcal{\hat{G}}^{\prime}|\ge|D_{k}|\left[1+c_{4}\left(\frac{k_{0}-k}{T-k}\right)\cdot
N\right]\ge|D_{k}|\left[1+\frac{c_{4}(k_{0}-k)}{(1-\tau_{1})T}\cdot N\right]\ \ \mathrm{w.p.a.}1\]
 as $N,T\to\infty.$

Comparing $|D_{k}|$ and $|\hat{\Sigma}_{2}^{0}|$, we have \begin{eqnarray*}
|D_{k}|-|\hat{\Sigma}_{2}^{0}| & = &
\left|\frac{T-k_{0}}{T-k}\left(\frac{1}{T-k_{0}}\sum_{t=k_{0}+1}^{T}\hat{g}_{t}\hat{g}_{t}^{\prime}\right)\right|-\left|\frac{1}{T-k_{0}}\sum_{t=k_{0}+1}^{T}\hat{g}_{t}\hat{g}_{t}^{\prime}\right|=\left[\left(1-\frac{k_{0}-k}{T-k}\right)^{r}-1\right]|\hat{\Sigma}_{2}^{0}|\\
 & = & \left[-\frac{r(k_{0}-k)}{T-k}+\frac{r(r-1)}{2}\left(\frac{k_{0}-k}{T-k}\right)^{2}+...\right]|\hat{\Sigma}_{2}^{0}|\\
 &=&-\frac{c_{5}(k_{0}-k)}{T-k}|\hat{\Sigma}_{2}^{0}|\end{eqnarray*}
for some positive constant $c_{5}>0$. In addition, \[
|D_{k}|/|\hat{\Sigma}_{2}^{0}|=\left(\frac{T-k_{0}}{T-k}\right)^{r}\ge\left(\frac{1-\tau_{0}}{1-\tau_{1}}\right)^{r}.\]
Thus, \begin{align*}
\frac{|\hat{\Sigma}_{2}(k)|-|D_{k}|+|D_{k}|-|\hat{\Sigma}_{2}^{0}|}{|\hat{\Sigma}_{2}^{0}|} & \ge\frac{|D_{k}|}{|\hat{\Sigma}_{2}^{0}|}\frac{c_{4}(k_{0}-k)}{(1-\tau_{1})T}\cdot
N-\frac{c_{5}(k_{0}-k)}{T-k}\ \ \mathrm{w.p.a.}1\\
 & \ge\underbrace{\left(\frac{1-\tau_{0}}{1-\tau_{1}}\right)^{r}\frac{c_{4}(k_{0}-k)N}{(1-\tau_{1})T}}_{leading\ term}-\frac{c_{5}(k_{0}-k)}{T-k}\ \ \mathrm{w.p.a.}1,\end{align*}
as $N,T\to\infty$, which implies the desired result under the condition
$N\propto T$.  $\Box$

\subsection*{Proof of Theorem \ref{consistency}}

We first prove the consistency of $\hat{\tau}$,
then $\hat{k}-k_{0}=O_{p}(1)$, and finally, $\hat{k}-k_{0}=o_{p}(1)$. Again, it suffices to study
the case of $k<k_{0}$.

To prove $\hat{\tau}-\tau_{0}=o_{p}(1)$, we need to show that
for any $\varepsilon>0$ and $\eta>0$, $P(|\hat{\tau}-\tau_{0}|>\eta)<\varepsilon$
as $N,T\rightarrow\infty$. For any given $0<\eta\le\min(\tau_{0},1-\tau_{0})$,
define $D_{\eta}=\{k:(\tau_{0}-\eta)T\leq k\leq(\tau_{0}+\eta)T\}$
and $D_{\eta}^{c}$ as the complement of $D_{\eta}$. Similar
to the proof for the consistency of $\hat{\tau}$ when $B$ and $C$
are nonsingular, we need to show that $P(\hat{k}\in D_{\eta}^{c})<\varepsilon$.
Recalling \eqref{eq:equiv sets} and \eqref{eq:P}, we have \begin{eqnarray}
 &  & P(\min\limits _{k\in D_{\eta}^{c},k<k_{0}}U(k)-U(k_{0})\leq0)=P(\min\limits _{k\in D_{\eta}^{c},k<k_{0}}\frac{U(k)-U(k_{0})}{k_{0}-k}\leq0),\nonumber \\
 &  &\le P(\min\limits _{k\in D_{\eta}^{c},k<k_{0}}\frac{k}{k_{0}-k}\log|\hat{\Sigma}_{1}\hat{\Sigma}_{1}^{0-1}|+\min\limits _{k\in
 D_{\eta}^{c},k<k_{0}}\frac{T-k}{k_{0}-k}\log|\hat{\Sigma}_{2}\hat{\Sigma}_{2}^{0-1}|-\log|\hat{\Sigma}_{1}^{0}\hat{\Sigma}_{2}^{0-1}|\leq0)\label{eq:sec3.3 eq1}\end{eqnarray}

(1). Consider the first term $\frac{k}{k_{0}-k}\log|\hat{\Sigma}_{1}\hat{\Sigma}_{1}^{0-1}|$.
When $\Sigma_{1}$ is of full rank, it follows that
\begin{equation}
\min\limits _{k\in D_{\eta}^{c},k<k_{0}}\frac{k}{k_{0}-k}\log|\hat{\Sigma}_{1}\hat{\Sigma}_{1}^{0-1}|=o_{p}(1)\label{eq:term 1 is o1}\end{equation}
by the argument used in \eqref{eq:P} and Lemma
\ref{Baltagi_lemma}. When $\Sigma_{1}$ is singular, we can obtain
\begin{eqnarray}
\big|\min\limits _{k\in D_{\eta}^{c},k<k_{0}}\frac{k}{k_{0}-k}\log|\hat{\Sigma}_{1}\hat{\Sigma}_{1}^{0-1}|\big| & = & \big|\min\limits _{k\in
D_{\eta}^{c},k<k_{0}}\frac{k}{k_{0}-k}\log(\frac{|\hat{\Sigma}_{1}(k)|-|\hat{\Sigma}_{1}(k_{0})|}{|\hat{\Sigma}_{1}(k_{0})|}+1)\big|\nonumber \\
 & = & \big|\min\limits _{k\in D_{\eta}^{c},k<k_{0}}\frac{k}{k_{0}-k}\log(O_{p}(T^{-1}(k_{0}-k))+1)\big|\nonumber \\
 & = & O_{p}(1),\label{eq:term 1 is O1}\end{eqnarray}
where the second line follows from Lemma \ref{differ}
and the third line is based on the fact that $|k_{0}-k|>\eta T$.

(2). For the second and third terms, let \begin{align}
f(\hat{\Sigma}_{1}^{0},\hat{\Sigma}_{2}^{0}) & =\min\limits _{k\in
D_{\eta}^{c},k<k_{0}}\frac{T-k}{k_{0}-k}\log|\hat{\Sigma}_{2}\hat{\Sigma}_{2}^{0-1}|-\log|\hat{\Sigma}_{1}^{0}\hat{\Sigma}_{2}^{0-1}|\nonumber \\
 & =\min\limits _{k\in
 D_{\eta}^{c},k<k_{0}}\frac{T-k}{k_{0}-k}\sum_{i=1}^{r}\log\rho_{i}(\hat{\Sigma}_{2}\hat{\Sigma}_{2}^{0-1})-\sum_{i=1}^{r}\log\rho_{i}(\hat{\Sigma}_{1}^{0}\hat{\Sigma}_{2}^{0-1}).\label{eq:f
 function}\end{align}
 We show that $f(\hat{\Sigma}_{1}^{0},\hat{\Sigma}_{2}^{0})\to+\infty$
at the rate $\log T$.

When $\Sigma_{1}$ is singular and $\Sigma_{2}$
is a positive definite matrix, \eqref{eq:Sigmahat rewrite} implies
\begin{equation}
\hat{\Sigma}_{2}=\frac{k_{0}-k}{T-k}\Sigma_{1}+\frac{T-k_{0}}{T-k}\Sigma_{2}+o_{p}(1),\label{eq:Sigmahat2}\end{equation}
 where the $o_{p}(1)$ term is uniform over $k\in D_{\eta}^{c}$ for
$k<k_{0}$ by Lemma \ref{Baltagi_lemma} $(iv)$ and
$(vii)$. Together with $\hat{\Sigma}_{2}^{0-1}\to_{p}\Sigma_{2}^{-1}>0$,
\eqref{eq:Sigmahat2} implies that $\rho_{i}(\hat{\Sigma}_{2}\hat{\Sigma}_{2}^{0-1})$
is uniformly $O_{p}(1)$ and bounded away from zero; thus,
\[
\big|\frac{T-k}{k_{0}-k}\sum_{i=1}^{r}\log\rho_{i}(\hat{\Sigma}_{2}\hat{\Sigma}_{2}^{0-1})\big|=O_{p}(1)\]
 uniformly over $k\in D_{\eta}^{c}$ for $k<k_{0}$. In addition, we have
$\rho_{i}(\hat{\Sigma}_{1}^{0}\hat{\Sigma}_{2}^{0-1})=O_{p}(T^{-1})$
uniformly over $k$ for $i=r_{1}+1,...,r$ by proposition \ref{low_bound}
when $N\propto T$; thus, $\log\rho_{i}(\hat{\Sigma}_{1}^{0}\hat{\Sigma}_{2}^{0-1})\to-\infty$
at the rate of $\log T$ for $i=r_{1}+1,...,r$. Therefore, \eqref{eq:f function}
can be rewritten as\[
f(\hat{\Sigma}_{1}^{0},\hat{\Sigma}_{2}^{0})=\underbrace{-\sum_{i=1}^{r}\log\rho_{i}(\hat{\Sigma}_{1}^{0}\hat{\Sigma}_{2}^{0-1})}_{\to+\infty\ \mathrm{at\ the\ rate}\ \log T}+O_{p}(1).\]

When $\Sigma_{2}$ is singular and $\Sigma_{1}$
is singular or nonsingular, we can bound \eqref{eq:f function}
by\begin{align*}
f(\hat{\Sigma}_{1}^{0},\hat{\Sigma}_{2}^{0}) & \ge\min\limits _{k\in
D_{\eta}^{c},k<k_{0}}\frac{T-k}{k_{0}-k}\sum_{i=1}^{r}\log\rho_{r}(\hat{\Sigma}_{2})\rho_{i}(\hat{\Sigma}_{2}^{0-1})-\sum_{i=1}^{r}\log\rho_{1}(\hat{\Sigma}_{1}^{0})\rho_{i}(\hat{\Sigma}_{2}^{0-1})\\
 & =\min\limits _{k\in
 D_{\eta}^{c},k<k_{0}}\frac{T-k}{k_{0}-k}r\log\rho_{r}(\hat{\Sigma}_{2})+\underbrace{\frac{T-k_{0}}{k_{0}-k}\sum_{i=1}^{r}\log\rho_{i}(\hat{\Sigma}_{2}^{0-1})-r\log\rho_{1}(\hat{\Sigma}_{1}^{0})}_{\to+\infty\
 \mathrm{at\ the\ rate}\ \log T},\end{align*}
 where the first line uses inequalities $\rho_{i}(\hat{\Sigma}_{2}\hat{\Sigma}_{2}^{0-1})\ge\rho_{r}(\hat{\Sigma}_{2})\rho_{i}(\hat{\Sigma}_{2}^{0-1})$
and $\rho_{i}(\hat{\Sigma}_{1}^{0}\hat{\Sigma}_{2}^{0-1})\le\rho_{1}(\hat{\Sigma}_{1}^{0})\rho_{i}(\hat{\Sigma}_{2}^{0-1})$.
Note that $\rho_{i}(\hat{\Sigma}_{2}^{0-1})>0$ diverges at the rate
$T$ by proposition \ref{low_bound} for $i=1,...,r-r_{2}$; thus, $\sum_{i=1}^{r}\log\rho_{i}(\hat{\Sigma}_{2}^{0-1})\to+\infty$
at the rate $\log T$. In addition, when $\rho_{1}(\Sigma_{1})>0$, we have
$\rho_{1}(\hat{\Sigma}_{1}^{0})\to_{p}\rho_{1}(\Sigma_{1})>0$; thus,
$r\log\rho_{1}(\hat{\Sigma}_{1}^{0})=O_{p}(1)$. When $\rho_{1}(\Sigma_{1})=0$
(i.e., $r_{1}=0$), we have $\rho_{1}(\hat{\Sigma}_{1}^{0})\cdot T\ge c>0$
w.p.a.1 for $c>0$ by proposition \ref{low_bound}; thus, $-r\log\rho_{1}(\hat{\Sigma}_{1}^{0})\to+\infty$
at the rate $\log T$. For $\rho_{r}(\hat{\Sigma}_{2})$, rearranging
the terms in \eqref{eq:Sigmahat2} yields \begin{align*}
\hat{\Sigma}_{2} & =\frac{(k_{0}-k)T}{k_{0}(T-k)}(\frac{k_{0}}{T}\Sigma_{1}+\frac{T-k_{0}}{T}\Sigma_{2})+\frac{k}{k_{0}}\frac{T-k_{0}}{T-k}\Sigma_{2}+o_{p}(1)\\
 & =\frac{(k_{0}-k)T}{k_{0}(T-k)}[\tau_{0}\Sigma_{1}+(1-\tau_{0})\Sigma_{2}]+\frac{k}{k_{0}}\frac{T-k_{0}}{T-k}\Sigma_{2}+o_{p}(1),\end{align*}
 where $\tau_{0}\Sigma_{1}+(1-\tau_{0})\Sigma_{2}$ is a positive
definite matrix under Assumption \ref{B_C_full_rank_project} (i). Thus,
$\rho_{r}(\hat{\Sigma}_{2})$ is $O_{p}(1)$ and bounded away from
zero w.p.a.1, and \[
\big|\frac{T-k}{k_{0}-k}r\log\rho_{r}(\hat{\Sigma}_{2})\big|=O_{p}(1)\]
 uniformly over $k\in D_{\eta}^{c}$ for $k<k_{0}$. Combining the
above results, we establish the following result: $f(\hat{\Sigma}_{1}^{0},\hat{\Sigma}_{2}^{0})\to+\infty$
at the rate $\log T$. Together with \eqref{eq:term 1 is o1} and
\eqref{eq:term 1 is O1}, we have \[
\min\limits _{k\in
D_{\eta}^{c},k<k_{0}}\frac{k}{k_{0}-k}\log|\hat{\Sigma}_{1}\hat{\Sigma}_{1}^{0-1}|+\frac{T-k}{k_{0}-k}\log|\hat{\Sigma}_{2}\hat{\Sigma}_{2}^{0-1}|-\log|\hat{\Sigma}_{1}^{0}\hat{\Sigma}_{2}^{0-1}|>0,\]
 w.p.a.1; thus, $P(\min\limits _{k\in D_{\eta}^{c},k<k_{0}}U(k)-U(k_{0})\leq0)\to0$
for any $\eta>0$, and hence, $\hat{\tau}\to_{p}\tau$.

Next, we show that $\hat{k}-k_{0}=O_{p}(1)$.

Similar to the proof of Theorem \ref{bound_theorem}, for given $\eta$
and $M$, define $D_{\eta,M}=\{k:(\tau_{0}-\eta)T\leq k\leq(\tau_{0}+\eta)T,\ |k_{0}-k|>M\}$,
such that $P(|\hat{k}-k_{0}|>M)=P(\hat{k}\in D_{\eta}^{c})+P(\hat{k}\in D_{\eta,M})$.
Hence, it suffices to show that for any $\varepsilon>0$ and $\eta>0$,
there exists an $M>0$ such that $P(\hat{k}\in D_{\eta,M})<\varepsilon$
as $(N,T)\rightarrow\infty$. Similar to \eqref{eq:p(Uk - Uk0 <0)}
and \eqref{eq:sec3.3 eq1}, it suffices to show that
for any given $\varepsilon>0$ and $\eta>0$, there exists
an $M>0$ such that\begin{align}
 & P(\min\limits _{k\in
 D_{\eta,M},k<k_{0}}\frac{k}{k_{0}-k}\log|\hat{\Sigma}_{1}\hat{\Sigma}_{1}^{0-1}|+\frac{T-k}{k_{0}-k}\log|\hat{\Sigma}_{2}\hat{\Sigma}_{2}^{0-1}|-\log|\hat{\Sigma}_{1}^{0}\hat{\Sigma}_{2}^{0-1}|\leq0)\nonumber
 \\
\le & P(\min\limits _{k\in D_{\eta,M},k<k_{0}}\frac{k}{k_{0}-k}\log|\hat{\Sigma}_{1}\hat{\Sigma}_{1}^{0-1}|+\min\limits _{k\in
D_{\eta,M},k<k_{0}}\frac{T-k}{k_{0}-k}\log|\hat{\Sigma}_{2}\hat{\Sigma}_{2}^{0-1}|-\log|\hat{\Sigma}_{1}^{0}\hat{\Sigma}_{2}^{0-1}|\le0)<\varepsilon\label{eq:Op(1) TBP}\end{align}

For the term $\frac{k}{k_{0}-k}\log|\hat{\Sigma}_{1}\hat{\Sigma}_{1}^{0-1}|$,
when $\Sigma_{1}$ is of full rank, we have \[
P(\Big|\min\limits _{k\in D_{\eta,M},k<k_{0}}\frac{k}{k_{0}-k}\log|\hat{\Sigma}_{1}\hat{\Sigma}_{1}^{0-1}|\Big|\le c_{\Delta})\ge P(\max\limits _{k\in
D_{\eta,M},k<k_{0}}\Big|\frac{k}{k_{0}-k}\log|\hat{\Sigma}_{1}\hat{\Sigma}_{1}^{0-1}|\Big|\le c_{\Delta})\ge1-\frac{C}{Mc_{\Delta}^{2}}\to1\]
 for a constant $C>0$ under the same arguments as those in \eqref{eq:Hajek2}
and \eqref{eq:bounded2}.

When $\Sigma_{1}$ is singular, \begin{eqnarray}\label{Sigma1_singular}
\min\limits _{k\in D_{\eta,M},k<k_{0}}\frac{k}{k_{0}-k}\log|\hat{\Sigma}_{1}\hat{\Sigma}_{1}^{0-1}| & = & \min\limits _{k\in
D_{\eta,M},k<k_{0}}\frac{k}{k_{0}-k}\log(\frac{|\hat{\Sigma}_{1}(k)|-|\hat{\Sigma}_{1}(k_{0})|}{|\hat{\Sigma}_{1}(k_{0})|}+1)\nonumber \\
 & = & \min\limits _{k\in D_{\eta,M},k<k_{0}}\frac{k}{k_{0}-k}\log(O_{p}(T^{-1}(k_{0}-k))+1)\nonumber \\
 & = & O_{p}(1),\end{eqnarray}
 where the second equation holds because of Lemma \ref{differ} and
the last equality holds because $\frac{k}{k_{0}-k}\log(O_{p}(T^{-1}(k_{0}-k))+1)=\frac{k}{k_{0}-k}tr(O_{p}(T^{-1}(k_{0}-k)))=O_{p}(1)$
whether $k_{0}-k$ is bounded or diverging.

For the second and third terms, we consider several cases.
\begin{itemize}
\item [(i).] When $\Sigma_{1}$ is singular and $\Sigma_{2}$
is positive definite, we have \begin{eqnarray}\label{Sigma2differ}
 &  & \min\limits _{k\in D_{\eta,M},k<k_{0}}\frac{T-k}{k_{0}-k}\log|\hat{\Sigma}_{2}\hat{\Sigma}_{2}^{0-1}|-\log|\hat{\Sigma}_{1}^{0}\hat{\Sigma}_{2}^{0-1}|\nonumber\\
 & = & \min\limits _{k\in
 D_{\eta,M},k<k_{0}}\frac{T-k}{k_{0}-k}\log|I+\frac{k-k_{0}}{T-k}I+\frac{k_{0}-k}{T-k}\hat{\Sigma}_{1}^{0}\hat{\Sigma}_{2}^{0-1}+\frac{k_{0}-k}{T-k}\Big(\frac{1}{k_{0}-k}\sum\limits
 _{t=k+1}^{k_{0}}\hat{g}_{t}\hat{g}_{t}^{'}-\hat{\Sigma}_{1}^{0}\Big)\hat{\Sigma}_{2}^{0-1}|-\log|\hat{\Sigma}_{1}^{0}\hat{\Sigma}_{2}^{0-1}|\nonumber\\
 & = & \min\limits _{k\in
 D_{\eta,M},k<k_{0}}\frac{T-k}{k_{0}-k}tr\left(\frac{k-k_{0}}{T-k}I+\frac{k_{0}-k}{T-k}\hat{\Sigma}_{1}^{0}\hat{\Sigma}_{2}^{0-1}+\frac{k_{0}-k}{T-k}\Big(\frac{1}{k_{0}-k}\sum\limits
 _{t=k+1}^{k_{0}}\hat{g}_{t}\hat{g}_{t}^{'}-\hat{\Sigma}_{1}^{0}\Big)\hat{\Sigma}_{2}^{0-1}\right)-\log|\hat{\Sigma}_{1}^{0}\hat{\Sigma}_{2}^{0-1}|+o_{p}(1)\nonumber\\
 & = & O_{p}(1)-\log|\hat{\Sigma}_{1}^{0}|+\log|\hat{\Sigma}_{2}^{0}|\to\infty\ \mathrm{at\ the\ rate}\ \log T\end{eqnarray}
where the second line is based on the fact that $\hat{\Sigma}_{2}=\frac{1}{T-k}\sum_{t=k+1}^{k_{0}}\hat{g}_{t}\hat{g}_{t}^{\prime}+\frac{T-k_{0}}{T-k}\hat{\Sigma}_{2}^{0}$,
the third line follows from the fact that $(k_{0}-k)/T\to0$ through the
consistency of $\hat{\tau}$ and the boundedness of $\frac{1}{k_{0}-k}\sum\limits _{t=k+1}^{k_{0}}\hat{g}_{t}\hat{g}_{t}^{'}-\hat{\Sigma}_{1}^{0}$
by \eqref{eq:gg-Sigmahat1} and \eqref{eq:gg HR}, and the divergence
rate in the last line follows from the fact that $-\log|\hat{\Sigma}_{1}^{0}|\ge\log(c_{1}T)$
for some $c_{1}>0$ by proposition \ref{low_bound} under the assumption
$N\propto T$ and the fact that $\log|\hat{\Sigma}_{2}^{0}|=O_{p}(1)$
because $\hat{\Sigma}_{2}^{0}\to_{p}\Sigma_{2}$ is positive definite.

\item [(ii).] When $\Sigma_{2}$ is singular
and $\Sigma_{1}$ is either singular or positive definite, we have
\begin{eqnarray}\label{consistent_proof_sigma2}
 &  & \min\limits _{k\in D_{\eta,M},k<k_{0}}\frac{T-k}{k_{0}-k}\log|\hat{\Sigma}_{2}\hat{\Sigma}_{2}^{0-1}|-\log|\hat{\Sigma}_{1}^{0}\hat{\Sigma}_{2}^{0-1}|\nonumber\\
 & = & \min\limits _{k\in
 D_{\eta,M},k<k_{0}}\frac{T-k}{k_{0}-k}\log|\frac{|\hat{\Sigma}_{2}|-|\hat{\Sigma}_{2}^{0}|}{|\hat{\Sigma}_{2}^{0}|}+1|+\log|\hat{\Sigma}_{2}^{0}|-\log|\hat{\Sigma}_{1}^{0}|\nonumber\\
 & \geq & \min\limits _{k\in D_{\eta,M},k<k_{0}}\frac{T-k}{k_{0}-k}\log(c(k_{0}-k)+1)+\underbrace{\log|\hat{\Sigma}_{2}^{0}|-\log|\hat{\Sigma}_{1}^{0}|}_{O_{p}(\log T)}\nonumber\\
 & \to & \infty,\end{eqnarray}
where the inequality in the third line holds because
of Lemma \ref{differ2}, the $O_{p}(\log T)$ term in the third line
follows from proposition \ref{low_bound}, and the divergence in the last
line evidently holds when $k_{0}-k\to\infty$ and $(k_{0}-k)/T\to0$, because $\frac{T-k}{k_{0}-k}\frac{\log(k_{0}-k)}{\log(T)}>\frac{T-k}{k_{0}-k}/\log(\frac{T}{k_{0}-k})\rightarrow\infty$.
\end{itemize}
Thus, we have shown that the second and third terms dominate
the first term, and hence, \eqref{eq:Op(1) TBP} holds.

To indicate the consistency of $\hat{k}$, we will show that
for any $k<k_{0}$ and $k_{0}-k\leq M$, the objective function $V(k)=U(k)-U(k_{0})$
diverges to infinity as $N,T\rightarrow\infty$; thus, the minimum
$U(k)$ cannot be achieved at a point other than $k_{0}$. For the given $M$,
define $D_{M}=\{k:|k_{0}-k|\leq M\}$, then \begin{eqnarray}
\min\limits _{k\in D_{M},k<k_{0}}\frac{U(k)-U(k_{0})}{k_{0}-k} & = & \min\limits _{k\in
D_{M},k<k_{0}}\frac{k}{k_{0}-k}\log|\hat{\Sigma}_{1}\hat{\Sigma}_{1}^{0-1}|+\frac{T-k}{k_{0}-k}\log|\hat{\Sigma}_{2}\hat{\Sigma}_{2}^{0-1}|-\log|\hat{\Sigma}_{1}^{0}\hat{\Sigma}_{2}^{0-1}|.\label{objective}\end{eqnarray}
 When $\Sigma_{1}$ is of full rank, the first term in (\ref{objective})
is \begin{eqnarray*}
\min\limits _{k\in D_{M},k<k_{0}}\frac{k}{k_{0}-k}\log|\hat{\Sigma}_{1}\hat{\Sigma}_{1}^{0-1}| & = & \min\limits _{k\in
D_{M},k<k_{0}}\frac{k}{k_{0}-k}\log\left|(\hat{\Sigma}_{1}-\hat{\Sigma}_{1}^{0})\hat{\Sigma}_{1}^{0-1}+I\right|\\
 & = & \min\limits _{k\in D_{M},k<k_{0}}\frac{k}{k_{0}-k}\log\left|\left(\frac{k_{0}-k}{kk_{0}}\sum\limits _{t=1}^{k}(\xi_{t}+\zeta_{t})-\frac{1}{k_{0}}\sum\limits
 _{t=k+1}^{k_{0}}(\xi_{t}+\zeta_{t})\right)\hat{\Sigma}_{1}^{0-1}+I\right|\\
 & = & \min\limits _{k\in D_{M},k<k_{0}}tr\left(\frac{1}{k_{0}}\sum\limits _{t=1}^{k}(\xi_{t}+\zeta_{t})-\frac{k}{k_{0}(k_{0}-k)}\sum\limits
 _{t=k+1}^{k_{0}}(\xi_{t}+\zeta_{t})\right)\hat{\Sigma}_{1}^{0-1}+o_{p}(1)\\
 & = & O_{p}(1).\end{eqnarray*}
 Similar to (\ref{Sigma1_singular}), when $\Sigma_{1}$ is singular,
the first term in (\ref{objective}) is \begin{eqnarray*}
\min\limits _{k\in D_{M},k<k_{0}}\frac{k}{k_{0}-k}\log|\hat{\Sigma}_{1}\hat{\Sigma}_{1}^{0-1}|=\min\limits _{k\in
D_{M},k<k_{0}}\frac{k}{k_{0}-k}\log(\frac{|\hat{\Sigma}_{1}|-|\hat{\Sigma}_{1}^{0-1}|}{|\hat{\Sigma}_{1}^{0-1}|}+1)=O_{p}(1).\end{eqnarray*}

The second and third terms are discussed below.
\begin{itemize}

\item [(i).] When $\Sigma_{1}$ is a singular matrix and $\Sigma_{2}$
is a positive matrix, \begin{eqnarray*}
\frac{T-k}{k_{0}-k}\log|\hat{\Sigma}_{2}\hat{\Sigma}_{2}^{0-1}|-\log|\hat{\Sigma}_{1}^{0}\hat{\Sigma}_{2}^{0-1}|=O_{p}(1)+\log(T)\rightarrow\infty,\end{eqnarray*}
 where $\frac{T-k}{k_{0}-k}\log|\hat{\Sigma}_{2}\hat{\Sigma}_{2}^{0-1}|=O_{p}(1)$
is similar to (\ref{Sigma2differ}).

\item [(ii).] When $\Sigma_{2}$ is singular
and $\Sigma_{1}$ is either singular or positive definite, similar to (\ref{consistent_proof_sigma2}), we have
\begin{eqnarray*}
 &  & \min\limits _{k\in D_{\eta,M},k<k_{0}}\frac{T-k}{k_{0}-k}\log|\hat{\Sigma}_{2}\hat{\Sigma}_{2}^{0-1}|-\log|\hat{\Sigma}_{1}^{0}\hat{\Sigma}_{2}^{0-1}|\\
 & = & \min\limits _{k\in D_{\eta,M},k<k_{0}}\frac{T-k}{k_{0}-k}\log|\frac{|\hat{\Sigma}_{2}|-|\hat{\Sigma}_{2}^{0}|}{|\hat{\Sigma}_{2}^{0}|}+1|+\log|\hat{\Sigma}_{2}^{0}|-\log|\hat{\Sigma}_{1}^{0}|\\
 & \geq & \min\limits _{k\in D_{\eta,M},k<k_{0}}\frac{T-k}{k_{0}-k}\log(c(k_{0}-k)+1)+\underbrace{\log|\hat{\Sigma}_{2}^{0}|-\log|\hat{\Sigma}_{1}^{0}|}_{O_{p}(\log T)}\\
 & \to & \infty,\end{eqnarray*}
where the inequality in the third line holds because
of Lemma \ref{differ2}, the $O_{p}(\log T)$ term in the third line
follows from proposition \ref{low_bound}, and the divergence in the last
line evidently holds when $k_{0}-k$ is bounded, because $c(k_{0}-k)+1>1$ and by the same argument in (\ref{consistent_proof_sigma2}).
\end{itemize}

In summary, we can determine $U(k)\rightarrow\infty$ when $k<k_{0}$ and
$k_{0}-k<M$ as $N,T\rightarrow\infty$. Thus, we prove the consistency
of $\hat{k}$. $\Box$

\end{document}